\documentclass[11pt]{article}
\usepackage[affil-it]{authblk}

\usepackage{color}
\usepackage{amsmath,amsfonts,amssymb,mathrsfs,amsthm}
\usepackage{graphicx}
\usepackage{listings}
\usepackage{enumerate,enumitem}
\usepackage{mathtools}
\usepackage[a4paper, total={6in, 8in}]{geometry}
\usepackage{cite}
\usepackage[colorlinks,linkcolor=blue,citecolor=blue]{hyperref}
\usepackage{empheq}
\usepackage{pdfpages}
\usepackage[titletoc]{appendix}
\usepackage[utf8]{inputenc}
\usepackage{xspace}
\numberwithin{equation}{section}

\newcommand{\Eqref}[1]{Eq.~\eqref{#1}}
\newcommand{\Eqsref}[1]{Eqs.~\eqref{#1}}

\newcommand{\Figref}[1]{Fig.~\ref{#1}}
\newcommand{\Sectionref}[1]{Section~\ref{#1}}
\newcommand{\subfig}[2]{ Fig.~\hyperref[#1]{\ref{#1}#2}}

\newcounter{mnotecount}[section]
\let\oldmarginpar\marginpar
\renewcommand\marginpar[1]{\-\oldmarginpar[\raggedleft\footnotesize #1]%
	{\raggedright\footnotesize #1}}

\title{Turing instability and pattern formation on directed networks}
\author{J. Ritchie\footnote{Email: jritchie@maths.otago.ac.nz}}
\affil{Department of Mathematics and Statistics, University of Otago, New Zealand.}
\date{\today}

\begin{document}
	\maketitle
	
	
	\begin{abstract}
		Pattern formation, arising from systems of autonomous reaction-diffusion equations, on networks has become a common topic of study in the scientific literature. In this work we focus primarily on directed networks. Although some work prior has been done to understand how patterns arise on directed networks, these works have restricted their attentions to networks for whom the Laplacian matrix (corresponding to the network) is diagonalizable. Here, we address the question ``how does one detect pattern formation if the Laplacian matrix is \emph{not} diagonalizable?'' To this end, we find it is useful to also address the related problem of pattern formation arising from systems of reaction-diffusion equations with non-local (global) reaction kinetics. These results are then generalized to include non-autonomous systems as well as temporal networks, i.e., networks whose topology is allowed to change in time. 
	\end{abstract}
	\section{Introduction}
	Pattern formation occurs when a spatially homogeneous solution is perturbed in a spatially inhomogeneous way leading to finitely many of the spatial modes, corresponding the Laplacian operator, destabilizing. The resulting solution evolves into a spatially inhomogeneous or `patterned' state. This is the so called \emph{Turing instability}. Pattern formation via the Turing mechanism was first introduced by Alan Turing in \cite{Turing:1952}, to provide a mathematical description of morphogenesis. In his work, Turning considered an autonomous system of reaction-diffusion equations. That is, a system of partial differential equations (PDE's) that do not possess an explicit dependence on their time or space variables. Since then, such systems have become a popular tool for understanding patterning in both biological and chemical systems \cite{Tompkins:2014,Ouyang:1991,Lengyel:1992,Maini:2006,Seul:1995,Maini:1997,Baker:2008,Marcon:2012,VanGorder:2020_2}.  
	
	Although initially introduced for autonomous PDE-systems on continuous manifolds the Turing mechanism has also been applied to networks \cite{Kouvaris:2015,Asllani:2014_2,Ermentrout:1998,PastorSatorras:2010,Padirac:2013,Kouvaris:2016,McCullen:2016}. Recall that a \emph{network} is a collection of nodes, edges (that connect the nodes), and edge weights (that convey information about ``how well'' two nodes are connected). Throughout this work we often imagine that one has a known network and the corresponding adjacency matrix $A$, which contains all of the edge weights. The components of $A$ are $A_{ij}$, where $A_{ij}$ is the edge-weight connecting node $i$ to node $j$. If $A_{ij}=0$ then the two nodes are not connected. Given such a matrix, the corresponding Laplacian can be defined. Here, one studies systems of ordinary differential equations (ODE's), instead of PDE's. If the adjacency matrix is symmetric (i.e., $A_{ij}=A_{ji}$) then the underlying graph is referred to as an \emph{undirected network}. Works about pattern forming systems on networks often restrict their attention to undirected networks. This restriction is made for practical purposes as it ensures the Laplacian is symmetric, which in turn means that there exists a coordinate system such that the eigenvectors of the Laplacian define a complete orthonormal basis and hence the autonomous ODE system is symmerizable. In this setting, pattern formation is well understood \cite{Kouvaris:2015,Asllani:2014_2,Ermentrout:1998,PastorSatorras:2010,Padirac:2013,Kouvaris:2016,McCullen:2016}. One particularly interesting property of pattern formation on an undirected networks is that the underlying topology can influence the emergence and suppression of patterning \cite{Mimar:2019,Asllani:2016}.
	
	In some applications one may wish to consider networks for which the adjacency matrix $A$, is \emph{not} symmetric (i.e., $A_{ij}\ne A_{ji}$). If the adjacency matrix is not symmetric then the underlying graph is referred to as a \emph{directed network}. For this type of graph, edges are allowed to have direction. Directed networks themselves have many applications in the literature \cite{DiPatti:2017,Dhingra:2019,Park:2006,Son:2012}, including pattern formation \cite{Asllani:2014,Blanchini:2018}. Of particular interest to this work is \cite{Asllani:2014}. In their work \cite{Asllani:2014}, Asllani et. al. perform a linear instability analysis (which is how pattern forming systems are ``detected'') on a class of directed networks. For this they restricted their attentions to the class of directed networks for which the corresponding Laplacian is diagonalizable. In this work here we always refer to networks with this property as \emph{diagonalizable networks}. Similarly, if a network has a corresponding Laplacian matrix that is non-diagonalizable then we say that the network is \emph{non-diagonalizable}. By restricting to diagonalizable networks, Asllani et. al. \cite{Asllani:2014} are able to ensure that there exists a coordinate system such that the eigenvectors of the Laplacian form a complete orthonormal basis. Given this they were able to perform the traditional Turing analysis with only a small adjustment to allow for the now complex eigenvalues. However, the analysis presented in \cite{Asllani:2014} cannot be applied to non-diagonalizable networks. To the best of the authors knowledge there is currently no method for detecting pattern formation on non-diagonalizable networks. 
	 
	Pattern formation is also known to occur in systems of \emph{non-autonomous} equations. That is ODE (or PDE) systems of reaction diffusion equations that depend explicitly on their time or space coordinates. In order to determine whether or not patterning occurs one must first perform a linear instability analysis. It is tempting to proceed as in the autonomous setting and simply calculate the (now time dependent) eigenvalues of the linearised system. However, it turns out that in the non-autonomous setting one cannot determine instability from the eigenvalues. Indeed there exist counter examples in the literature \cite{Wu:1974,Knobloch:1992,Josic:2008,Mierczynski:2017}. There have been works addressing this issue \cite{Holme:2012,Ward:2014,Li:2017}. Perhaps the most successful of these was presented by Van Gorder in \cite{VanGorder:2020_3} (see also \cite{VanGorder:Evolvingmanifolds}). In \cite{VanGorder:2020_3} Van Gorder uses a second order comparison principal to determine whether or not an instability occurs for a two-species system of non-autonomous reaction diffusion equations on manifolds. In \cite{VanGorder:2020} Van Gorder used his comparison principal to study pattern formation on \emph{undirected temporal networks}, i.e., undirected networks whose topologies are allowed to change in time. To the best of the authors knowledge, \emph{directed temporal networks} have not been studied in the context of pattern formation before.   
	
	The approach used by Van Gorder in \cite{VanGorder:2020_3,VanGorder:Evolvingmanifolds,VanGorder:2020} has the drawback that it can only be applied to two-species systems of first order reaction diffusion equations. Moreover, in the case of a continuum, the considered systems are not allowed to depend on their spatial coordinates. These issues were addressed, again by Van Gorder, in \cite{VanGorder:2021}. Although networks themselves do not possess any `spatial' coordinates one can think of an analogous scenario in a system of reaction-diffusion equations with \emph{global reaction kinetics}. That is, systems whose reaction kinetics change at each node. Systems (of ODE's) with global (non-local) reaction kinetics have been studied before on networks \cite{Ding:2021,Carletti:2022}. To the best of the authors knowledge, systems with global reaction kinetics have \emph{not} been studied before in the context of pattern formation on directed networks. However, such systems have been considered in the context of pattern formation on \emph{hypergraphs} \cite{Carletti:2022}. Note that global reaction kinetics are not only allowed to change functional form at each node, but they are also allowed depend on the unknowns (that you are solving the ODE's for) value at any \emph{other} node in the network.   
	
	In this work we investigate the following questions: How does one perform a linear instability analysis, to detect pattern formation, in systems of reaction diffusion equations that are: (1) defined over a non-diagonalizable network?; (2) has terms that explicitly depend on time and/or are defined over a directed temporal network?; or (3) have global reaction functions? We address these questions in two parts. 
	
	First, we establish a generic method for analysing linear instability in systems of \emph{autonomous} reaction-diffusion equations on static directed networks with (possibly) global reaction kinetics. To do this we \emph{split} the adjacency matrix into its symmetric and anti-symmetric components. We then used the symmetrized adjacency matrix, and its corresponding Laplacian, to find a complete orthonormal basis. This basis is then used to facilitate our linear instability analysis.
	
	Second, we investigate non-autonomous systems of reaction diffusion equations defined over directed temporal networks. For this, we do not perform a basis decomposition. Instead we solve the linearised equations directly.

	This paper is outlined as follows: In \Sectionref{Sec:Pattern_formation_on_networks} we discuss techniques for detecting patterning on networks. To this end, in \Sectionref{SubSec:Setting_up_the_problem_1}, we first give an overview of the standard Turing analysis. In \Sectionref{SubSec:Linear_instability_analysis_for_undirected_networks} we then generalize this approach for systems of autonomous reaction-diffusion equations on networks with (or without) global reaction kinetics. Then, in \Sectionref{Sec:Linear_instability_analysis_on_temporal_networks}, we discuss how a linear instability analysis could be performed on networks, for systems of \emph{non-autonomous} reaction-diffusion equations. In \Sectionref{Sec:Numerical_examples_of_pattern_formation_on_static_directed_networks} we provide numerical examples of our theory for autonomous reaction-diffusion equations on static networks, and in \Sectionref{Sec:Numerical_examples_of_pattern_formation_of_temporal_directed_networks} we provide numerical examples of our theory for non-autonomous reaction-diffusion equations on temporal networks. Finally, in \Sectionref{Sec:Conclusions} we summarize our findings.

	\section{Pattern formation on networks}
	\label{Sec:Pattern_formation_on_networks}
	\subsection{Turing instability on undirected networks with autonomous reaction diffusion equations}
	\label{SubSec:Setting_up_the_problem_1}
	We begin by first reviewing the Turing instability on undirected networks. To this end we consider the two-species reaction-diffusion system
	\begin{align}
		\frac{d u_{i}}{d t} =\, d_{1}\sum_{j=1}^{n}A_{ij}\left( u_{j} - u_{i} \right) + f\left(u_{i},v_{i}\right),
		\quad
		\frac{d v_{i}}{d t} =\, d_{2}\sum_{j=1}^{n}A_{ij}\left( v_{j} - v_{i} \right) + g\left(u_{i},v_{i}\right),
		\label{graphsystem_heterogenous_1}
	\end{align}
	defined over a network, with $n$ nodes, corresponding to the symmetric adjacency matrix $A_{ij}$. Here $d_{1}$ and $d_2$ are global diffusion rates and the functions $f,g$ are assumed to have continuous derivatives in all arguments. The elements of $A$ measure the connectedness of a pair of nodes $i$ and $j$, and we assume these can take any values $0\le A_{ij} \le 1$, with a value equal to $0$ denoting two nodes which are not connected, while a fractional value strictly between $0$ and $1$ means that the local rate of diffusion between nodes is less than the maximal rate possible over the network for that species. In this section here we assume that the underlying network is \emph{undirected}, in which case we have that the adjacency matrix is symmetric (i.e., $A_{ij}=A_{ji}$). We now define the Laplacian matrix, ${L}$, associated with ${A}$ in the standard way, i.e, ${L}_{ij} = \delta_{ij}\sum_{k=1}^{n}{A}_{ik}-{A}_{ij}$, where $\delta_{kj}$ denotes the Kronecker delta. For the special case in which the adjacency matrix $A$ is symmetric (i.e, $A_{ij}=A_{ji}$) then the spectral properties of ${L}$ are well understood: A Laplacian matrix ${L}$ admits a collection of $n$ distinct eigenvectors, say $\mathbf{\Phi}_1, \mathbf{\Phi}_2, \dots , \mathbf{\Phi}_n \in \mathbb{R}^n$. To each eigenvector $\mathbf{\Phi}_\ell$ there exists an eigenvalue $\rho_\ell \geq 0$ such that ${L}\mathbf{\Phi}_\ell = \rho_\ell \mathbf{\Phi}_\ell$. Furthermore, we have the eigenvalue $\rho_1 =0$ and the corresponding eigenvector $\mathbf{\Phi}_1 = (1,1,\dots, 1)^T$. This is the analogue of the Neumann spectral parameter $\rho_0 = 0$ and Neumann eigenfunction $\Psi_0(\mathbf{x})=1$ in the case of a continuum domain. However, when the adjacency matrix $A$ is \emph{not} symmetric then the spectral properties are \emph{not} well-understood. In particular, there is no guarantee that the eigenvectors $\mathbf{\Phi}_\ell$ form a basis, nor even that there are $n$ of them. We shall return to this point shortly. 
	
	We now assume that there exists solutions $u_i=u_\star$ and $v_i=v_\star$ for all $i=1,2,\dots,n$ to the algebraic system ${f}\left(u_{\star},v_{\star}\right)={g}\left(u_{\star},v_{\star}\right)=0$, which is a constant solution of \Eqref{graphsystem_heterogenous_1}. A diffusive instability, such as the Turing instability, occurs when a perturbation destabilizes such a solution, resulting in a new, spatially inhomogeneous, solution. In order to study instability of such a solution, we consider perturbations of the form $\mathbf{u} = (u_\star,\dots,u_\star)^T +\epsilon \mathbf{U}$ and $\mathbf{v} = (v_\star,\dots,v_\star)^T +\epsilon \mathbf{V}$ where $\mathbf{U}=[U_1,\dots,U_n]^T \in \mathbb{R}^{n}$ and $\mathbf{V}=[V_1,\dots,V_n]^T \in \mathbb{R}^{n}$. Putting these expansions into \Eqref{graphsystem_heterogenous_1} and retaining $\mathcal{O}(\epsilon)$ terms we obtain
	\begin{align}
		\frac{d \mathbf{U} }{d t} = d_{11}{L}\mathbf{U} + d_{12}{L}\mathbf{V} + J_{11}\mathbf{U} + J_{12}\mathbf{V},
		\\
		\frac{d \mathbf{V} }{d t} = d_{21}{L}\mathbf{U} + d_{22}{L}\mathbf{V} + J_{21}\mathbf{U} + J_{22}\mathbf{V},
		\label{Eq:Diagonalizable_Linear}
	\end{align}
	where 
	\begin{align}
		J_{11}=\frac{\partial f}{\partial u}(u_\star,v_\star),
		\quad
		J_{12}=\frac{\partial f}{\partial v}(u_\star,v_\star),
		\quad
		J_{21}=\frac{\partial g}{\partial u}(u_\star,v_\star),
		\quad
		J_{22}=\frac{\partial g}{\partial v}(u_\star,v_\star).
		\label{Eq:JacobiDef}
	\end{align}
	In the case of an undirected network the eigenvectors of $L$ form an orthonormal basis i.e., ${\mathbf{\Phi}}_{\ell}^{T}{\mathbf{\Phi}}_{k}=\delta_{\ell k}$. In order to carry forward our analysis we we consider the following eigenfunction expansions 
	\begin{align}
		\mathbf{U} = \sum_{\ell=1}^{n} A_\ell (t) {\mathbf{\Phi}}_\ell,
		\quad
		\mathbf{V}  = \sum_{\ell=1}^{n} B_\ell (t) {\mathbf{\Phi}}_\ell,
		\label{Eq:EigenDecomp_Special}
	\end{align}
	where the $A_\ell$ and $B_\ell$ are unknown functions. Then, for each $k=1,\dots,n$, we obtain the following system of equations
	\begin{align}
		\frac{d}{dt}\left(
		\begin{array}{c}
			A_{k}(t) \\
			B_{k}(t)
		\end{array}
		\right)=
		M_{k}\left(
		\begin{array}{c}
			A_{k}(t) \\
			B_{k}(t)
		\end{array}
		\right),
		\quad
		M_{k}=J-
		\rho_{k}D,
		\label{Eq:SimplifiedProblem}
	\end{align}
	where
	\begin{align}
		J=\left(
		\begin{array}{cc}
			J_{11} & J_{12} \\
			J_{21} & J_{22}
		\end{array}
		\right),
		\quad
		D = \left(
		\begin{array}{cc}
			d_{1} & 0 \\
			0 & d_{2}
		\end{array}
		\right).
	\end{align}
	It is well known that the stability of the solutions $A_{k}, B_{k}$ depends on the eigenvalues of $M_{k}$. In particular we have that the solutions $A_{k}, B_{k}$ are stable if and only if all of the eigenvalues of $M_k$ have negative real part. If, on the other hand, at least one of the eigenvalues of $M_{k}$ have positive real part we find that the solutions $A_{k}, B_{k}$ are unstable. The eigenvalues of $M_{k}$ are named $\lambda_{k}$ and are determined as solutions of the polynomial equation 
	\begin{align}
		\lambda_{k}^{2} + \left( \text{tr}(D)\rho_{k}-\text{tr}(J) \right)\lambda_{k} + \det(D)\rho_{k}^{2}-(J_{11}d_{2}+J_{22}d_{1})\rho_{k}+\det(J)=0.
	\end{align}
	Using the Ruth-Huruwitz stability criterion we find that $\text{Re}(\lambda_{k})<0$ if and only if 
	\begin{align}
		\text{tr}(D)\rho_{k}-\text{tr}(J)>0,
		\label{Eq:Unstable_Undirected_1}
		\\
		\det(D)\rho_{k}^{2}-(J_{11}d_{2}+J_{22}d_{1})\rho_{k}+\det(J)>0.
		\label{Eq:Unstable_Undirected_2}
	\end{align}
	Recall now that we require $u_\star,v_{\star}$ to be stable solutions of. This means that the solutions should be stable in the absence of diffusion. i.e., when $\rho_{\ell}=0$. We therefore impose the following restrictions:
	\begin{align}
		-\text{tr}(J)>0,
		\quad
		\det(J)>0.
	\end{align}
	These are the standard stability restrictions imposed when studying the Turing instability on undirected graphs. Recall now that, by assumption, we have $\text{tr}(D)>0$. It follows then that it is \emph{not} possible to violate \Eqref{Eq:Unstable_Undirected_1} at any point throughout the evolution. However, it \emph{is} possible to violate \Eqref{Eq:Unstable_Undirected_2}. This occurs if and only if
	\begin{align}
		J_{11}d_{2}+J_{22}d_{1}>2\sqrt{\det(D)\det(J)}.
	\end{align}
	For such
	a case, the spatially uniform state is unstable under a	perturbation with the $\ell$th mode corresponding to $\rho_\ell$ provided that $0 < \rho_- < \rho_\ell < \rho_+ < \infty$, where
	\begin{align}
		\begin{split}
			\rho_\pm=\frac{J_{11}d_{2}+J_{22}d_{1}}{2\det(D)}
			\pm\frac{\sqrt{(J_{11}d_{2}+J_{22}d_{1})^2-4\det(D)\det(J)}}{2\det(D)}.
		\end{split}
	\end{align}

	\subsection{Generic instability mechanism on static networks}
	\label{SubSec:Linear_instability_analysis_for_undirected_networks}
	The instability analysis presented in \Sectionref{SubSec:Setting_up_the_problem_1} provides a convenient framework that can be utilized to study instabilities on diagonalizable networks. This approach can even be extended to include more than two unknowns. However, this approach quickly becomes cumbersome as the number of unknowns is increased. Moreover, it is unclear how such an approach can be extended to include non-diagonalizable networks. The goal of this subsection here is to present a method that (1) can be used to determine instability and (2) reduces to the standard Turing analysis in the case of an undirected network with local reaction kinetics. In \Sectionref{Sec:General_method} we present our instability analysis for systems (of ODEs) with $m$-unknowns defined over an $n$-node network. This level of generality means that the method we present here is significantly more complicated than the analysis presented in \Sectionref{SubSec:Setting_up_the_problem_1}. Here, care should be taken with the notation. In \Sectionref{Sec:Reductions} we consider various reductions, of the analysis presented in \Sectionref{Sec:General_method}, to less general scenarios. In one of these reductions we demonstrate that, under suitable conditions, the approach we consider here reduces to the standard Turing analysis. 
	\subsubsection{General method}
	\label{Sec:General_method}
	We now consider systems (of ODEs) with $m$-unknowns defined over an $n$-node network. As mentioned above, the notation we use here is more complicated than what was presented in \Sectionref{SubSec:Setting_up_the_problem_1}. In this subsection all unknowns come equipped with two indices; one Greek and one Latin. Greek indices are used to indicate which unknown is being considered, and therefore run from $1$ to $m$, while lower-case-Latin indices indicate which node the unknown is defined on, and therefore run from $1$ to $n$. 
	
	Let us now consider the system 
	\begin{align}
		\frac{d u_{\alpha,i}}{d t} =\, & \sum_{\beta=1}^{m} d_{\alpha\beta}\sum_{j=1}^{n}A_{ij}\left( u_{\beta,j} - u_{\beta,i} \right)+f_{\alpha,i}\left(\mathbf{u}\right),
		\quad
		\alpha=1,\dots,m,
		\quad
		i=1,\dots,n,
		\label{Eq:graphsystem_adjacency_1}
	\end{align}
	where $\mathbf{u}=(u_{1,1},\dots,u_{1,n},u_{2,1},\dots,u_{2,n},\dots,u_{m,1},\dots,u_{m,n})^T$, and
	\begin{align}
		f_{\alpha,i}\left(\mathbf{u}\right)=f_{\alpha,i}(u_{1,1},\dots,u_{1,n},u_{2,1},\dots,u_{2,n},\dots,u_{m,1},\dots,u_{m,n}),
	\end{align} 
	are global reaction kinetics. If the functions $f_{\alpha,i}\left(\mathbf{u}\right)$ describe local reaction kinetics then we must have $f_{\alpha,i}\left(\mathbf{u}\right)=f_{\alpha}(u_{1,i},\dots,u_{m,i})$, for some functions $f_{\alpha}:\mathbb{R}^m\rightarrow \mathbb{R}$. For physical applications (of global reaction kinetics) one might expect the functions $f_{\alpha,i}\left(\mathbf{u}\right)$ to retain their functional form, and vary only in their reaction rates, i.e., $f_{\alpha,i}\left(\mathbf{u}\right)=\beta_{i}f(\mathbf{u}_i)$ with $\mathbf{u}_i=(u_{1,i},\dots,u_{m,i})^T$, and where $\beta_{i}=$constant for each $i=1,\dots,n$. In this scenario one imagines that reaction rates $\beta_{i}$ change at each node to account for physical effects, such as temperature. Although this is the type of example we have in mind when discussing global reaction kinetics we nevertheless present our analysis for more general functions $f_{\alpha,i}\left(\mathbf{u}\right)$, as it does not significantly change the methodology or indeed the results.   
	
	Let us now \emph{split} the adjacency matrix $A$ into its symmetric $\tilde{A}$ and antisymmetric $\hat{A}$ pieces. The components of which are defined as
	\begin{align}
		\tilde{A}_{ij}=\frac{1}{2}\left( A_{ij} + A_{ji} \right),
		\quad
		\hat{A}_{ij}=\frac{1}{2}\left( A_{ij} - A_{ji} \right).
	\end{align}
	The adjacency matrix $A$ is easily reconstructed as $A=\tilde{A}+\hat{A}$. The matrix $\tilde{A}_{ij}$ can be thought of as the adjacency matrix corresponding to an \emph{undirected} network with $n$ nodes. Taking this view point allows us to define the Laplacian matrix $\tilde{L}$, associated with $\tilde{A}$, as $\tilde{L}_{ij} = \delta_{ij}\sum_{k=1}^{n}\tilde{A}_{ik}-\tilde{A}_{ij}$. Using these definitions we write the reaction-diffusion system \Eqref{Eq:graphsystem_adjacency_1} as,
	\begin{align}
		\frac{d u_{\alpha,i}}{d t} =\, & -\sum_{\beta=1}^{m} d_{\alpha\beta}\sum_{j=1}^{n}\tilde{L}_{ij}u_{\beta,j} + \hat{f}_{\alpha,i}\left(\mathbf{u}\right),
		\quad
		\alpha=1,\dots,m,
		\quad
		i=1,\dots,n,
		\label{Eq:graphsystem_heterogenous}
	\end{align}
	where 
	\begin{align}
		\hat{f}_{\alpha,i}\left(\mathbf{u}\right)=\sum_{\beta=1}^{m} d_{\alpha\beta}\sum_{j=1}^{n}\hat{A}_{ij}\left( u_{\beta,j} - u_{\beta,i} \right) + {f}_{\alpha,i}\left(\mathbf{u}\right).
		\label{Eq:ModifiedReactionKinetics}
	\end{align}
	From here we see that one may think of the system \Eqref{Eq:graphsystem_adjacency_1}, defined on an directed network, as being equivalent to a problem defined on a \emph{undirected} network, with `modified' reaction kinetics. Similar to what was discussed in \Sectionref{SubSec:Setting_up_the_problem_1}, the spectral properties of $\tilde{L}$ are well understood, and hence this is the appropriate setting for our instability analysis.
	
	Suppose now that there exist constants $u_{\alpha,i}^\star,\alpha=1,\dots,m,i=1,\dots,n$ which are solutions of the algebraic system
	\begin{align}
		-\sum_{\beta=1}^{m} d_{\alpha\beta}\sum_{j=1}^{n}\tilde{L}_{ij}u_{\beta,j} + \hat{f}_{\alpha,i}\left(\mathbf{u}\right)=0,
		\quad
		\alpha=1,\dots,m,
		\quad
		i=1,\dots,n.
		\label{Eq:Algebraic_2}
	\end{align}
	Then $u_{\alpha,i}=u_{\alpha,i}^\star$ is a constant solution of \Eqref{Eq:graphsystem_adjacency_1}. In order to study the stability of such a solution, we consider perturbations of the form 
	\begin{align}
		u_{\alpha,i} = u_{\alpha,i}^{\star} + \epsilon U_{\alpha,i}, 
	\end{align} 
	where $U_{\alpha,i},\alpha=1,\dots,m,i=1,\dots,n$ and $\epsilon\ll 1$ is a small parameter. Putting these expansions into \Eqref{Eq:graphsystem_heterogenous}, expansing in $\epsilon$, and retaining $\mathcal{O}(\epsilon)$ terms we obtain the linearised equations
	\begin{align}
		\frac{d\mathbf{U}_{\alpha}}{dt}=-\sum_{\beta=1}^{m}d_{\alpha\beta}\tilde{L}\mathbf{U}_{\beta} + \sum_{\beta=1}^{m}J_{\alpha\beta}\mathbf{U}_{\beta},
		\quad
		\alpha=1,\dots,m,
	\end{align}
	where $\mathbf{U}_{\alpha}=(U_{\alpha,1},\dots,U_{\alpha,n})^T$ and where each $J_{\alpha\beta},\alpha,\beta=1,\dots,m$ is an $n\times n$ matrix with entries 
	\begin{align}
		(J_{\alpha\beta})_{ij}=\frac{\partial \hat{f}_{\alpha,i}}{\partial u_{\beta,j}}(\mathbf{u}^\star).
		\label{Eq:JDeff}
	\end{align}
	The eigenvectors of $\tilde{L}$ are $\mathbf\Phi_{\ell}$ and the corresponding eigenvalues are $\sigma_{\ell}$. Let $S$ be the matrix of eigenvectors $\mathbf\Phi_{\ell}$ and define 
	\begin{align}
		\mathbf{v}_\alpha=S\, \mathbf{U}_\alpha,
		\quad
		\alpha = 1,\dots,m.
	\end{align}
	Then, one obtains the following evolution equations for $\mathbf{v}_\alpha$:
	\begin{align}
		\frac{d\mathbf{v}_{\alpha}}{dt}=-\sum_{\beta=1}^{m}d_{\alpha\beta}P\mathbf{v}_{\beta} + \sum_{\beta=1}^{m}\hat{J}_{\alpha\beta}\mathbf{v}_{\beta},
		\quad
		\alpha=1,\dots,m,
	\end{align}
	where $P=S\,\tilde{L}S^{-1}$ is the diagonalization of $\tilde{L}$ and $\hat{J}_{\alpha\beta}=S\, {J}_{\alpha\beta}S^{-1}$. Before continuing it is useful to briefly discuss why we diagonalise here. In \Sectionref{SubSec:Setting_up_the_problem_1} we not did diagonalise, as the resulting expressions depended only on the eigenvalues (and not the eigen\emph{basis}) and as such one does ``need'' to diagonalise. However, similar to \cite{VanGorder:2021}, the expression that we obtain here \emph{does} explicitly depend on the eigenbasis. Let us now return to the matrix $P$. In this coordinate system, the eigenvectors of $P$ are $\mathbf{e}_{i},i=1,\dots,n$ with $(\mathbf{e}_{i})_j = \delta_{ij}$. Clearly, the collection of vectors $\mathbf{e}_{i}$ form an orthonormal basis over $\mathbb{R}^n$. Now, in order to carry our analysis forward we consider the following eigenvector decomposition: 
	\begin{align}
		\mathbf{v}_{\alpha} = \sum_{\ell=1}^{n} B_{\alpha,\ell} {\mathbf{e}}_\ell,
		\quad
		\alpha = 1, \dots, m,
	\end{align}
	where $B_{\alpha,\ell},\alpha=1,\dots,m,\ell=1,\dots,n$ are unknown functions of time. The resulting system of equations is
	\begin{align}
		\sum_{\ell=1}^{n}\left( \frac{d B_{\alpha,\ell} }{d t} + \sigma_\ell \sum_{\beta=1}^{m}d_{\alpha\beta}B_{\alpha,\ell} \right){\mathbf{e}}_{\ell} 
		=& \sum_{\ell=1}^{n}\sum_{\beta=1}^m \hat{J}_{\alpha\beta}B_{\beta,\ell}\mathbf{e}_\ell,
		\quad
		\alpha=1,\dots,m.
		\label{Eq:AB_Decomp_1}
	\end{align}
	To make further progress we note that, for each $\alpha,\beta=1,\dots,m$ and $\ell=1,\dots,n$ the quantity $\hat{J}_{\alpha\beta}\mathbf{e}_\ell$ is a vector and can therefore be written as 
	\begin{align}
		\hat{J}_{\alpha\beta}\mathbf{e}_\ell = \sum_{r=1}^{n} \gamma_{r,\ell,\alpha,\beta}\mathbf{e}_{r},
		\label{Eq:Gamma_Def}
	\end{align}
	where, for each $\alpha,\beta = 1,\dots,m$ and $k,\ell = 1,\dots,n$, we have $\gamma_{k,\ell,\alpha,\beta} = \mathbf{e}^T_{k}\hat{J}_{\alpha\beta}\mathbf{e}_{\ell}$.
	Moreover, for $k,\ell=1,\dots,n$ we define the $m\times m$ matrices $\Gamma_{k,\ell}$ and $D$ as $(\Gamma_{k,\ell})_{\alpha\beta}=\gamma_{k,\ell,\alpha,\beta}$ and $(D)_{\alpha\beta}=d_{\alpha\beta}$. Returning now to \Eqsref{Eq:AB_Decomp_1} we multiply by ${\mathbf{e}}_{k}^{T}$, and sum over $k$, to obtain 
	\begin{align}
		\frac{d \mathbf{B}_{k} }{d t} +\sigma_{k}D \mathbf{B}_{k}=\sum_{\ell=1}^n \Gamma_{k,\ell}\mathbf{B}_\ell,
		\quad
		k=1,\dots,n
		\label{Eq:FinalLinear_1}
	\end{align}
	where we have introduced the notation $\mathbf{B}_{k}=(B_{1,k},\dots,B_{m,k})^T$. Notice that, unlike in \Sectionref{SubSec:Setting_up_the_problem_1}, the modes have \emph{not} decoupled. Because of this it is, in general, not possible to derive an explicit inequality that determines stability.
	
	In order to understand when an instability occurs, we now define $\mathbf{w}=\left(\mathbf{B}_{1},\dots,\mathbf{B}_{n}\right)^{T}$ so that \Eqsref{Eq:FinalLinear_1} can be written in matrix form as
	\begin{align}
		\frac{d\mathbf{w}}{dt}= M\mathbf{w},
		\label{Eq:GeneralLinearEq}
	\end{align}
	where $M$ is an $(nm)\times(nm)$ constant matrix with each $n\times n$ element $M_{k\ell}$ an $m\times m$ matrix of the form $M_{k\ell}=-D \sigma_{k}\delta_{k\ell} + \Gamma_{k,\ell}$.
	
	In order to understand whether or not an instability occurs one must study the eigenvalues of $M$. Throughout this work we often represent the eigenvalues of $M$ as $\lambda_{\ell}$. If one of the eigenvalues has positive real part then an instability (such as the Turing instability) occurs, i.e., the steady state solution $u_{\alpha,i}^\star$ is unstable if $\max_{\ell}\text{Re}(\lambda_{\ell})>0$. This is our instability criterion. Although it is not, in general, possible to find the eigenvalues of $M$ explicitly, we note that they are easily calculated \emph{numerically} using programs such as \emph{Python, Matlab}, or \emph{Mathematica}.

	\subsubsection{Reductions}
	\label{Sec:Reductions}
	We now consider two special cases of \Sectionref{Sec:General_method}. The goal here is to show that the analysis presented in \Sectionref{Sec:General_method} reduces to the standard Turing analysis (see \Sectionref{SubSec:Linear_instability_analysis_for_undirected_networks}), under the appropriate assumptions.
	\paragraph{Reduction for local reaction kinetics.}
	Suppose now that the reaction kinetics in \Eqref{Eq:graphsystem_adjacency_1} are \emph{local}, i.e., $f_{\alpha,i}(\mathbf{u})=f_{\alpha}(\mathbf{u}_i)$ for all $\alpha=1,\dots,m$ and $i=1,\dots,n$ where $\mathbf{u}_i = (u_{1,i},\dots,u_{m,i})^T$. In this special case we find that the modified reaction kinetics (defined in \Eqref{Eq:ModifiedReactionKinetics}) are 
	\begin{align}
		\hat{f}_{\alpha,i}\left(\mathbf{u}\right)=\sum_{\beta=1}^{m} d_{\alpha\beta}\sum_{j=1}^{n}\hat{A}_{ij}\left( u_{\beta,j} - u_{\beta,i} \right) + {f}_{\alpha}\left(\mathbf{u}_{i}\right).
	\end{align}
	Moreover, we find that the matrices $J_{\alpha\beta}$ are
	\begin{align}
		(J_{\alpha\beta})_{ij} = -d_{\alpha\beta}\hat{L}_{ij} + \mathcal{J}_{\alpha\beta}\delta_{ij},
	\end{align}
	where we have defined $\hat{L}_{ij}$ as $\hat{L}_{ij} = \delta_{ij}\sum_{k=1}^{n}\hat{A}_{ik}-\hat{A}_{ij}$, and the \emph{constants} $\mathcal{J}_{\alpha\beta}$ (for $\alpha,\beta=1,\dots,m$) as
	\begin{align}
		\mathcal{J}_{\alpha\beta} = \frac{\partial \hat{f}_{\alpha}}{\partial u_{\beta,i}}(\mathbf{u}^{\star}).
	\end{align}
	Moreover, we find that $\hat{J}_{\alpha\beta} = -d_{\alpha\beta}S\, \hat{L}\, S^{-1} + \mathcal{J}_{\alpha\beta} I_{n}$, where $I_{n}$ is the $n\times n$ identity matrix. Recall that $S$ is the matrix of eigenvectors of the symmetrised Laplacian $\tilde{L}$. Using this particular form $\hat{J}_{\alpha\beta}$ we calculate the constants $\gamma_{k,\ell,\alpha,\beta}$ as
	\begin{align}
		\gamma_{k,\ell,\alpha,\beta} = -d_{\alpha\beta}\Omega_{k\ell} + \mathcal{J}_{\alpha\beta}\delta_{k\ell},
		\quad 
		\Omega_{k\ell} = \mathbf{e}_{k}^{T}S\, \hat{L}\, S^{-1}\mathbf{e}_{\ell}.
	\end{align} 
	The $m\times m$ matrices $\Gamma_{k,\ell}$ are therefore $\Gamma_{k,\ell}=\Omega_{k\ell} D_{\alpha\beta} + \mathcal{J}\delta_{k\ell}$, where $D$ and $\mathcal{J}$ are $m\times m$ matrices with entries $d_{\alpha\beta}$ and $\mathcal{J}_{\alpha\beta}$, respectively. Finally, we find that the $(nm)\times(nm)$ matrix $M$ (see \Eqref{Eq:GeneralLinearEq}) has elements $M_{k\ell}=\left( -D \sigma_{k} + \mathcal{J} \right)\delta_{k\ell} - D\Omega_{k\ell}$.

	\paragraph{Reduction to the standard Turing analysis.}
	Suppose now that, in addition to local reaction kinetics, the adjacency matrix $A_{ij}$ represents an \emph{undirected} network (so that $A_{ij}=A_{ji}$). Then, one expects that the analysis presented in \Sectionref{SubSec:Setting_up_the_problem_1} is applicable (although now one has $m$ equations, instead of just two). To demonstrate this we use the results of the previous paragraph with $\hat{A}_{ij}=\hat{L}_{ij}=\Omega_{ij}=0$. In this case we find that $M$ is a block diagonal matrix with each $n\times n$ block taking the form, $M_{k\ell}=\left( -D \sigma_{k} + \mathcal{J} \right)\delta_{k\ell}$. One can therefore determine the eigenvalues of $M$ by calculating the eigenvalues of $M_{kk}$ for each $k=1,\dots,n$. 
	
	This is equivalent to the standard Turing analysis. To explicitly demonstrate this we now set $m=2$. Then,
	\begin{align}
		M_{kk}=
		\left(
		\begin{array}{cc}
			J_{11} & J_{12} \\
			J_{21} & J_{22}
		\end{array}
		\right)
		-
		\sigma_{k}\left(
		\begin{array}{cc}
			d_{11} & d_{12} \\
			d_{21} & d_{22}
		\end{array}
		\right).
	\end{align}
	Compare this to \Eqref{Eq:SimplifiedProblem} and note that the eigenvalues of $L$ and $\tilde{L}$ are equal (i.e., $\sigma_{\ell}=\rho_{\ell}$) since $A_{ij}=\tilde{A}_{ij}$ (and hence $L_{ij}=\tilde{L}_{ij}$). This suggests that our approach here is one possible generalisation of the standard Turing analysis.

	\subsection{Linear instability analysis on temporal networks}
	\label{Sec:Linear_instability_analysis_on_temporal_networks}
	Now that we are now able to detect instability for a general class of autonomous reaction-diffusion equations, it is natural to wonder to what extent these results can be extended to include \emph{non-autonomous} systems. The purpose of this subsection is to address exactly this issue. To this end, we now consider the following system of non-autonomous reaction-diffusion equations: 
	\begin{align}
		\frac{d u_{\alpha,i}}{d t} =\, & \sum_{\beta=1}^{m} d_{\alpha\beta}(t)\sum_{j=1}^{n}A_{ij}(t)\left( u_{\beta,j} - u_{\beta,i} \right)+f_{\alpha,i}\left(\mathbf{u},t\right),
		\quad
		\alpha=1,\dots,m,
		\quad
		i=1,\dots,n.
		\label{Eq:Gen_Temp}
	\end{align}
	Note that \Eqref{Eq:Gen_Temp} corresponds to a static network if and only if $dA_{ij}/dt=0$ for all $i,j=1,\dots,n$. Suppose now that there exists a solution $u_{\alpha,i}^{\star}(t),\alpha=1,\dots,m,i=,\dots,n$ of \Eqref{Eq:Gen_Temp}. Notice here that, unlike in \Sectionref{SubSec:Linear_instability_analysis_for_undirected_networks}, we do not require that the background solution $u_{\alpha,i}^{\star}(t)$ is constant in time. Then we consider perturbations of the form $u_{\alpha,i} = u_{\alpha,i}^{\star} + \epsilon U_{\alpha,i}$, where $U_{\alpha,i},\alpha=1,\dots,m,i=1,\dots,n$ and $\epsilon\ll 1$ is a small parameter. Putting these expansions into \Eqref{Eq:Gen_Temp} and retaining $\mathcal{O}(\epsilon)$ terms we obtain
	\begin{align}
		\frac{d U_{\alpha,i}}{d t} =\, & \sum_{\beta=1}^{m} \sum_{j=1}^{n}\left( d_{\alpha\beta}(t)A_{ij}(t)\left( U_{\beta,j} - U_{\beta,i} \right)+ (J_{\alpha\beta})_{ij}(t){U}_{\beta,j}\right),
		\quad
		\alpha=1,\dots,m,
		\quad
		i=1,\dots,n,
		\label{Eq:Gen_Temp_Lin}
	\end{align}
	where $J_{\alpha\beta},\alpha,\beta=1,\dots,m$ is defined as in \Eqref{Eq:JDeff}. From here one could proceed in much the same way as in \Sectionref{SubSec:Linear_instability_analysis_for_undirected_networks}. i.e., by symmetrising the adjacency matrix and performing an eigenbasis decomposition. The calculations that follow are near identical to what was done in \Sectionref{SubSec:Linear_instability_analysis_for_undirected_networks}, with the only significant difference coming from a modified definition of $\gamma_{k,\ell,\alpha,\beta}$. In this temporal setting, for $k,\ell=1,\dots n$, and $\alpha,\beta=1,\dots,m$, one defines the \emph{functions} $\gamma_{k,\ell,\alpha,\beta}:\mathbb{R}\rightarrow\mathbb{R}$ as 
	\begin{align}
		\gamma_{k,\ell,\alpha,\beta} = \mathbf{e}^T_{k}\left(\hat{J}_{\alpha\beta} - S\, \frac{d}{dt}S^{-1}\delta_{\alpha\beta} \right)\mathbf{e}_{\ell},
		\quad
		k,\ell = 1,\dots,n
		\quad
		\alpha,\beta = 1,\dots,m.
	\end{align}
	Doing this allows one to once again obtain a system of the form \Eqref{Eq:GeneralLinearEq}. From here one could proceed as in \cite{VanGorder:2021} by numerically solving the resulting equations, or some subsystem. There are some benefits to this approach. As one can easily define a subsystem i.e., a reduced system of ODE's also of the form \Eqref{Eq:GeneralLinearEq} where the `reduced $M$' has a lower dimension than the original system. However, this approach is not without its own drawbacks. For example, in order to calculate $\gamma_{k,\ell,\alpha,\beta}$ one must first find $S$ and $S^{-1}$. Recall here that $S$ is the matrix of eigenvectors of the symmetrized Laplacian $\tilde{L}$. Although numerically calculating $S$ (and $S^{-1}$) is easily done on a computer in the case of a static network (i.e., networks that satisfy $dA_{ij}/dt=0$ for all $i,j=1,\dots,n$), this becomes more complicated for a dynamical network as, in addition to $S$ and $S^{-1}$, one must now also numerically calculate $\frac{d}{dt}S^{-1}$. For this, one might consider using a central finite differencing approximation. However, care should be taken here as approximating the derivative in this way could be a significant source of numerical error, particularly if $S^{-1}$ changes rapidly. For this reason we do not consider this approach here. Instead we opt to numerically solve the linearized system \Eqref{Eq:Gen_Temp_Lin}, without any further reductions. One may ask ``why not numerically solve the non-linear system \Eqref{Eq:Gen_Temp} instead?'' This is a reasonable question as, in either case, one must numerically solve $nm$ ODE's. However, it is worth noting that it is less computationally expensive to solve linear equations, than it is to solve non-linear ones.

	\section{Numerical examples of pattern formation on static directed networks}
	\label{Sec:Numerical_examples_of_pattern_formation_on_static_directed_networks}
	We now provide numerical examples of the theory presented in \Sectionref{SubSec:Linear_instability_analysis_for_undirected_networks}. In particular we numerically solve equations of the form \Eqref{Eq:graphsystem_adjacency_1}. The goal of this section is twofold: On the one hand, we provide examples of pattern formation on static non-diagonalizable networks. On the other hand, we provide examples of pattern formation (on static networks) arising from global reaction kinetics. In this section we consider only two types of networks. Many of the results can of course be applied to other types of networks. However, we find that the ones considered here are sufficient to demonstrate the different aspects of our theory. 
	\subsection{Numerical methods}
	\label{Sec:Numerical_methods} 
	We numerically solve equations of the form \Eqref{Eq:graphsystem_adjacency_1}, using the Mathematica function \emph{NDSolve}\footnote{See \url{https://reference.wolfram.com/language/ref/NDSolve.html}}, which has an absolute error tolerance of $10^{-8}$. Moreover, suppose that $u_{\alpha,i}^{\star}$ for $\alpha=1,\dots,m$ and $i=1,\dots,n$ is a constant steady-state solution of \Eqref{Eq:graphsystem_adjacency_1}. Then, for initial data, we pick $u_{\alpha,i}(0)=u_{\alpha,i}^{\star}+\xi_{\alpha,i}$, where $\xi_{\alpha,i}$ is a random real number. Here, we calculate $\xi_{\alpha,i}$ using the Mathematica function\footnote{see \url{https://reference.wolfram.com/language/ref/RandomReal.html}} \emph{RandomReal[1]}.

	\subsection{A simple example of our instability criterion}
	To demonstrate the theory outlined in \Sectionref{SubSec:Linear_instability_analysis_for_undirected_networks}, we first consider systems with \emph{local reaction kinetics}. The goal of this subsection here is to give an explicit example of our instability criterion on a non-diagonalizable network. For this we study equations of the form 
	\begin{align}
	\frac{d u_{i}}{d t} =\, & \frac{1}{50}\sum_{j=1}^{n}A_{ij}\left( u_{j} - u_{i} \right) + \frac{1}{10}-u_{i}+u_{i}^{2}v_{i},
	\label{Eq:Local_u}
	\\
	\frac{d v_{i}}{d t} =\, & \sum_{j=1}^{n}A_{ij}\left( v_{j} - v_{i} \right) + 1 - u_{i}^{2}v_{i},
	\label{Eq:Local_v}
	\end{align}
	for $n=4$, where $A$ is given below. It is worth noting that the reaction kinetics we have chosen here are the Schnakenberg kinetics. A locally stable steady state is given by $(u_i,v_i)=(u^\star,v^\star)$ with $u^\star=11/10$ and $v^\star = 100/121$.
	
	Let us now consider a $4$-node directed network with the following adjacency matrix 
	\begin{align}
		\label{Eq:Adjacency_0}
		A = \frac{1}{2} \left( 
		\begin{array}{cccc}
			0 & 2 & 2 & 2 \\
			2 & 0 & 1 & 0 \\
			2 & 1 & 0 & 0 \\
			2 & 2 & 0 & 0
		\end{array}
		\right) 
		\implies 
		L = -\frac{1}{2} \left( 
		\begin{array}{cccc}
			-6 & 2 & 2 & 2 \\
			2 &-3 & 1 & 0 \\
			2 & 1 &-3 & 0 \\
			2 & 2 & 0 &-4
		\end{array}
		\right),
	\end{align}
	where $L$ is the Laplacian matrix corresponding to $A$. This network is directed since $A_{24}\ne A_{42}$. The eigenvectors corresponding to the Laplacian $L$ are $\Phi_{1}=\left( 1,1,1,1 \right)^T,\Phi_{2}=\left( 0,0,0,0 \right)^T,\Phi_{3}=\left( 1/3,1/3,1/3,1 \right)^T$ and $\Phi_{4}=\left( -3,1,1,1 \right)^T$ and the corresponding eigenvalues are $\rho_{1}=0,\rho_{2}=2,\rho_{3}=2,$ and $\rho_{4}=4$. Here we see that the eigenvectors clearly do not form a basis and hence the Laplacian $L$ is non-diagonalizable. Thus, in order to detect a Turning instability one must use the theory presented by us in \Sectionref{SubSec:Linear_instability_analysis_for_undirected_networks}. 
	To this end we now calculate the symmetrised adjacency matrix, and its associated Laplacian, as
	\begin{align}
		\tilde{A} = \frac{1}{2} \left( 
		\begin{array}{cccc}
			0 & 2 & 2 & 2 \\
			2 & 0 & 1 & 1 \\
			2 & 1 & 0 & 0 \\
			2 & 1 & 0 & 0
		\end{array}
		\right) 
		\implies 
		\tilde{L} = -\frac{1}{2} \left( 
		\begin{array}{cccc}
			-6 & 2 & 2 & 2 \\
			2 & -4 & 1 & 1 \\
			2 & 1 & -3 & 0 \\
			2 & 1 & 0 & -3
		\end{array}
		\right).
	\end{align}
	The eigenvalues of $\tilde{L}$ are $\sigma_{1}=0,\sigma_{2}=3/2,\sigma_{3}=5/2,$ and $\sigma_{4}=4$. From here it is straightforward to calculate the matrix $M$ as is described in \Sectionref{SubSec:Linear_instability_analysis_for_undirected_networks}. In doing so we obtain 
	\begin{align}
		M = 
		\left(
		\begin{array}{cccccccc}
			0.738 & 1.21 & -0.01 & 0 & -0.01 & 0 & 0 & 0 \\
			-1.81 & -5.21 & 0 & -0.5 & 0 & -0.5 & 0 & 0 \\
			0 & 0 & 0.773 & 1.21 & 0.005 & 0 & 0 & 0 \\ 
			0 & 0 & -1.818 & -3.46 & 0 & 0.25 & 0 & 0 \\
			0 & 0 & -0.005 & 0 & 0.783 & 1.21 & 0 & 0 \\
			0 & 0 & 0 & -0.25 & -1.818 & -2.96 & 0 & 0 \\
			0 & 0 & -0.01 & 0 & -0.01 & 0 & 0.818 & 1.21 \\
			0 & 0 & 0 & -0.5 & 0 & -0.5 & 1.818 & -1.21
		\end{array}
		\right).
	\end{align}
	The corresponding eigenvalues are $\lambda_{1}=-4.81,\lambda_{2}=\lambda_{3}=-2.58,\lambda_{4}=-0.19 + 1.08 \text{i},\lambda_{5}=-0.19 - 1.08 \text{i},\lambda_{6}=0.34,\lambda_{7}=\lambda_{8}=0.12$. Observe carefully that there are \emph{three} eigenvalues with positive real part, with $\max_{\ell}\text{Re}(\lambda_{\ell})=0.34>0$. We therefore expect this network to give rise to a patterned state. 
	\begin{figure}
		\centering
		\includegraphics[width=0.3\linewidth]{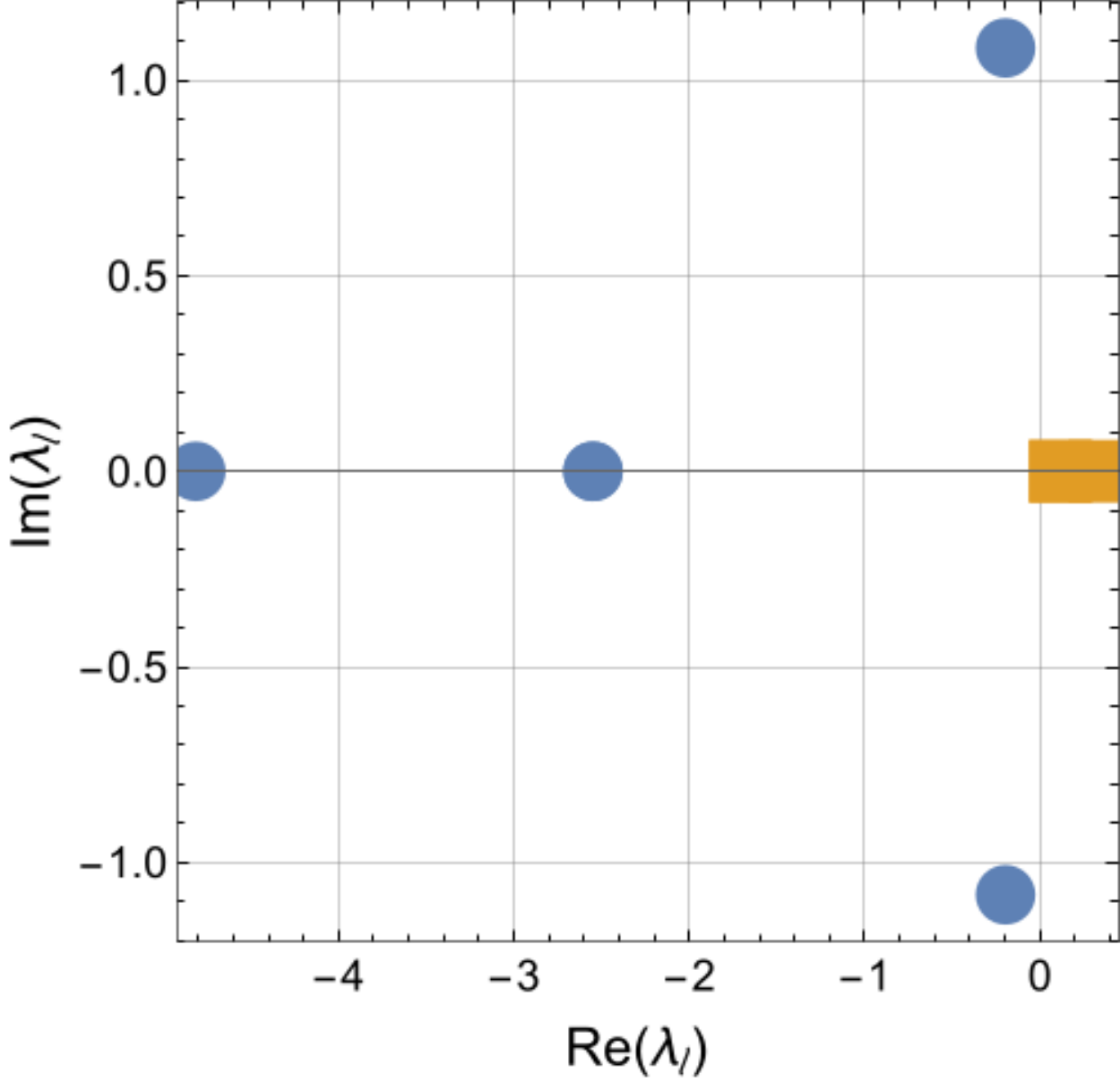}
		(a)
		\includegraphics[width=0.3\linewidth]{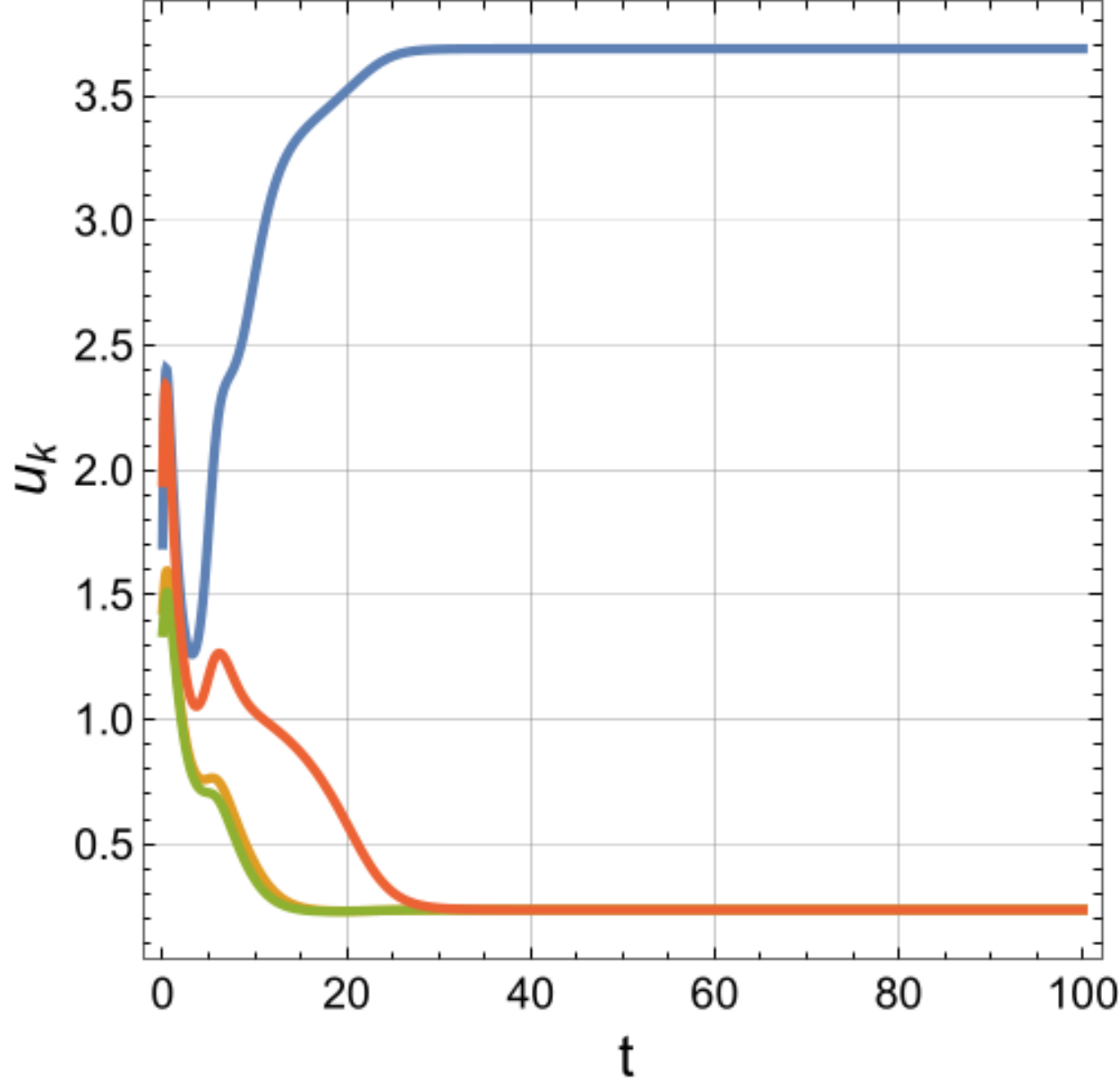}
		(b)
		\includegraphics[width=0.25\linewidth]{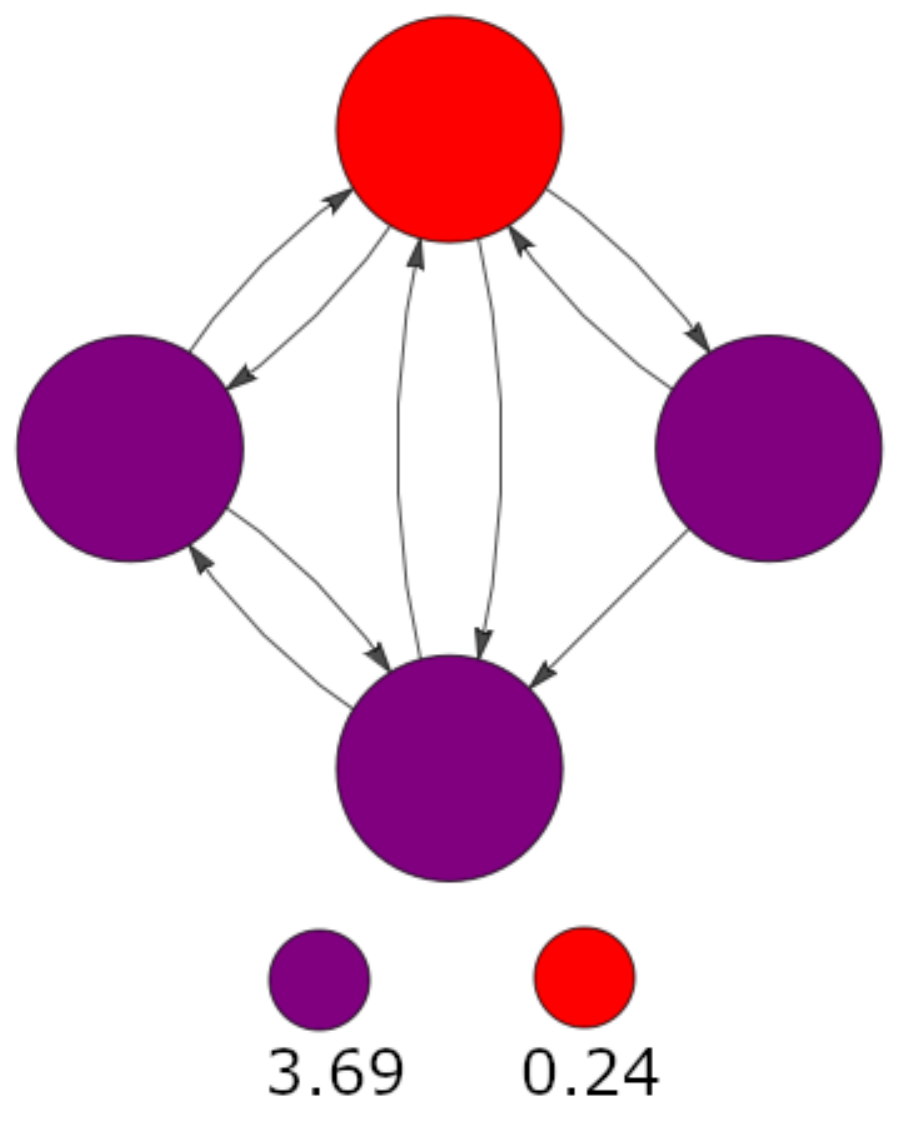}
		(c)
		\caption{Patterning corresponding to the adjacency matrix \Eqref{Eq:Adjacency_0}. In (a) we give the eigenvalues of the matrix $M$. In (b) we show the behaviour of the function $u$, for each of the nodes. Finally, in (c) we give the pattern for the function $u$ at $t=100$.}
		\label{fig:uall_0}
	\end{figure}
	The numerical results for this case are shown in \Figref{fig:uall_0}. In  \subfig{fig:uall_0}{(a)} we give the eigenvalues of the matrix $M$. The blue dots shows eigenvalues with negative real part and the orange squares show eigenvalues with positive real part. In \subfig{fig:uall_0}{(b)} we show each $u_{i}$ as a function of time $t$. Here we see that a stable pattern forms at around $t\approx 45$. In \subfig{fig:uall_0}{(c)} we show the pattern, emergent at $t=100$, corresponding to the values of $u_i$. Similar plots can be made for $v_i$.

	\subsection{Influence of directed edges on Turing pattern formation}
	\label{Pattern_formation_with_local_reaction_kinetics}
	For our next examples we once again consider \Eqsref{Eq:Local_u} and \eqref{Eq:Local_v}. However, this time we pick $n=16$ and set the adjacency matrix as
	\begin{align}
	A_{ij}=
	\begin{cases}
		1/2 \quad &\text{if}\quad i-j=1\quad \text{or}\quad i-j=15,
		\\
		a_{1} \quad &\text{if}\quad j-i=1\quad \text{or}\quad j-i=15,
		\\
		0 \quad &\text{if}\quad i-j=0
		\\
		a_{2} \quad &\text{otherwise},
	\end{cases}
	\label{Eq:Adjacency_1}
	\end{align}
	for $1\le i,j\le 16$. Here, $a_{1}$ and $a_{2}$ are freely specifiable constants with $a_{1},a_{2}\in[0,1]$. Note that $A_{ij}$ is symmetric if and only if $a_1=1/2$. Moreover, $A_{ij}$ describes a complete graph if $a_{1},a_{2}\ne 0$ and a cycle graph if $a_{2}=0$. To ease much of the upcoming discussions we now briefly introduce some terminology. In the special case $a_{1}=0$ and $a_{2}\ne 0$, we refer to the network as a \emph{complete cycle graph}. For these particular parameter choices each node is connected to all other nodes, excluding its left most neighbour. This network is of course not a complete network. nevertheless we find this terminology useful for our purposes here. Similarly, in the special case $a_{1}=a_{2}= 0$, we refer to the network as an \emph{incomplete cycle graph}. In setting each node is connected (via a directed edge) only to its left most neighbour. It is is worth noting that an incomplete cycle graph, regardless of the number of nodes, is a non-diagonalizable network. 
	\begin{figure}
		\centering
		\includegraphics[width=0.3\linewidth]{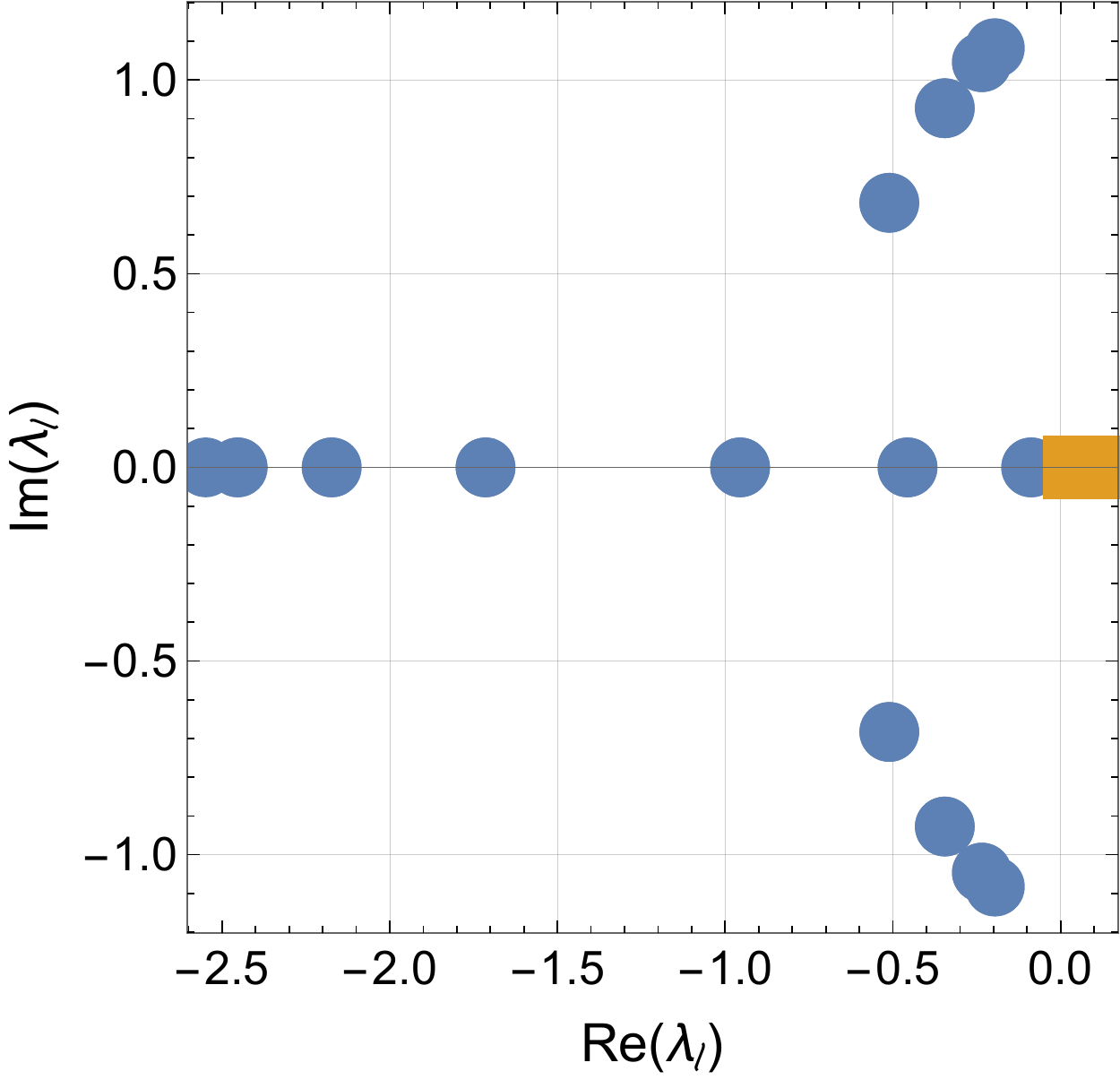}
		(a)
		\includegraphics[width=0.3\linewidth]{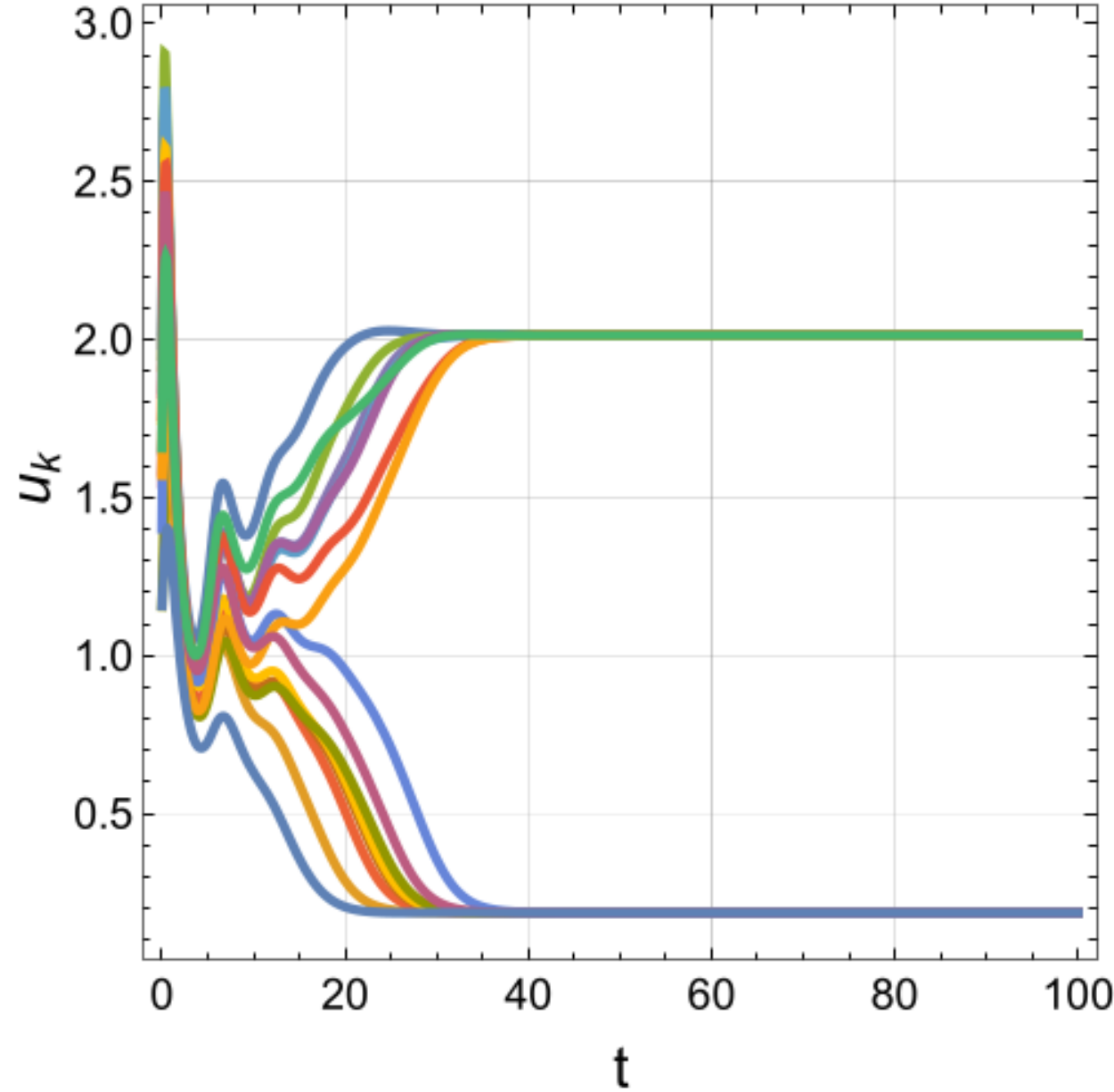}
		(b)
		\includegraphics[width=0.25\linewidth]{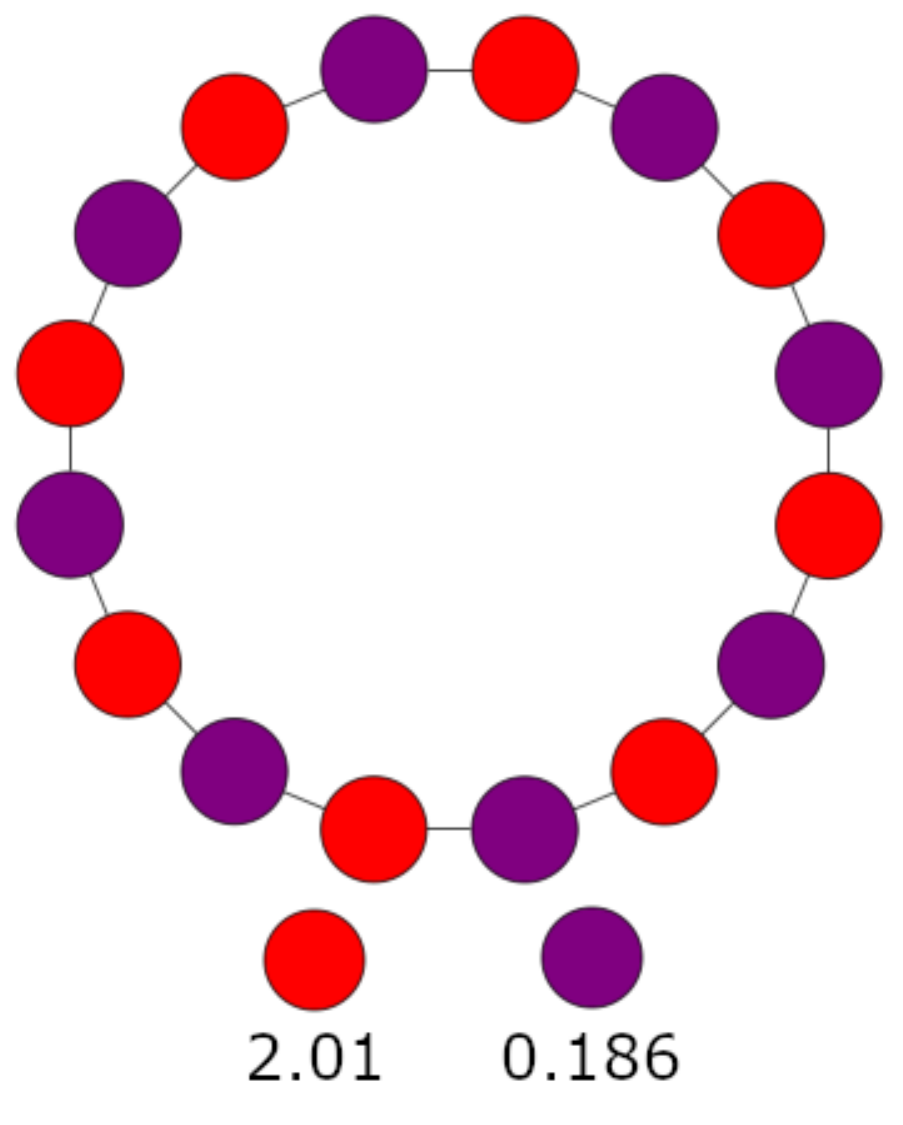}
		(c) \\
		\includegraphics[width=0.3\linewidth]{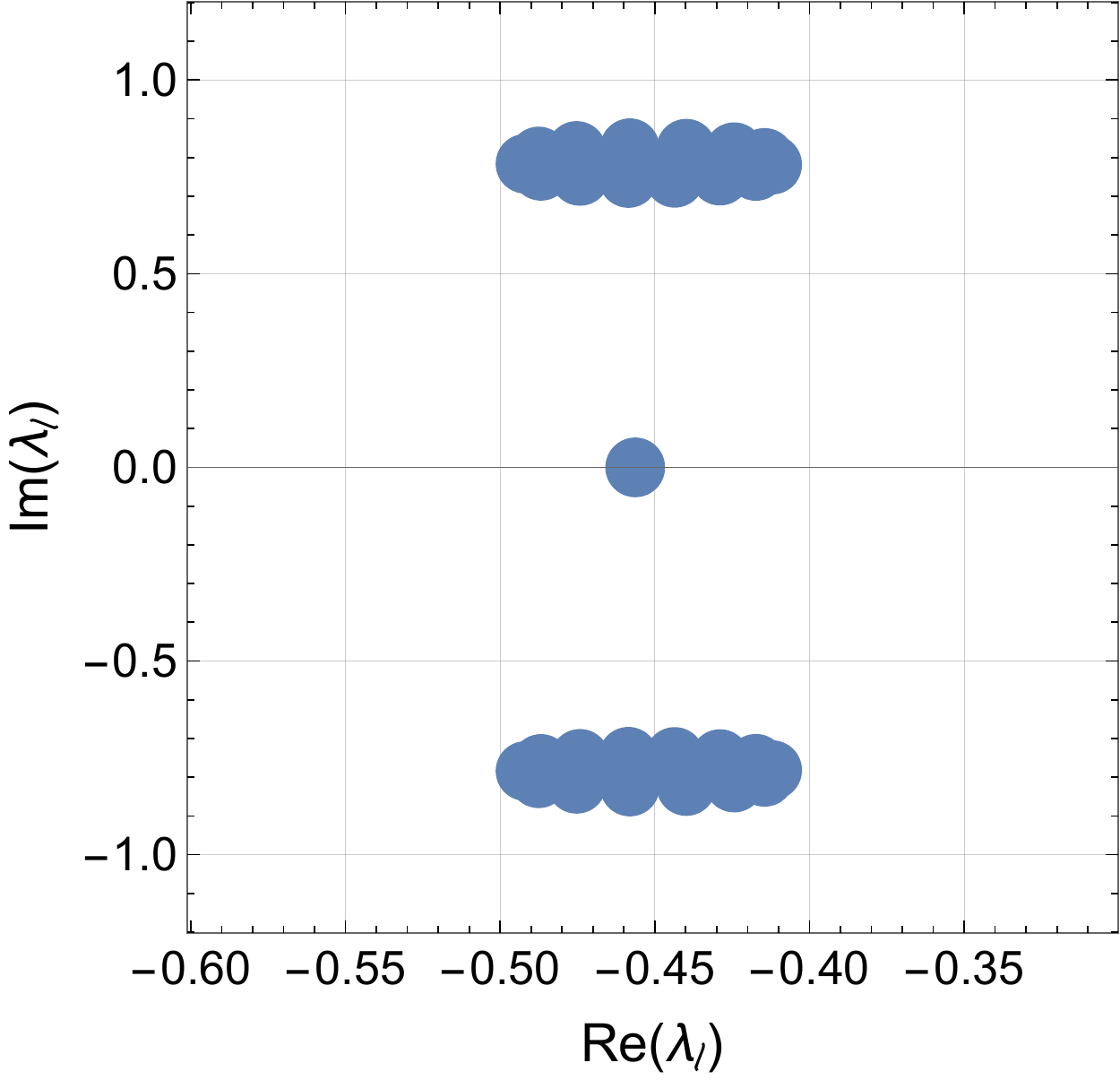}
		(d)
		\includegraphics[width=0.3\linewidth]{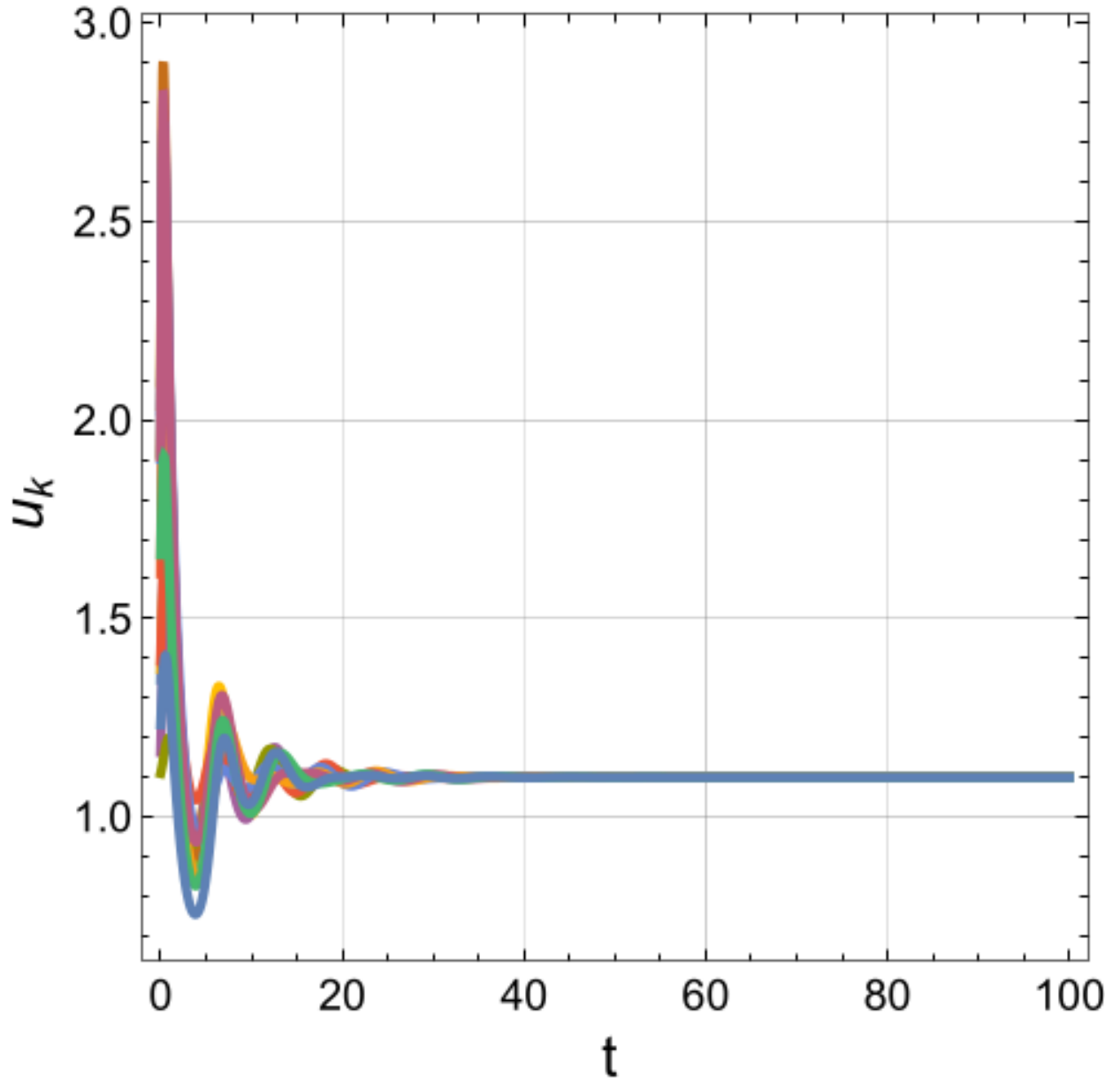}
		(e)
		\includegraphics[width=0.25\linewidth]{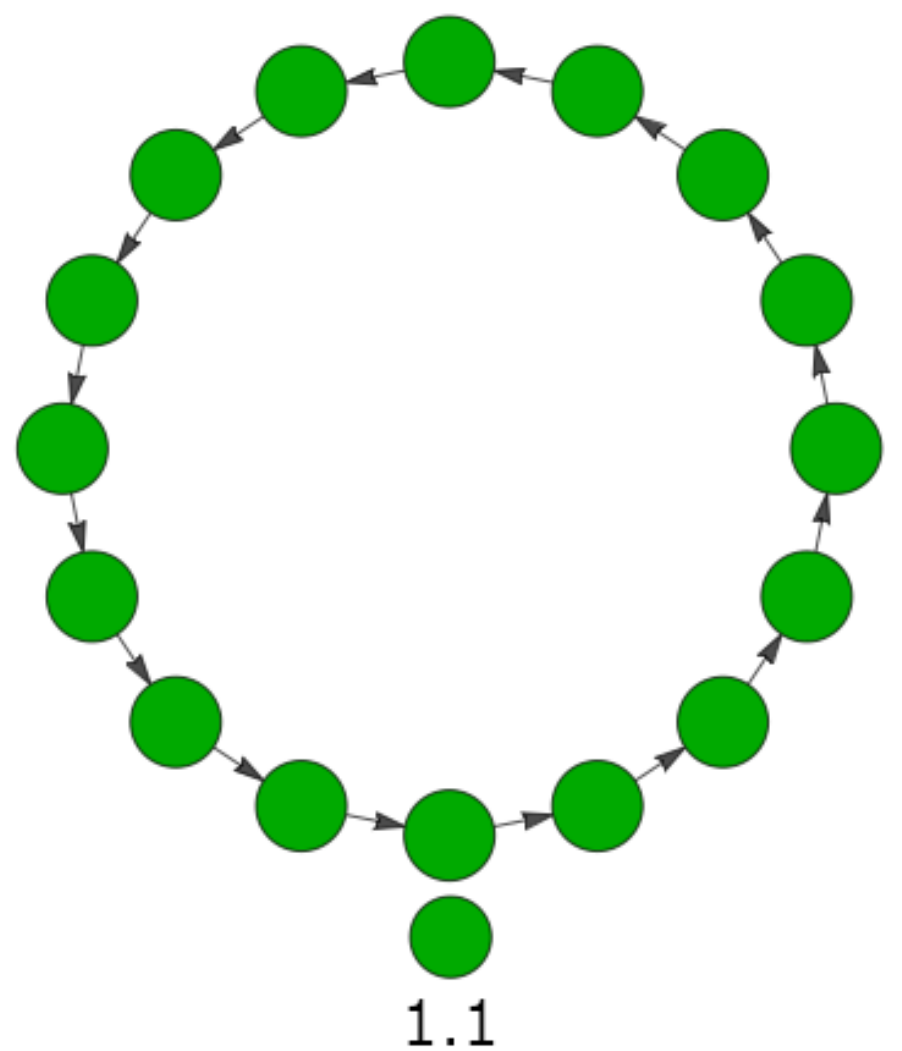}
		(f) \\
		\includegraphics[width=0.3\linewidth]{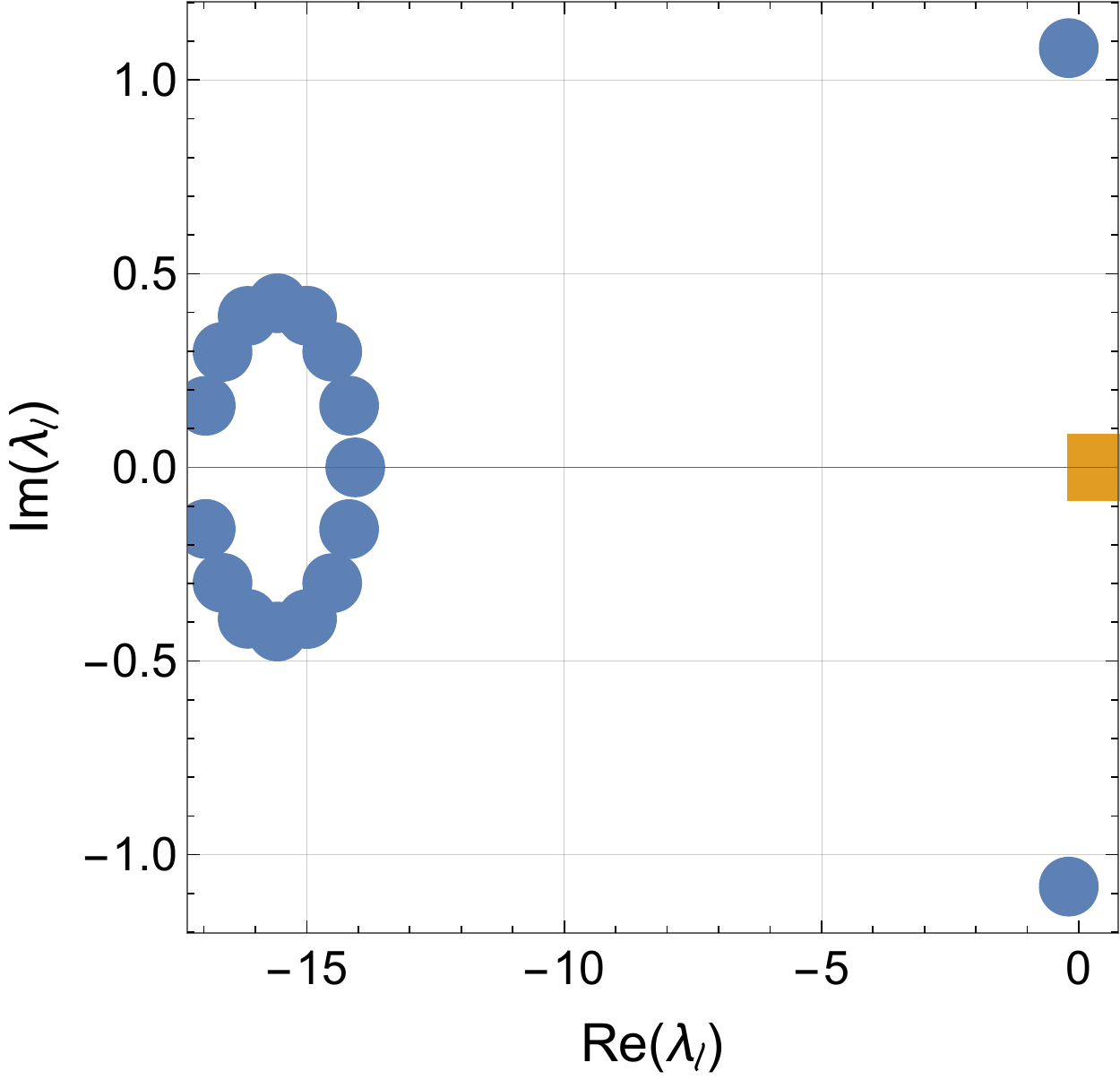} 
		(g)
		\includegraphics[width=0.3\linewidth]{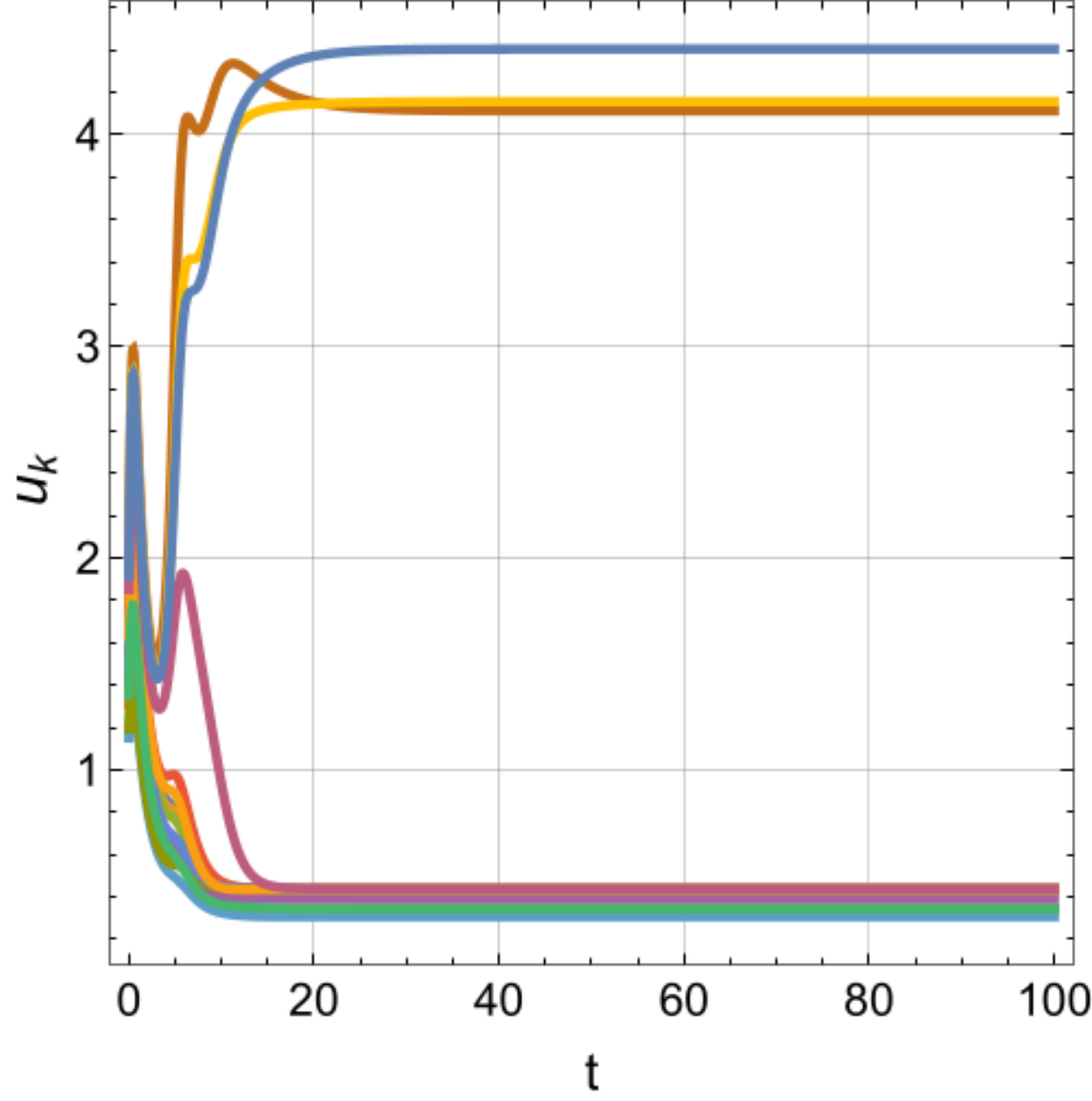}
		(h)
		\includegraphics[width=0.26\linewidth]{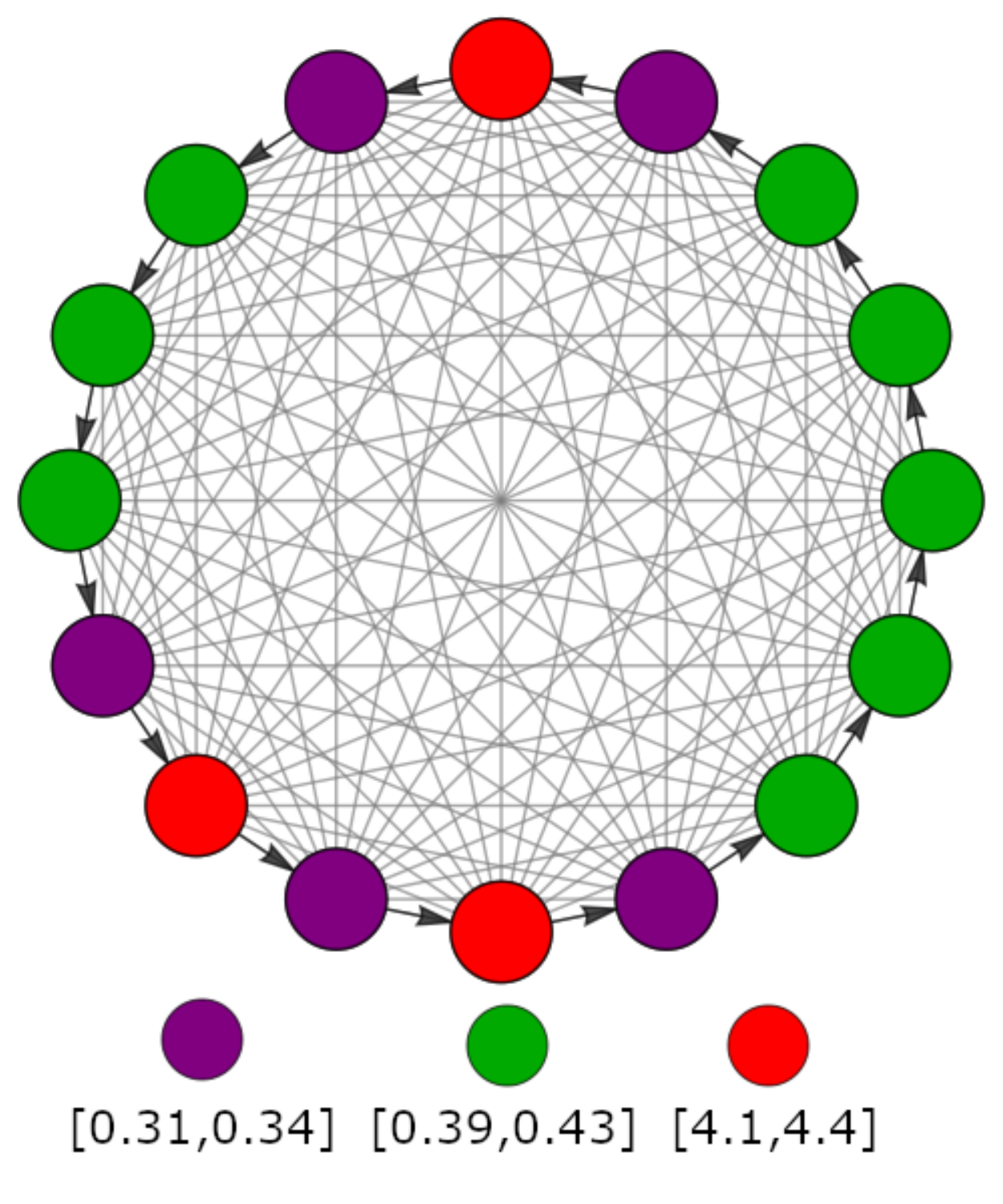}
		(i)
		\caption{Patterning corresponding to the adjacency matrix \Eqref{Eq:Adjacency_1}. The first row (plots (a)--(c)) shows the numerical results corresponding to setting $a_{1}=1/2$ and $a_2=0$. The second row (plots (d)--(f)) shows the numerical results corresponding to setting $a_{1}=a_2=0$, and the third row (plots (g)--(i)) shows the numerical results corresponding to setting $a_{1}=0$ and $a_2=1$. Plots (a),(d) and (g) give the eigenvalues of the matrix $M$ for each of the considered cases. Similarly, plots (b),(e), and (h) show the function $u$. Lastly, plots (c),(f) and (i) show the pattern corresponding to the function $u$ at $t=100$.}
		\label{fig:m}
	\end{figure}
	
	We now consider some numerical examples of patterning on networks described by  \Eqref{Eq:Adjacency_1}. Here, we focus primarily on what effect the parameters $a_{1}$ and $a_2$ have on the emergence of patterning. To this end, we first consider the choices $a_{1}=1/2$ and $a_{2}=0$. The numerical results corresponding to these choices are shown in the top row of \Figref{fig:m} (plots (a)--(c)). In this case the adjacency matrix is symmetric and hence the standard Turing analysis (presented in \Sectionref{SubSec:Setting_up_the_problem_1}) can be applied. In \subfig{fig:m}{(a)} we give the eigenvalues of the matrix $M$. As with the previous example, the blue dots shows eigenvalues with negative real part and the orange squares show eigenvalues with positive real part. We find that there are three (non-distinct) eigenvalues with negative real part. In particular, we find that $\text{Re}(\lambda_\ell)=0.12>0$ and hence we detect an instability. In \subfig{fig:m}{(b)} we show $u_{i}$ as a function of time $t$ for each $i=1,\dots,16$. Here see that a stable pattern forms at around $t\approx 40$. In \subfig{fig:m}{(c)} we show the pattern corresponding to the values of $u_i$ at $t=100$.

	Conversely, in the second row of \Figref{fig:m} (plots (d)--(f)), we show the results corresponding to the parameter choices $a_{1}=a_{2}=0$. In this setting the underlying network is an incomplete cycle graph. We find that all of the eigenvalues (of $M$) have negative real part (with $\text{Re}(\lambda_\ell)=-0.2<0$) and, as such, are linearly stable. This is clearly seen in \subfig{fig:m}{(e)}. The values of $u_i$, for $i=1,\dots,16$, at $t=100$ are shown in \subfig{fig:m}{(f)} on the incomplete cycle graph. It is interesting to note here that the presence of directed edges, at least in this particular example, have the effect of suppressing the formation of patterning. This shows that, as with the undirected networks, the underlying topology (of the network) plays an important role in the generation of a pattern. In particular, we see that the presence of directed edges can actually \emph{stabilize} the background solution.

	Of course, the presence of directed edges, does not \emph{always} suppress the emergence of a pattern. To see this we consider a complete cycle graph (described by the choices $a_{1}=0$ and $a_{2}=1$). The numerical results, corresponding to these parameter choices, are given in the last row of \Figref{fig:m} (plots (g)--(i)). In \subfig{fig:m}{(g)} we show the eigenvalues of the matrix $M$. In this case we find that there are $8$ distinct eigenvalues with positive real part. In particular, we find that $\text{Re}(\lambda_\ell)=0.41>0$ and hence we detect an instability. In \subfig{fig:m}{(h)} we plot the functions $u_i$, for $i=1,\dots,n$, as a function of time. Here we find that a stable pattern emerges at around $t\approx25$. The resulting pattern, corresponding to $u_i,i=1,\dots,16$ at $t=100$, is shown in \subfig{fig:m}{(i)}.

	\subsection{Pattern formation from hyperbolic reaction-diffusion equations}
	\label{Sec:Pattern_formation_from_hyperbolic_reaction_diffusion_equations}
	Having now seen some examples of patterning on static networks for two-species system of reaction-diffusion equations, it is interesting to now consider an example of patterning (on a static network) with \emph{more} than two equations. Recall that \Sectionref{SubSec:Linear_instability_analysis_for_undirected_networks} allows for arbitrarily (but finitely) many equations. To this end we consider the following system:
	\begin{align}
		\frac{d^2 u_{i}}{dt^2}+\frac{du_{i}}{dt} &= 3\sum_{j=1}^{10}A_{ij}\left( u_{j}-u_{i} \right) + 5 - 10u_{i} + u_{i}^{2}v_{i},
		\label{Eq:HypEqs_u}
		\\
		\frac{d^2 v_{i}}{dt^2}+\frac{dv_{i}}{dt} &=  \sum_{j=1}^{10}A_{ij}\left( v_{j} - v_{i} \right) + 9u_{i} - u_{i}^{2}v_{i},
		\label{Eq:HypEqs_v}
	\end{align}
	where 
	\begin{align}
		A_{ij}=
		\begin{cases}
			1 \quad &\text{if}\quad i=1\quad \text{and}\quad j\ne 1\quad \text{or}\quad j=1\quad \text{and}\quad i\ne 1,
			\\
			1/2 \quad &\text{if}\quad i=2\quad \text{and}\quad j =3\quad \text{or}\quad i=3\quad \text{and}\quad j =2,
			\\
			1 \quad &\text{if}\quad i=4\quad \text{and}\quad j =2,
			\\
			0 \quad &\text{otherwise},
		\end{cases}
		\label{Eq:Adjacency_Hyp_1}
	\end{align}
	for $1\le i,j\le 10$. This adjacency matrix describes a modified star graph (with $10$ nodes) where, in addition to the standard star graph connections, nodes two and three are connected by an undirected edge and node four is connected to node two via a directed edge. Although these modifications (to the star graph) may seem like a small change, it turns out that this adjacency matrix describes a non-diagonalizable network. Observe carefully that, written in this way, \Eqsref{Eq:HypEqs_u} and \eqref{Eq:HypEqs_u} form a system of \emph{hyperbolic reaction-diffusion} equations. Systems of this type have been investigated before in the context of pattern formation in \cite{AlGhoul:1996,Zemskov:2016,Ritchie:2022}. Moreover, we note that we have chosen to use the Brusselator reaction-kinetics which has a locally stable steady state is given by $(u_i,v_i)=(u^\star,v^\star)$ with $u^\star=5$ and $v^\star = 9/5$.

	In order to use the theory presented in \Sectionref{SubSec:Linear_instability_analysis_for_undirected_networks} we now define $u_{1,i}=u_{i},u_{2,i}={du_i}/{dt},u_{3,i} = v_i$ and $u_{4,i}={dv_i}/{dt}$. The resulting evolution equations for $u_{2,i}$ and $u_{4,i}$ are
	\begin{align} 
		\frac{d u_{2,i}}{dt} + u_{2,i} &= 3\sum_{j=1}^{n}A_{ij}\left( u_{1,j}-u_{1,i} \right) + 5 - 10u_{1,i} + u_{1,i}^{2}u_{3,i}
		\label{Eg_Hyp:udef}
		\\
		\frac{d u_{4,i}}{dt} + u_{4,i} &=  \sum_{j=1}^{n}A_{ij}\left( u_{3,j} - u_{3,i} \right) + 9u_{1,i} - u_{1,i}^{2}u_{3,i}.
		\label{Eg_Hyp:u4}
	\end{align}
	Together \Eqsref{Eg_Hyp:udef}--\eqref{Eg_Hyp:u4} take the form of \Eqsref{Eq:graphsystem_adjacency_1} with $m=4$, $d_{21}=3,d_{43}=1$ and $d_{\alpha\beta}=0$ otherwise. Moreover, the reaction kinetics are 
	\begin{align*}
		f_{1,i}(\mathbf{u})&=u_{2,i},
		\quad
		f_{2,i}(\mathbf{u})=5 - 10u_{1,i} + u_{1,i}^{2}u_{3,i} - u_{2,i},
		\\
		f_{3,i}(\mathbf{u})&=u_{4,i},
		\quad
		f_{4,i}(\mathbf{u})=9u_{1,i} - u_{1,i}^{2}u_{3,i} - u_{4,i}.
	\end{align*}

\begin{figure}
		\centering
		\includegraphics[width=0.26\linewidth]{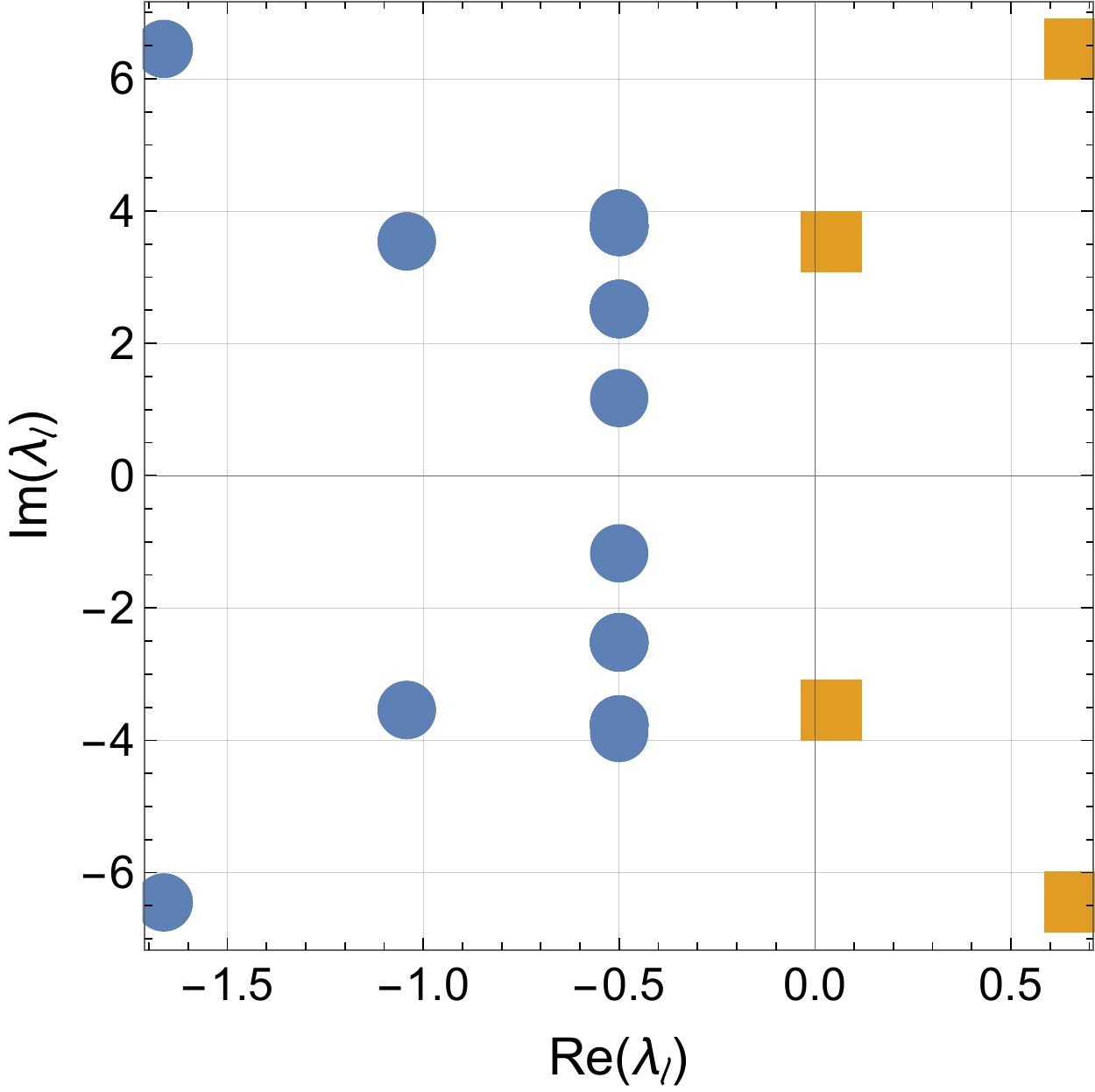}
		(a)
		\includegraphics[width=0.27\linewidth]{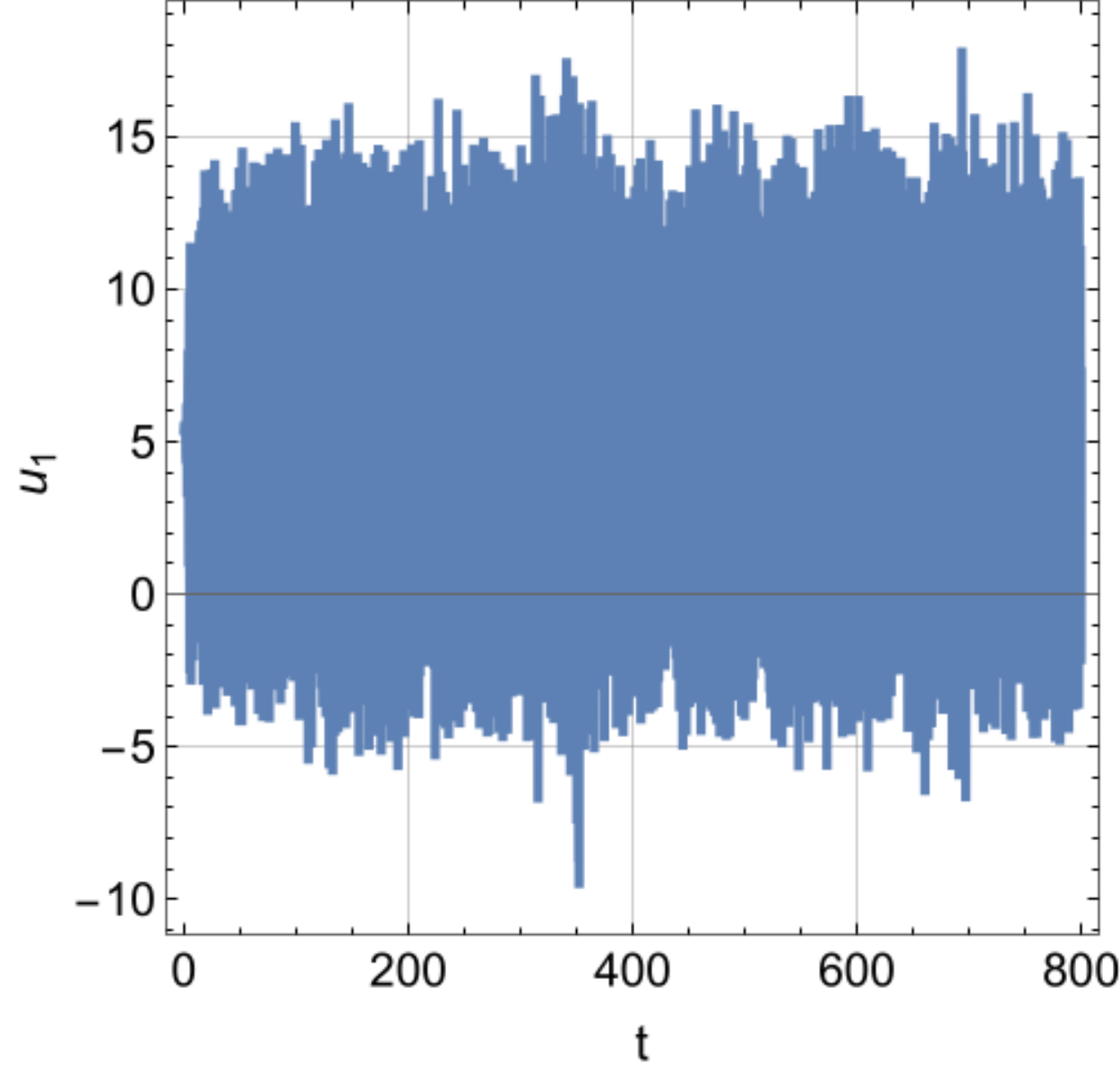}
		(b)
		\includegraphics[width=0.27\linewidth]{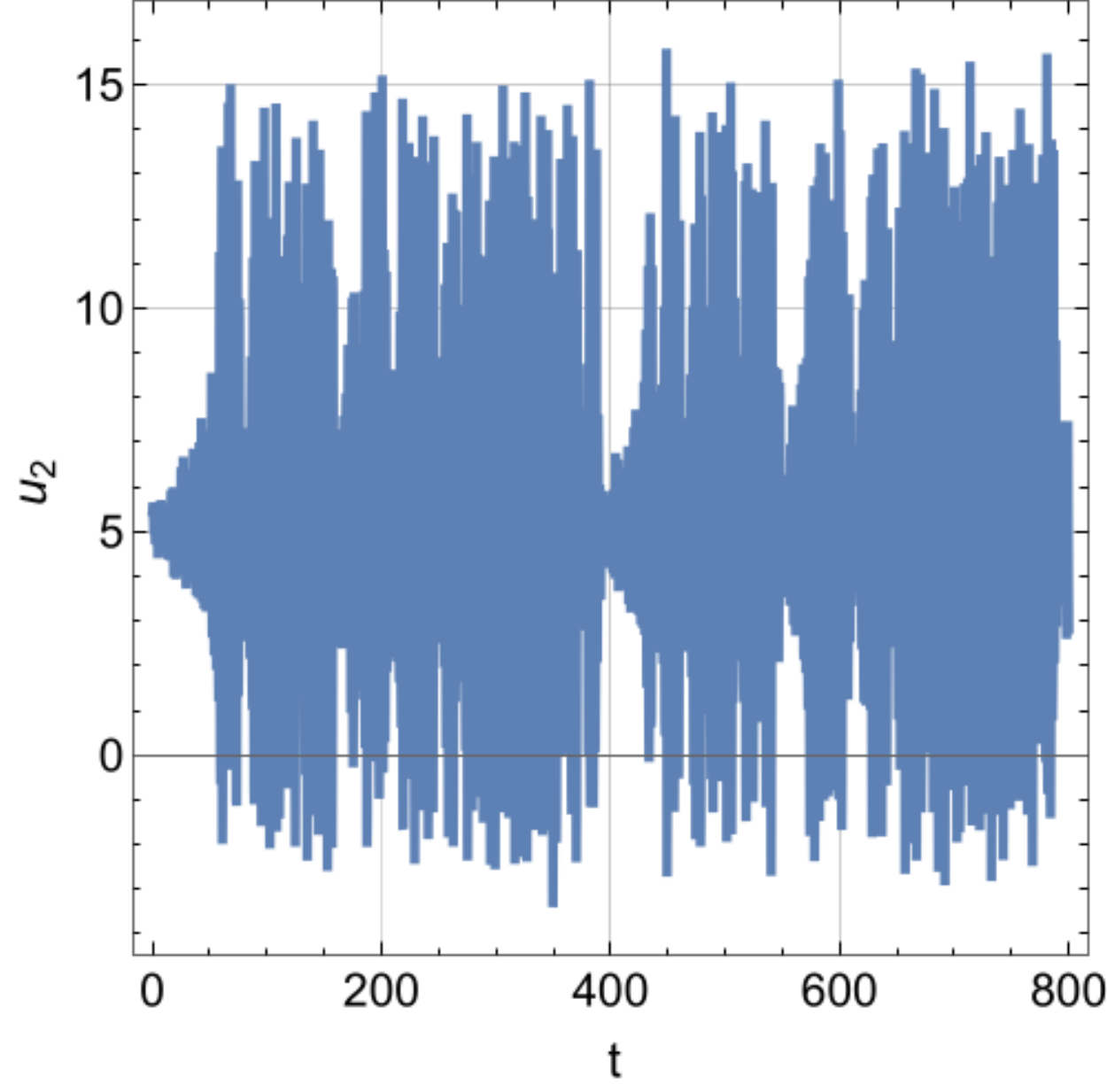}
		(c) \\
		\includegraphics[width=0.27\linewidth]{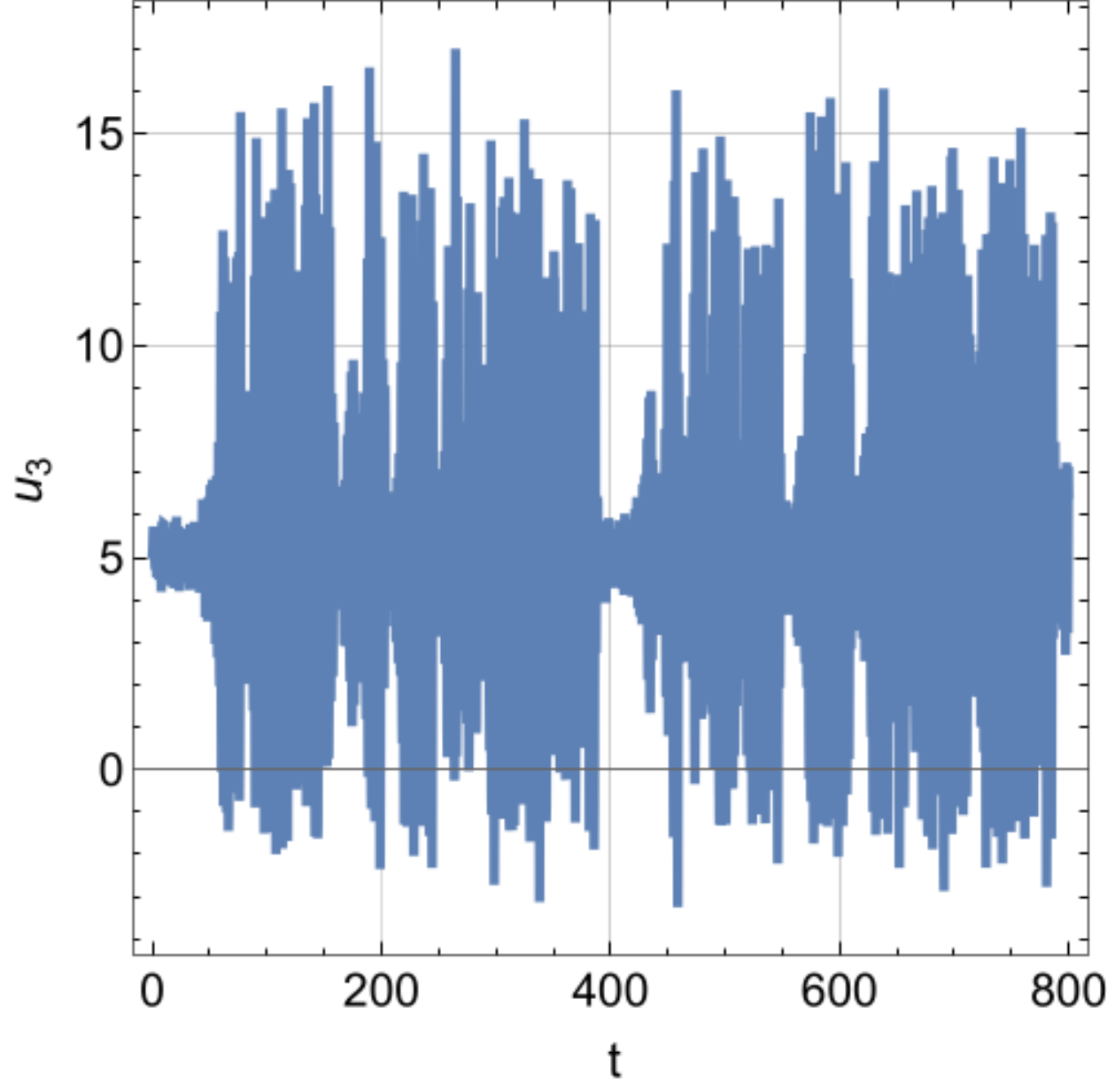}
		(d)
		\includegraphics[width=0.27\linewidth]{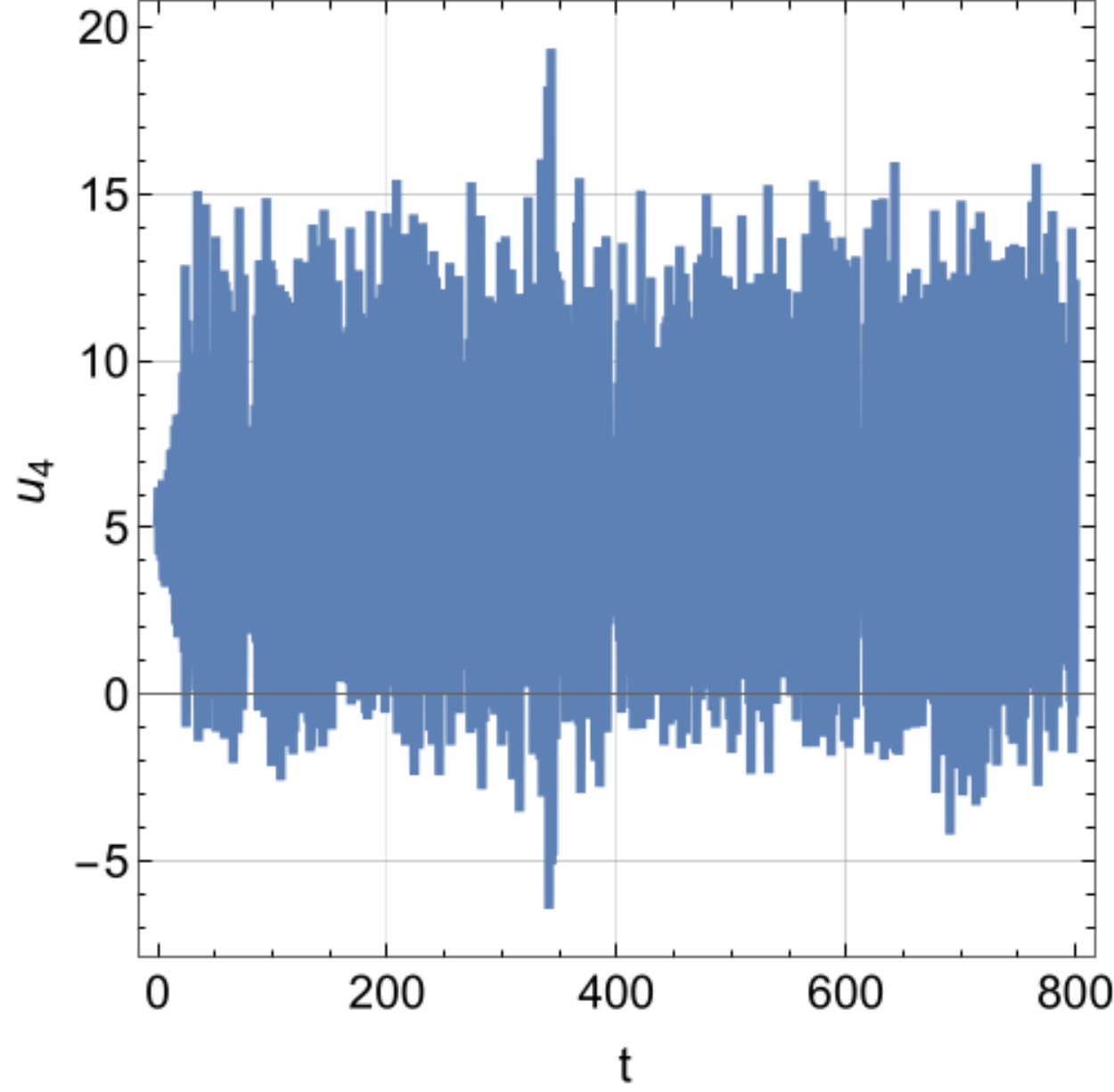}
		(e)
		\includegraphics[width=0.27\linewidth]{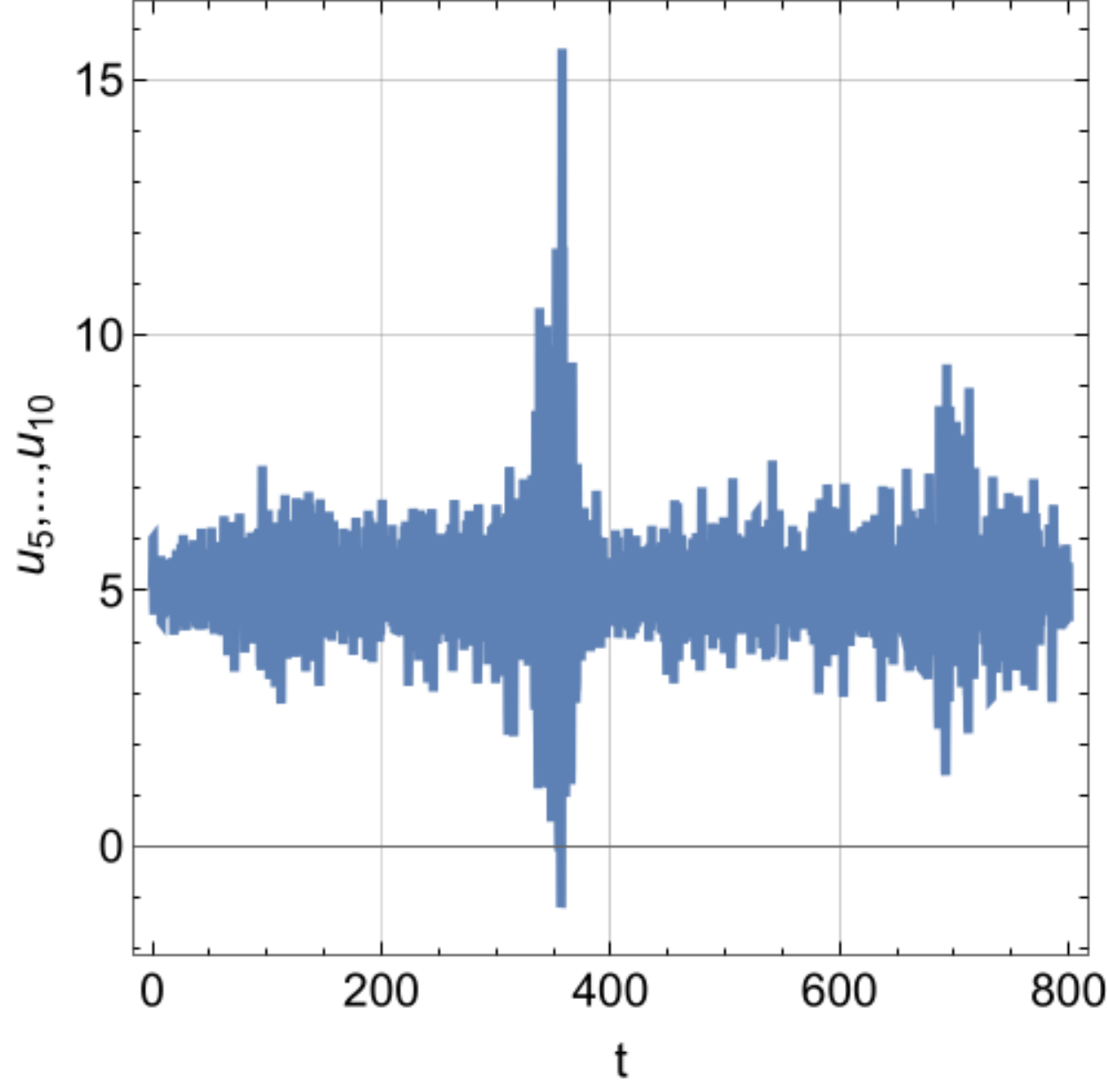}
		(f) \\
		\includegraphics[width=0.22\linewidth]{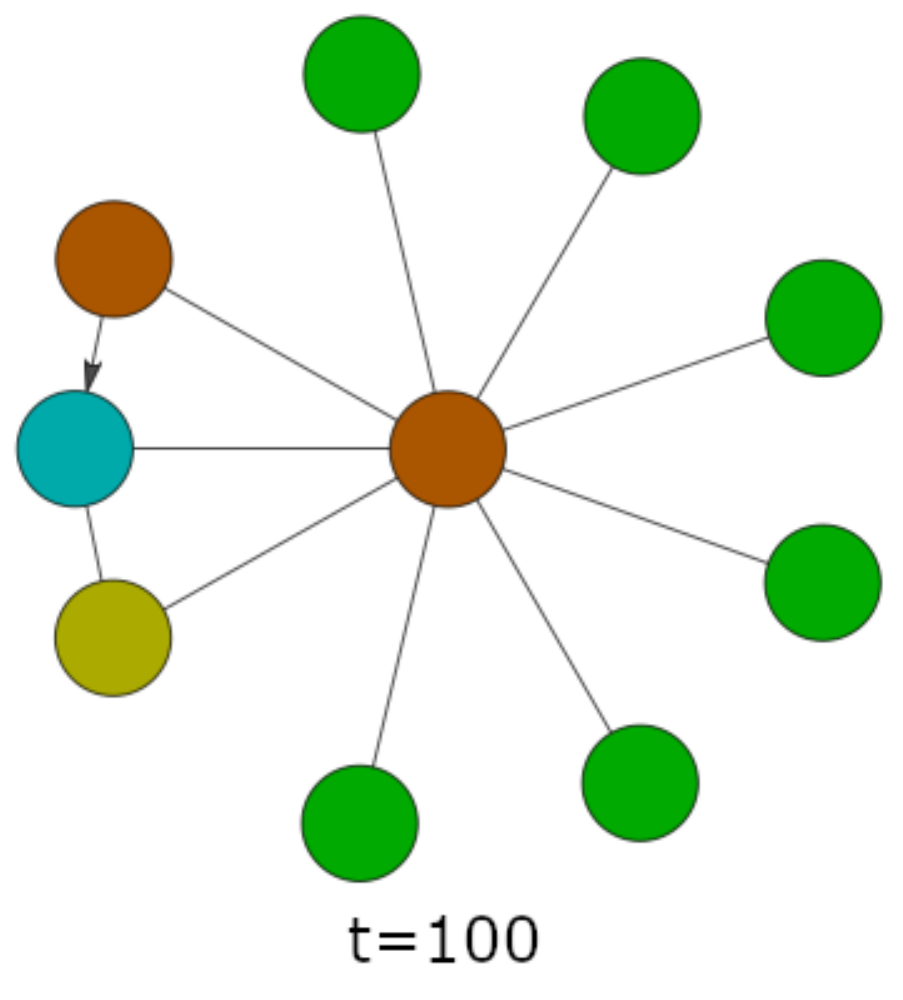}
		\includegraphics[width=0.22\linewidth]{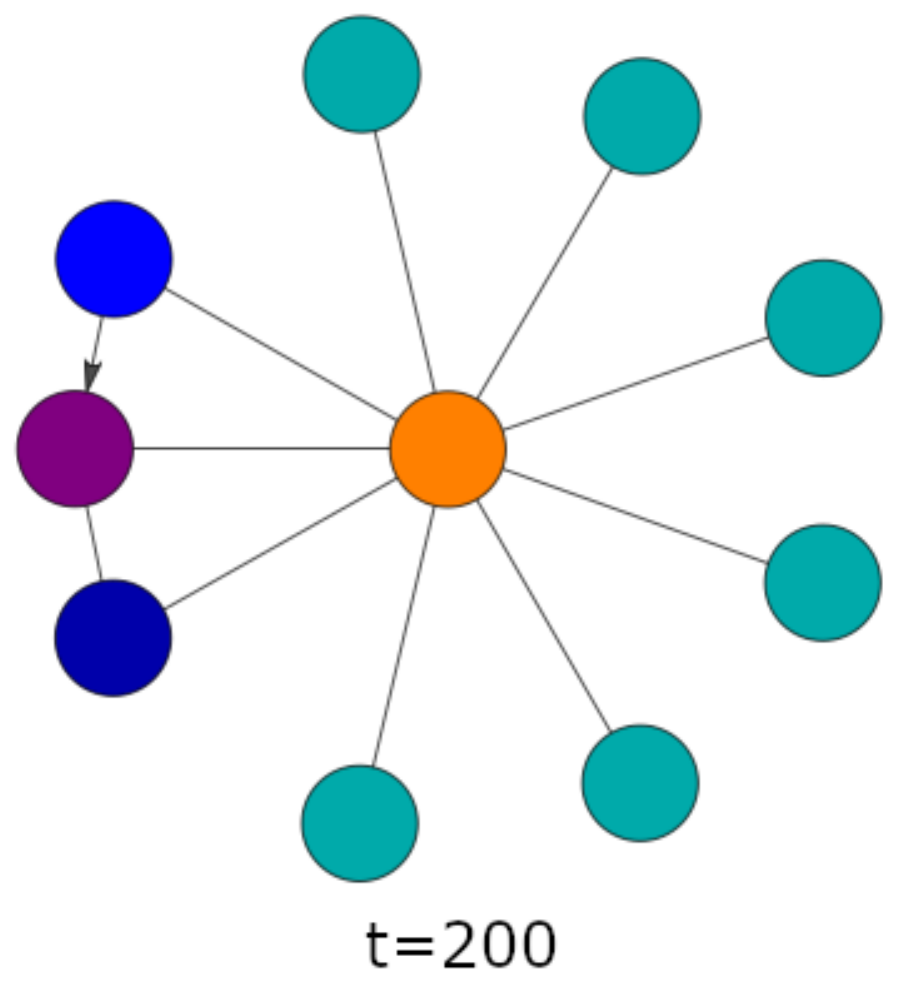}
		\includegraphics[width=0.22\linewidth]{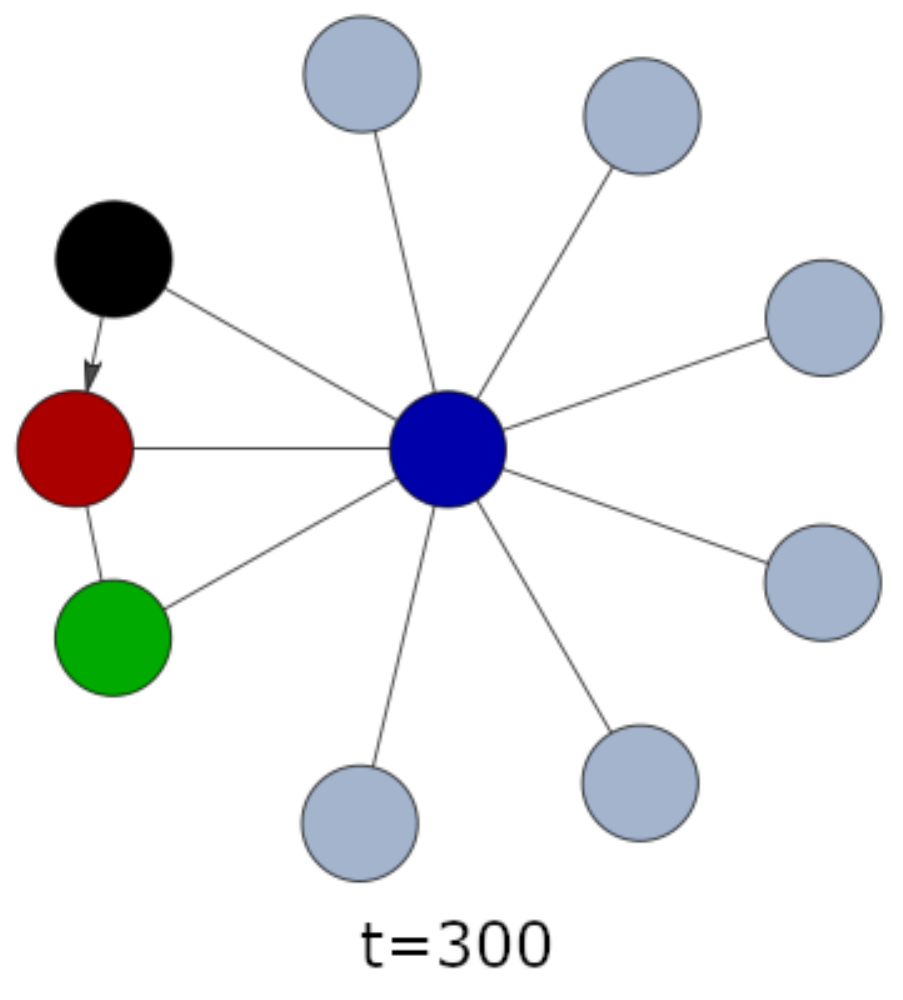}
		\includegraphics[width=0.22\linewidth]{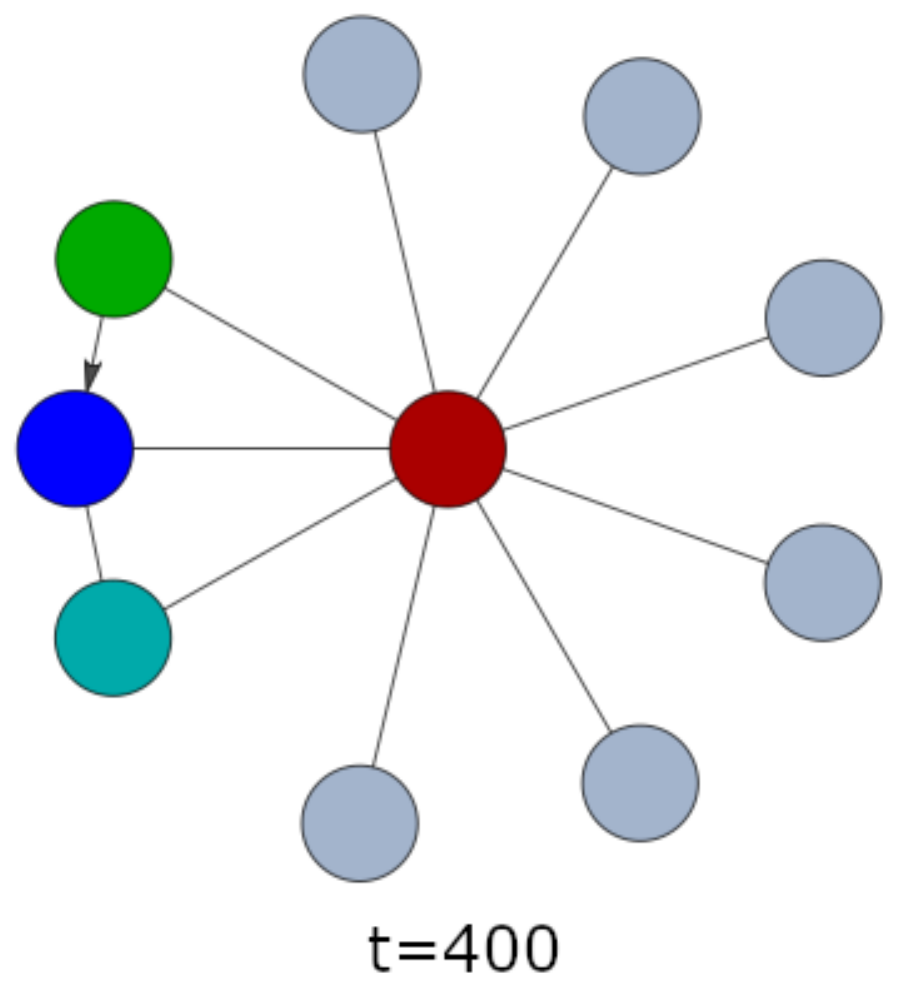}
		{\color{white}{(g)}}
		\\
		\includegraphics[width=0.22\linewidth]{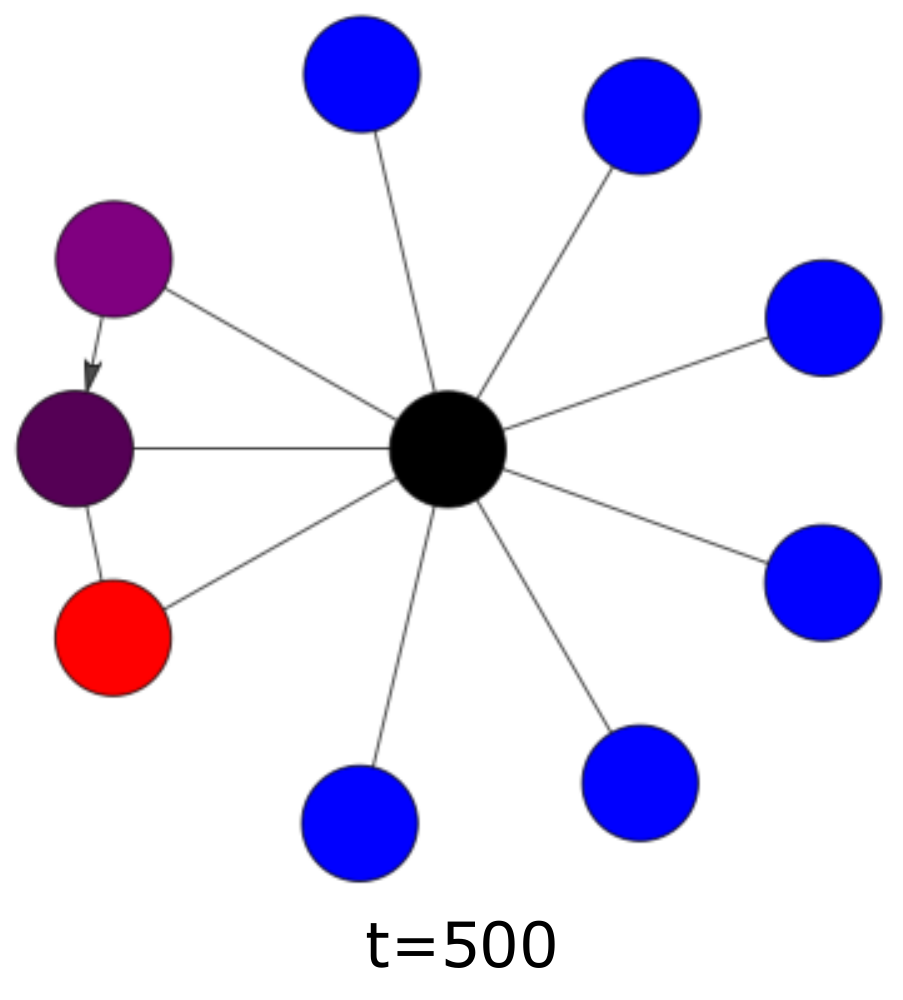}
		\includegraphics[width=0.22\linewidth]{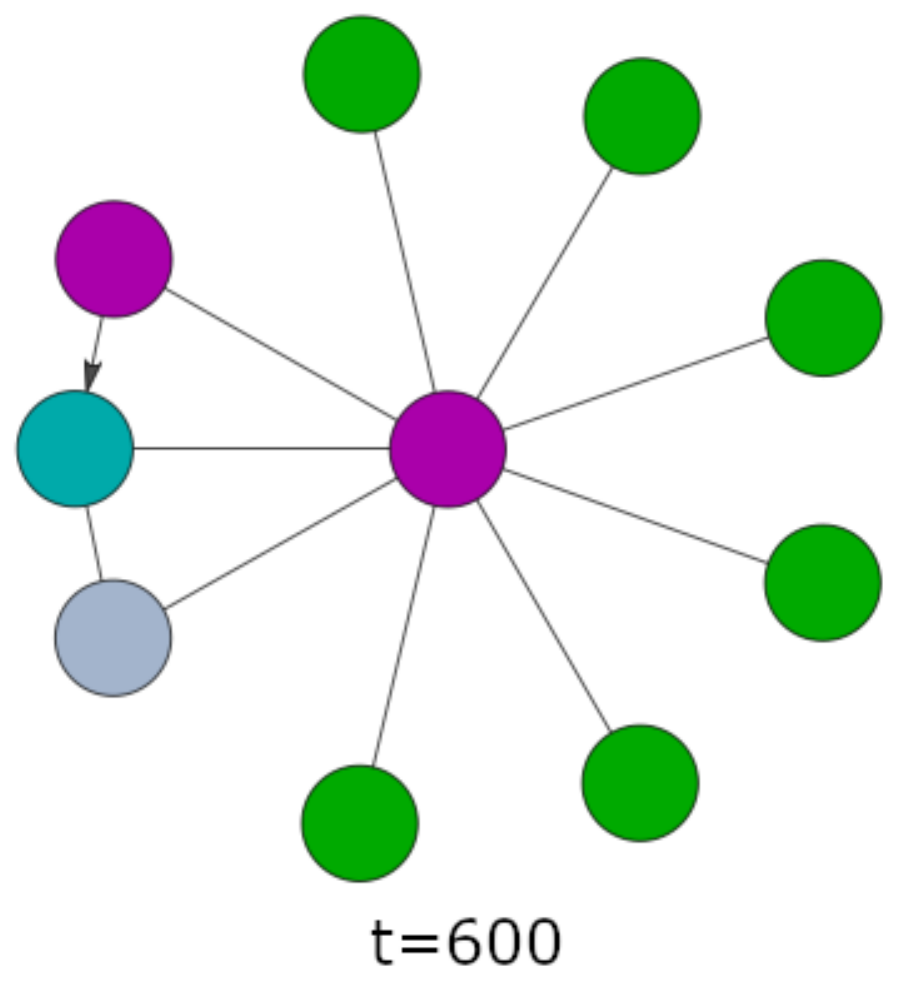}
		\includegraphics[width=0.22\linewidth]{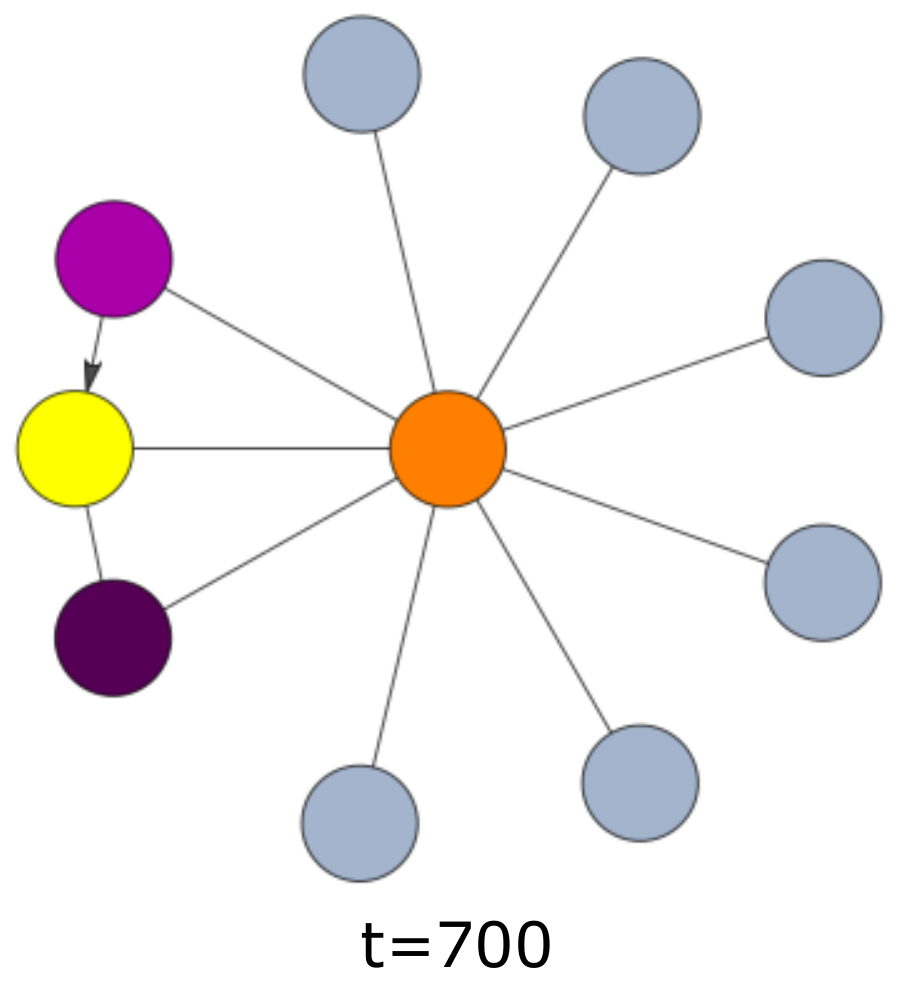}
		\includegraphics[width=0.22\linewidth]{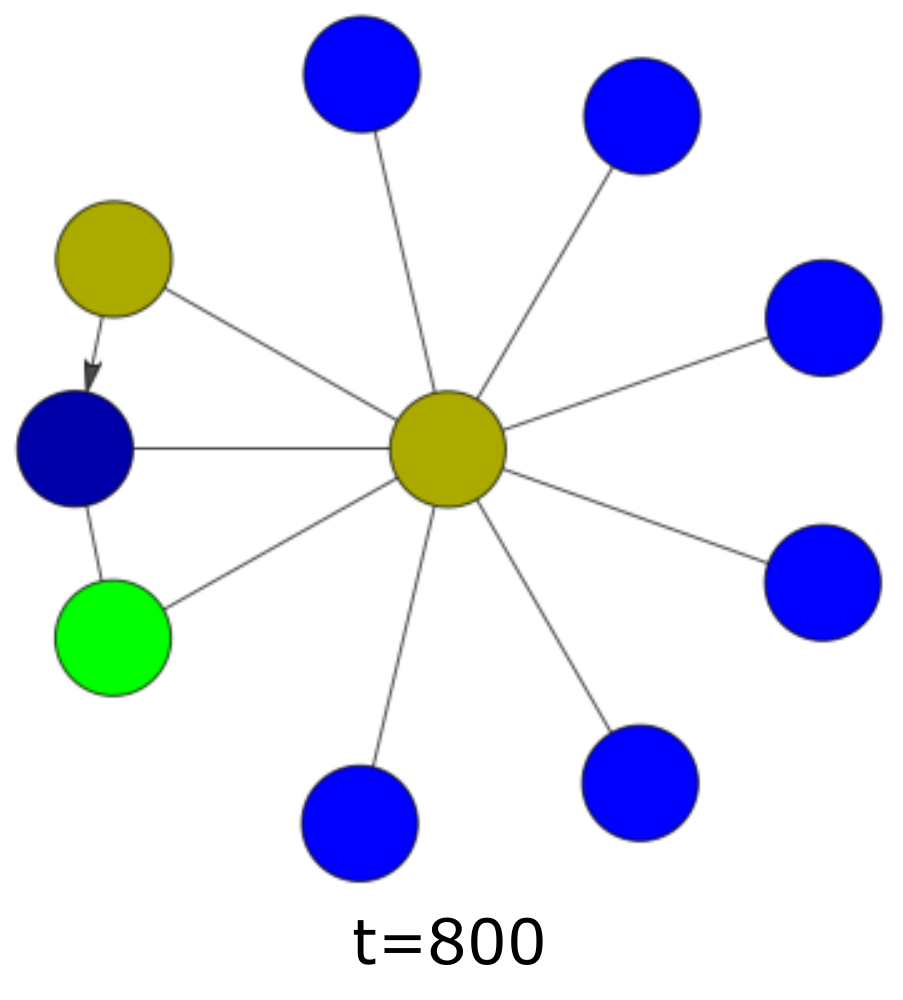}
		(g) \\
		\includegraphics[width=0.55\linewidth]{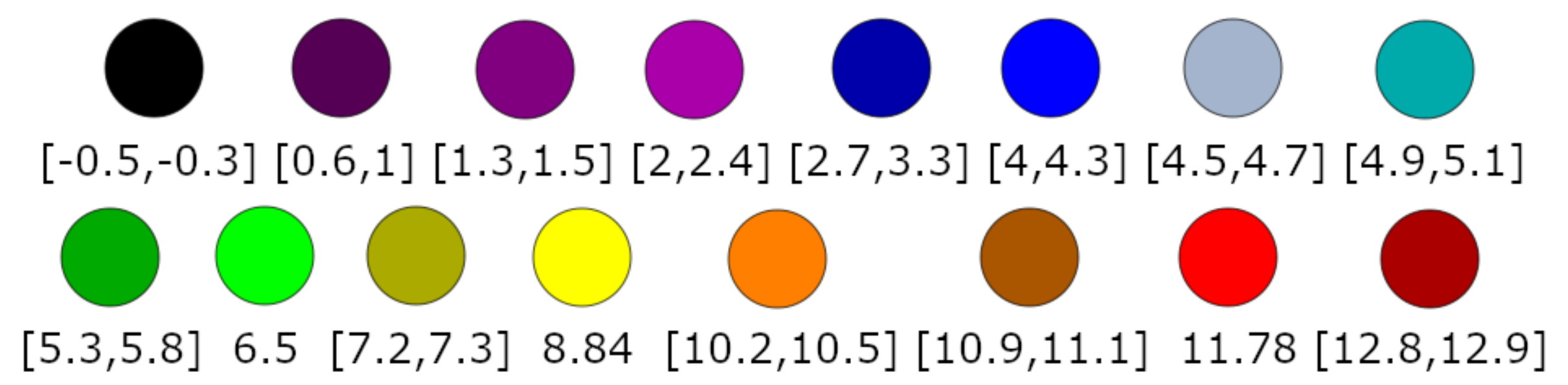}
		\caption{Patterning corresponding to the adjacency matrix \Eqref{Eq:Adjacency_Hyp_1}. In (a) we plot the eigenvalues of the matrix $M$. In (b),(c),(d) and (e) we plot $u_1,u_2,u_3$, and $u_4$ respectively, and, in (f), we plot $u_{5},\dots,u_{10}$. Finally, in (g) we give the pattern for the function $u$ at $t=100,200,300,400,500,600,700,800$.}
		\label{fig:M}
	\end{figure}

	A numerical example, of pattern formation resulting from \Eqsref{Eg_Hyp:udef}--\eqref{Eg_Hyp:u4}, is shown in \Figref{fig:M}. 
	In \subfig{fig:M}{(a)} we give the eigenvalues of the matrix $M$, corresponding to \Eqref{Eq:Adjacency_Hyp_1}. Here, we see that their are four eigenvalues (two conjugate pairs) with positive real part. In particular we find that $\text{Re}(\lambda_\ell)=0.74>0$, and hence we detect and instability. Interestingly, all of these eigenvalues also have a non-zero imaginary part. Recall that if the dominant eigenvalue (i.e., the eigenvalue with the largest real part) has a non-zero imaginary part then it is indicative of the \emph{Turing-wave instability} (also often just called a \emph{wave}-instability) \cite{Zhabotinsky:1995}. This means that we do not expect the unknowns to approach a time independent steady state. This behaviour is confirmed in \subfig{fig:M}{(b)--(f)}, where we see that the functions $u_{i}$ ($u_{1,i}$) \emph{does not} become constant in time. In \Figref{fig:M} we see that the values of $u_i$ at each node naturally separate into five groups, each of which oscillate with the same frequency and amplitude. The plots for each of these groups are given in \subfig{fig:M}{(b)--(f)}. In \subfig{fig:M}{(g)} we show the patterning resulting from $u_{i}$ with $i=1,\dots,n$ at different times.

	\subsection{Pattern formation with global reaction kinetics}
	\label{Sec:Pattern_formation_with_global_reaction_kinetics}
	In addition to being able to identify instabilities for non-diagonalizable networks, the theory outlined in \Sectionref{SubSec:Linear_instability_analysis_for_undirected_networks} can be applied when studying systems with \emph{global} reaction kinetics. In such a system the reaction kinetics can not only differ for each unknown but also at each node. The goal of this subsection here is to investigate an example of such system. For this, we consider the following equations
	\begin{align}
		\frac{d u_{i}}{d t} =\, & \frac{2}{100}\sum_{j=1}^{n}A_{ij}\left( u_{j} - u_{i} \right) + \beta_{i}\left( \frac{1}{10} - u_{i} + u_{i}^{2}v_{i} \right),
		\label{Eq:Global_u}
		\\
		\frac{d v_{i}}{d t} =\, & \sum_{j=1}^{n}A_{ij}\left( v_{j} - v_{i} \right) + \beta_{i}\left( 1 - u_{i}^{2}v_{i} \right),
		\label{Eq:Global_v}
	\end{align}
	where $\beta_i,i=1,\dots,n$ are freely specifiable constants. This particular choice (of global reaction kinetics) the reaction kinetics themselves maintain their function form but allow for different reaction rates at each node i.e., different $\beta_{i}$'s. In this subsection here we assume that the adjacency matrix $A_{ij}$ is given by \Eqref{Eq:Adjacency_1}. Moreover, in this subsection we restrict ourselves to $n=16$ nodes. Finally, it is worth noting here that one can apply the analysis presented in \cite{Asllani:2014} only in the special case $\beta_{i}=1$ (provided the underlying network is diagonalizable).  
	\begin{figure}[t!]
		\centering
		\includegraphics[width=0.3\linewidth]{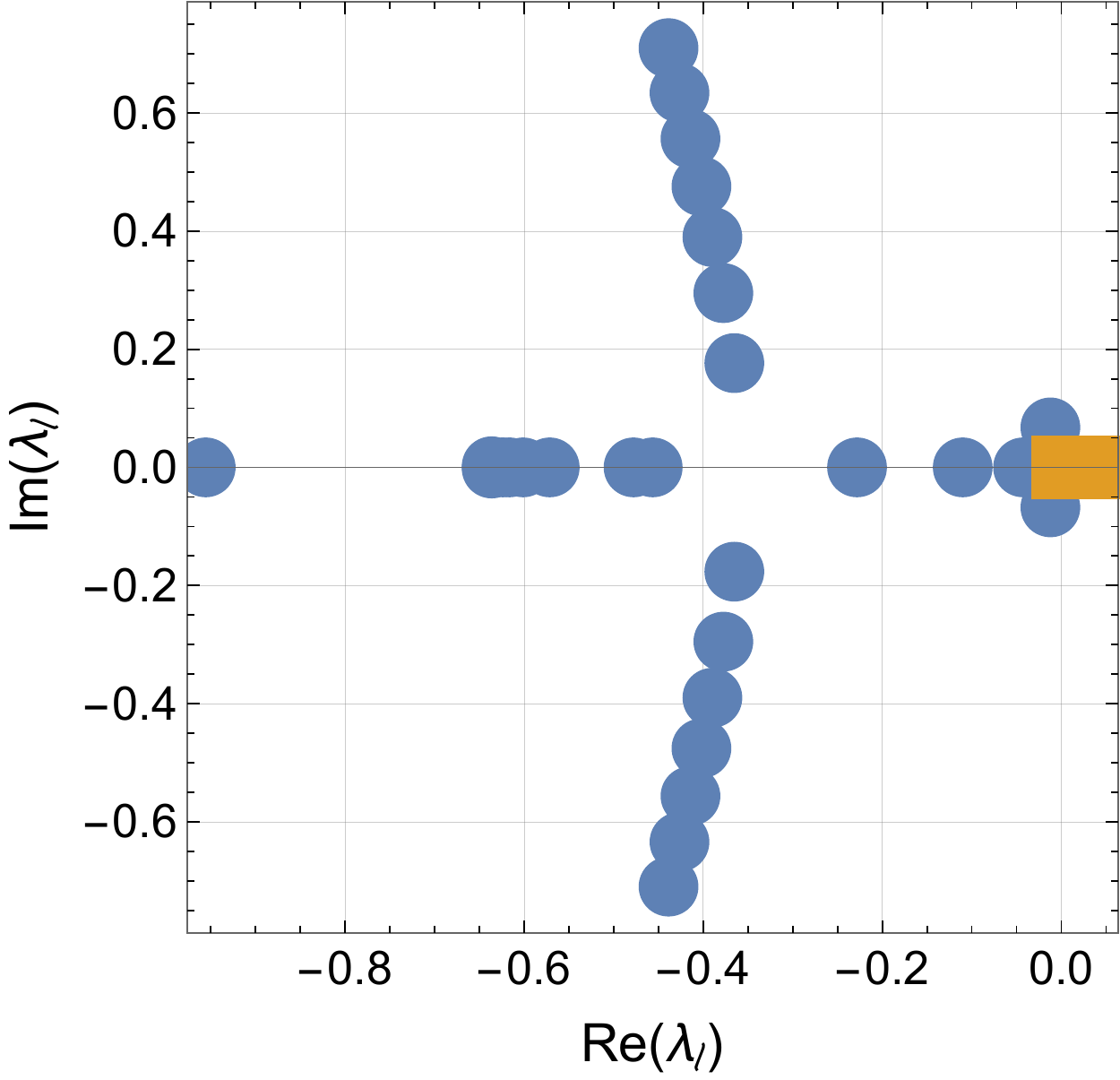}
		(a)
		\includegraphics[width=0.3\linewidth]{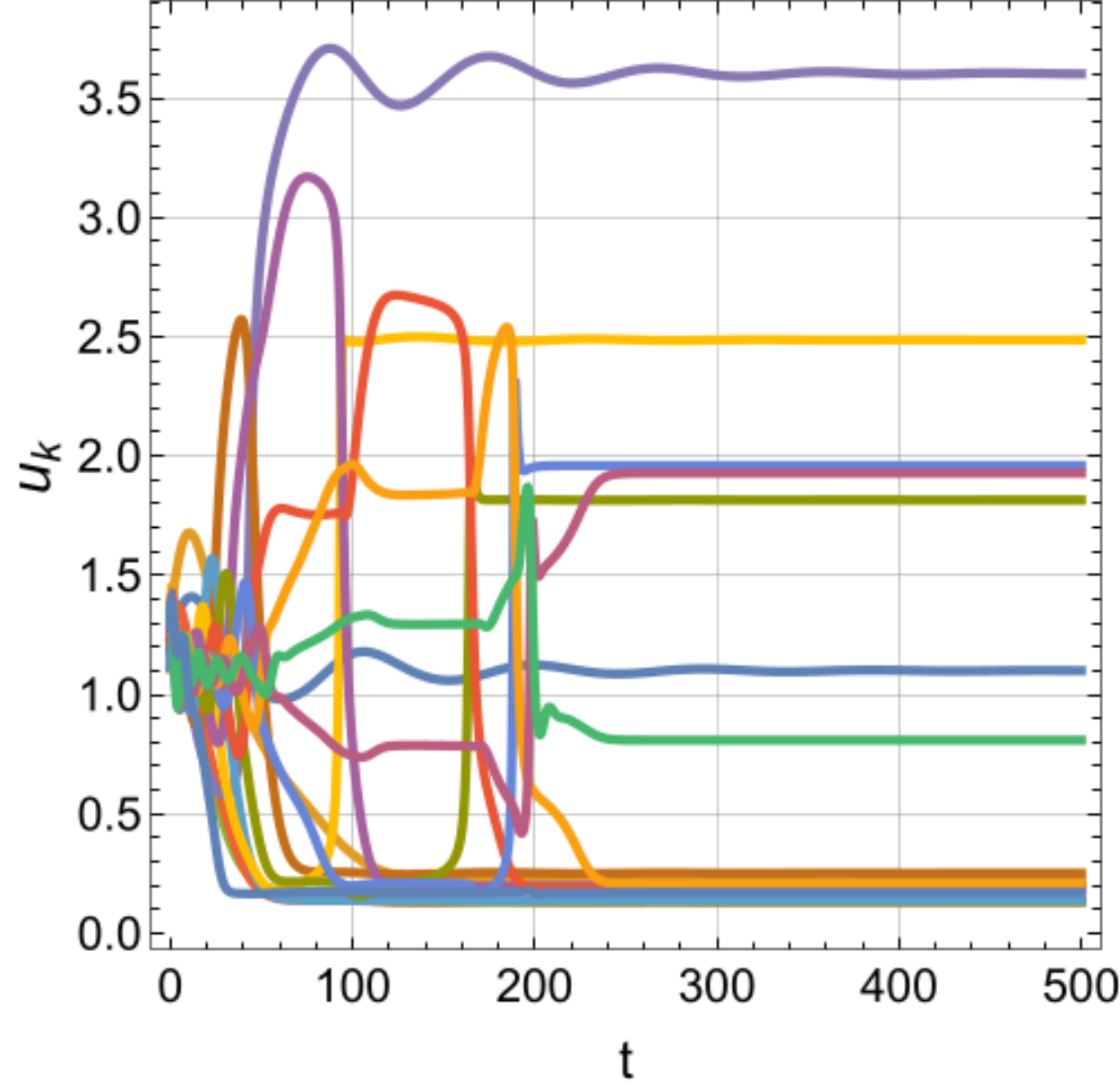}
		(b)
		\includegraphics[width=0.25\linewidth]{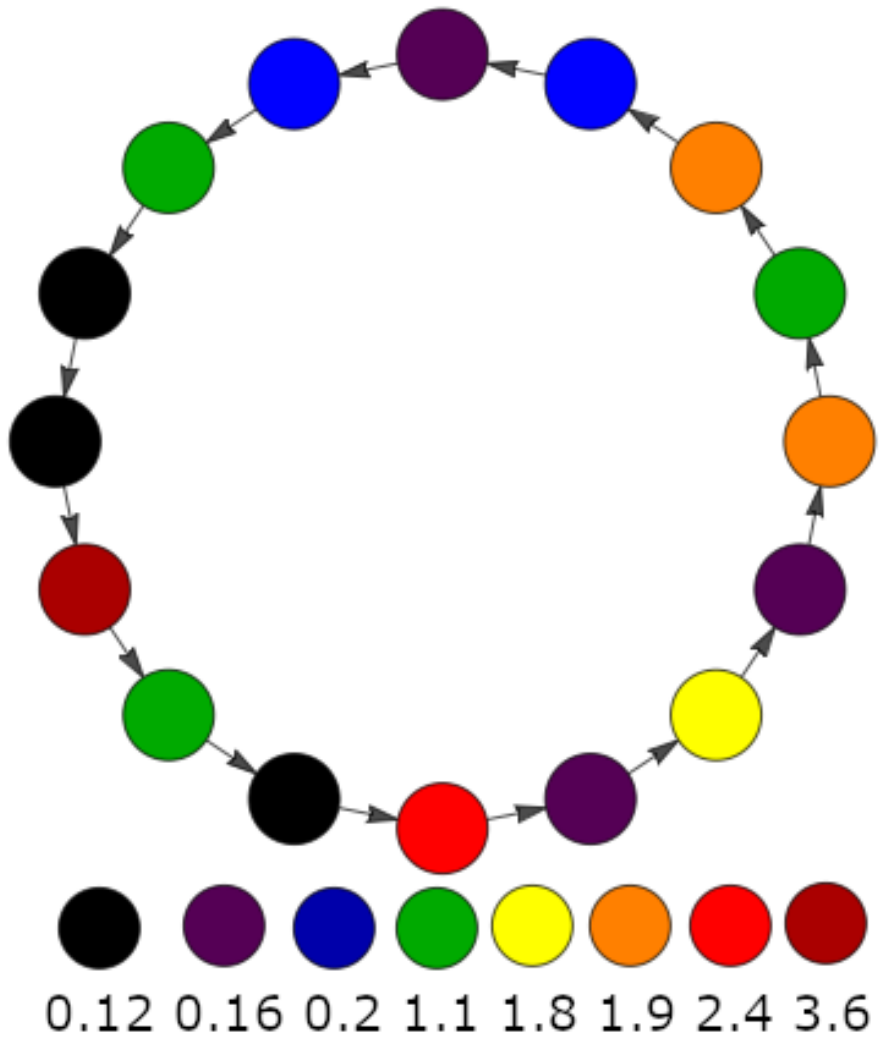}
		(c)
		\caption{Patterning corresponding to the adjacency matrix \Eqref{Eq:Adjacency_1} with $a_{1}=0,a_{2}=1$, and where the constants $\alpha_{i},i=1,\dots,16$ are given by \Eqref{Eq:Global_i_n}. In (a) we give the eigenvalues of the matrix M. In (b) we show the behaviour of the function $u$, for each of the nodes. Finally, in (c) we give the pattern for the function $u$ at $t=500$.}
		\label{fig:i_n}
	\end{figure}

	Let us now consider two specific choices of the constants $\beta_i$ with $i=1,\dots,n$. We first consider the choice
	\begin{align}
		\beta_{i}=\frac{i}{n}=\frac{i}{16},
		\label{Eq:Global_i_n}
	\end{align}
	where we have used that $n=16$. The numerical results corresponding to \Eqref{Eq:Global_i_n} are shown in \Figref{fig:i_n} for the choices $a_{1}=a_{2}=0$. For these values of $a_1$ and $a_2$ the adjacency matrix \Eqref{Eq:Adjacency_1} describes an incomplete cycle graph. In \subfig{fig:i_n}{(a)} we give the eigenvalues of the matrix $M$. In this case there are four eigenvalues with positive real part. In particular, we find that $\text{Re}(\lambda_\ell)=0.04>0$ and hence we detect an instability. Here, we see that, for the choices we have made, the system \Eqsref{Eq:Global_u}--\ref{Eq:Global_v} generates a patterned state. This is confirmed in \subfig{fig:i_n}{(b)}, where we see that the functions $u_i$ form a stable pattern at around $t\approx 45$. The resulting pattern is shown in \subfig{fig:i_n}{(c)}. It is interesting to compare \Figref{fig:i_n} to the second row (plots (d)--(f)) of \Figref{fig:m}, where we found that a pattern did not emerge. This tells us that the pattern shown in \Figref{fig:i_n} is purely a consequence of the global reaction kinetics. 
	
	\begin{figure}[t!]
		\centering
		\includegraphics[width=0.4\linewidth]{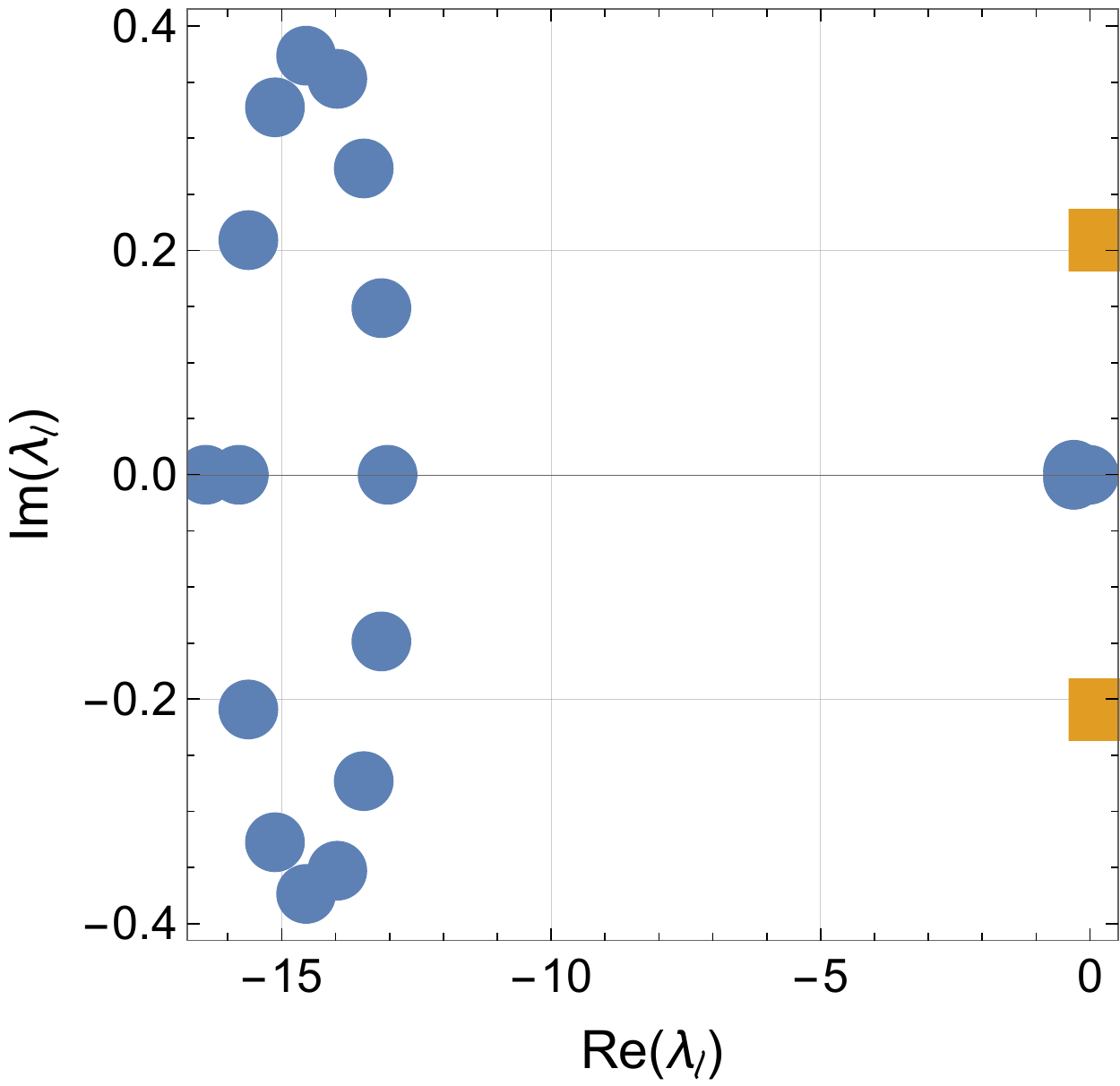}
		\caption{The eigenvalues of the matrix $M$ corresponding to the adjacency matrix \Eqref{Eq:Adjacency_1} with $a_{1}=0,a_{2}=1$, and where the constants $\alpha_{i},i=1,\dots,16$ are given by \Eqref{Eq:Global_PieceWise}.}
		\label{fig:m_glob}
	\end{figure}
	Let us now consider a different choice of the constants $\beta_{i},i=1,\dots,16$. In particular we consider the choice
	\begin{align}
		\beta_{i}=
		\begin{cases}
			1, &\quad\text{if}\quad i=1,
			\\
			0, &\quad\text{otherwise}.
		\end{cases}
		\label{Eq:Global_PieceWise}
	\end{align}
	Here, the reaction kinetics are non-zero at node-$1$ only, with no interaction between the species $u_i$ and $v_i$ occurring at any of the other nodes. The eigenvalues of the matrix $M$ corresponding to \Eqref{Eq:Global_PieceWise} are shown in \Figref{fig:m_glob} for the choices $a_{1}=0$ and $a_{2}=1$. Here, there are two eigenvalues (one conjugate pair) with positive real part $\text{Re}(\lambda_\ell)=0.17>0$. In this case the underlying network is a complete cycle graph. In \Figref{fig:m_glob} we see that the matrix $M$ has two eigenvalues with positive real parts. Interestingly, both of these eigenvalues also have a non-zero real part, in fact the two eigenvalues are conjugate pairs. Recall that if the dominant eigenvalue (i.e., the eigenvalue with the largest real part) has a non-zero imaginary part then it is indicative of \emph{Turing-wave instability}. This means that we do not expect the unknowns to approach a time independent solution.  
	
\begin{figure}[t!]
		\centering
		\includegraphics[width=0.27\linewidth]{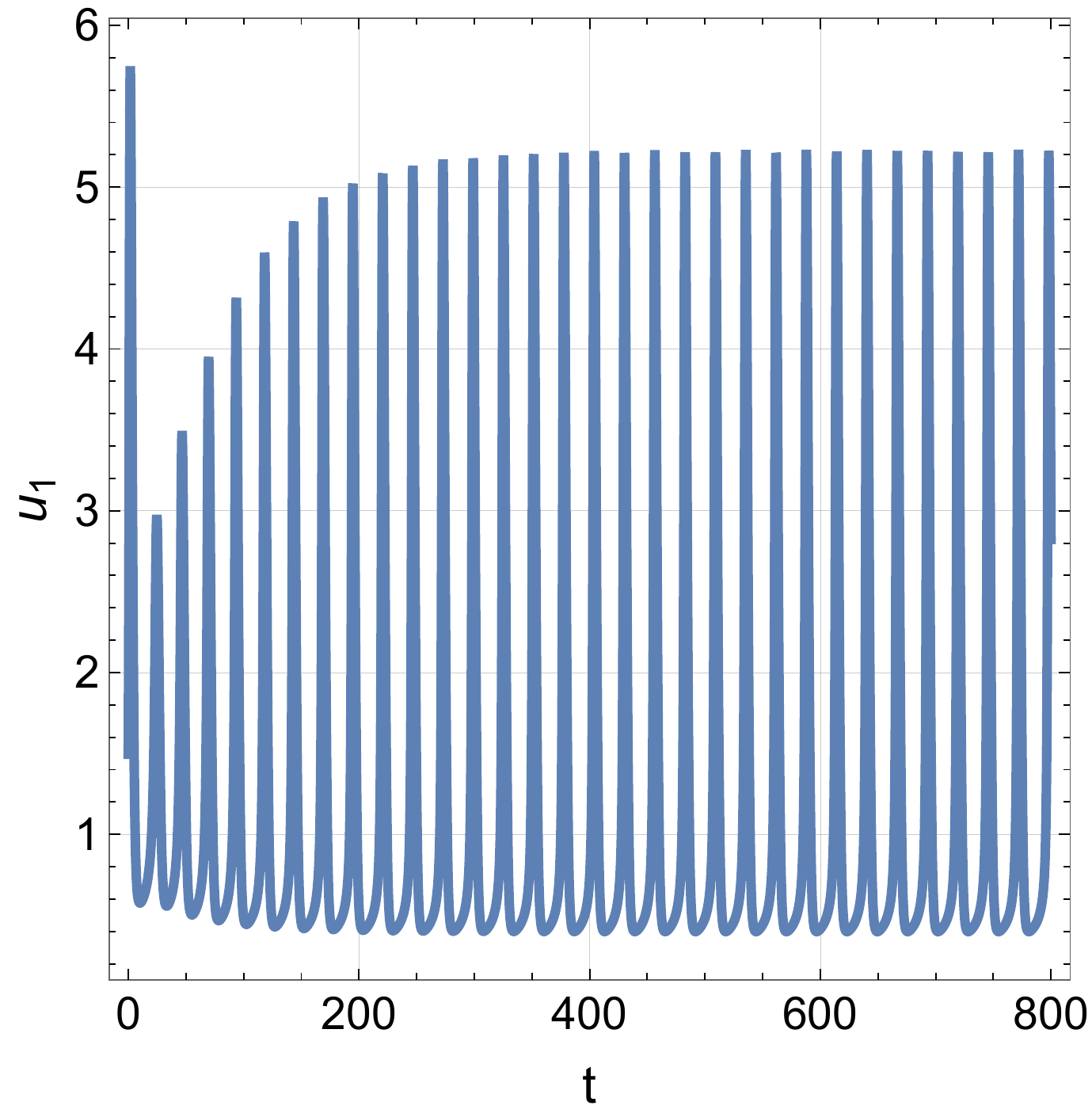}
		(a)
		\includegraphics[width=0.28\linewidth]{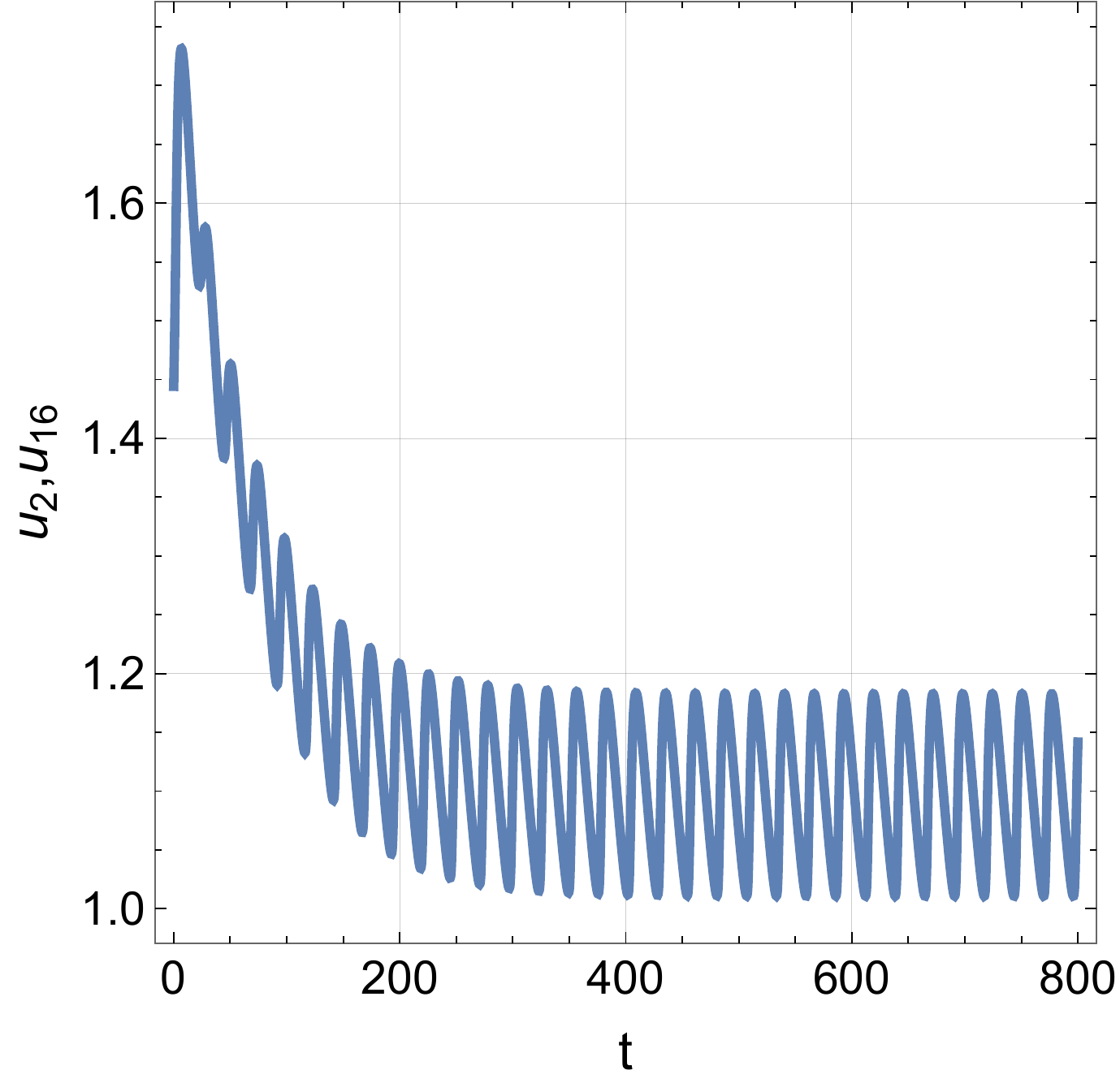}
		(b)
		\includegraphics[width=0.28\linewidth]{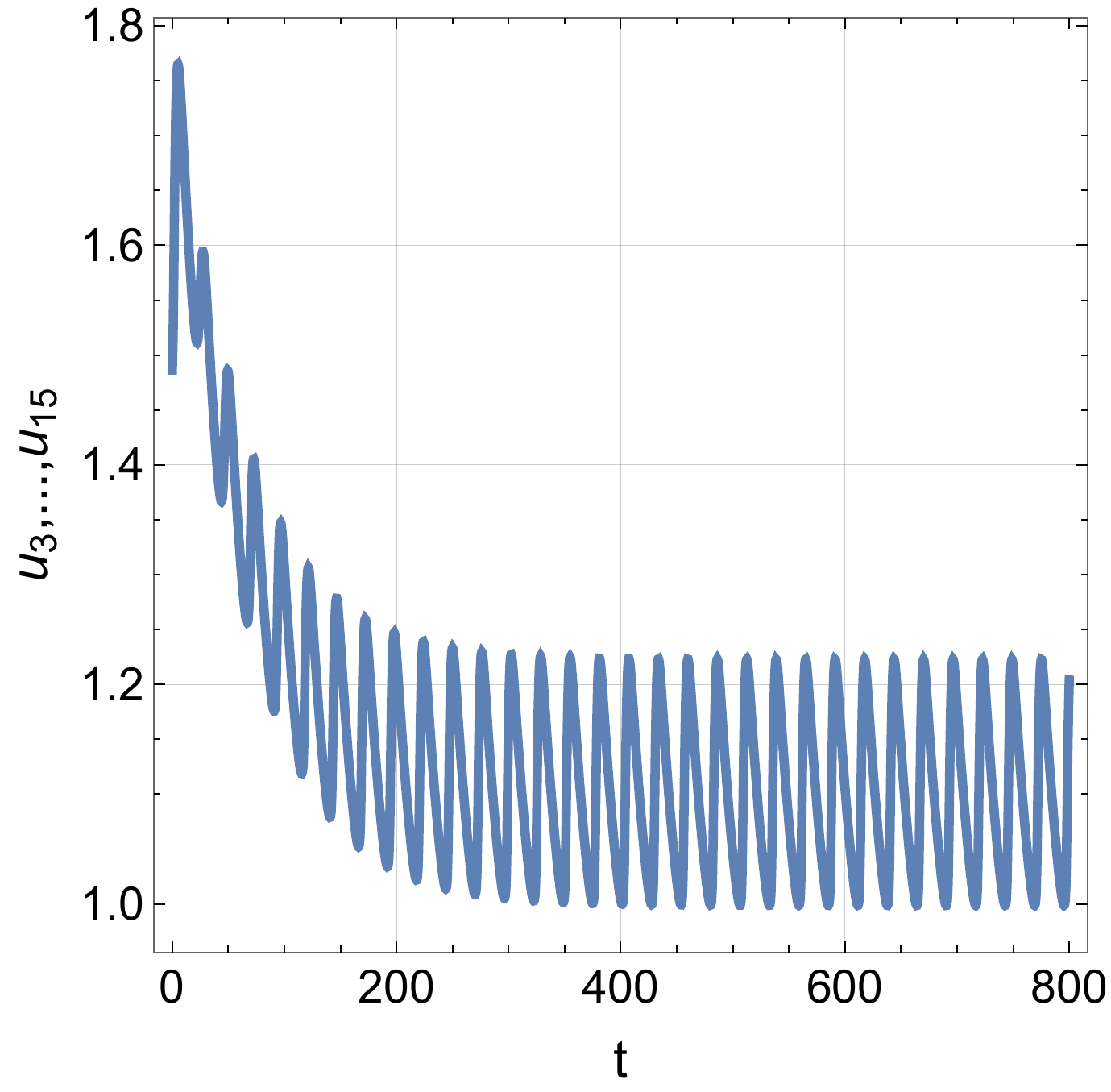}
		(c) \\
		\includegraphics[width=0.23\linewidth]{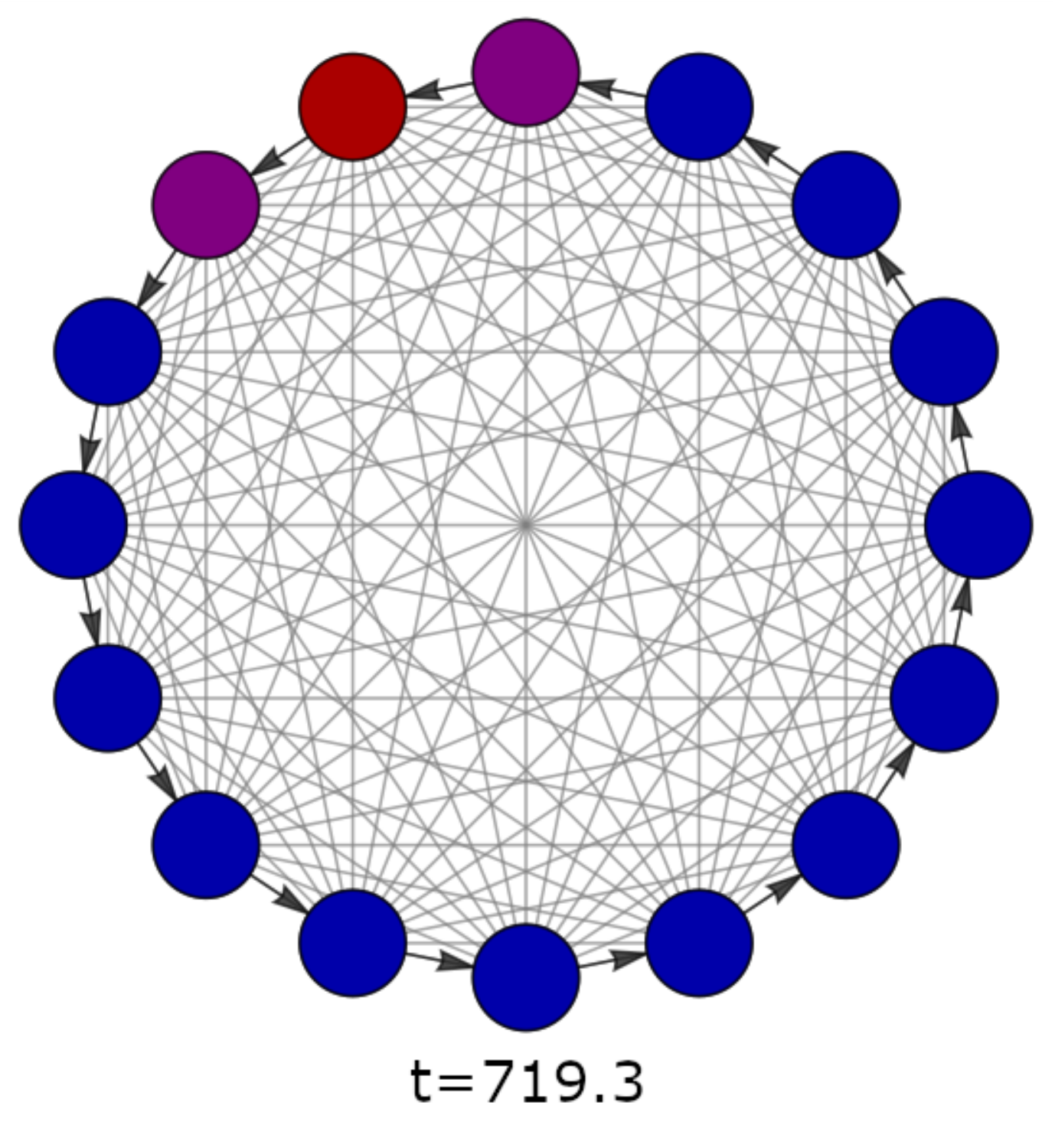}
		\includegraphics[width=0.23\linewidth]{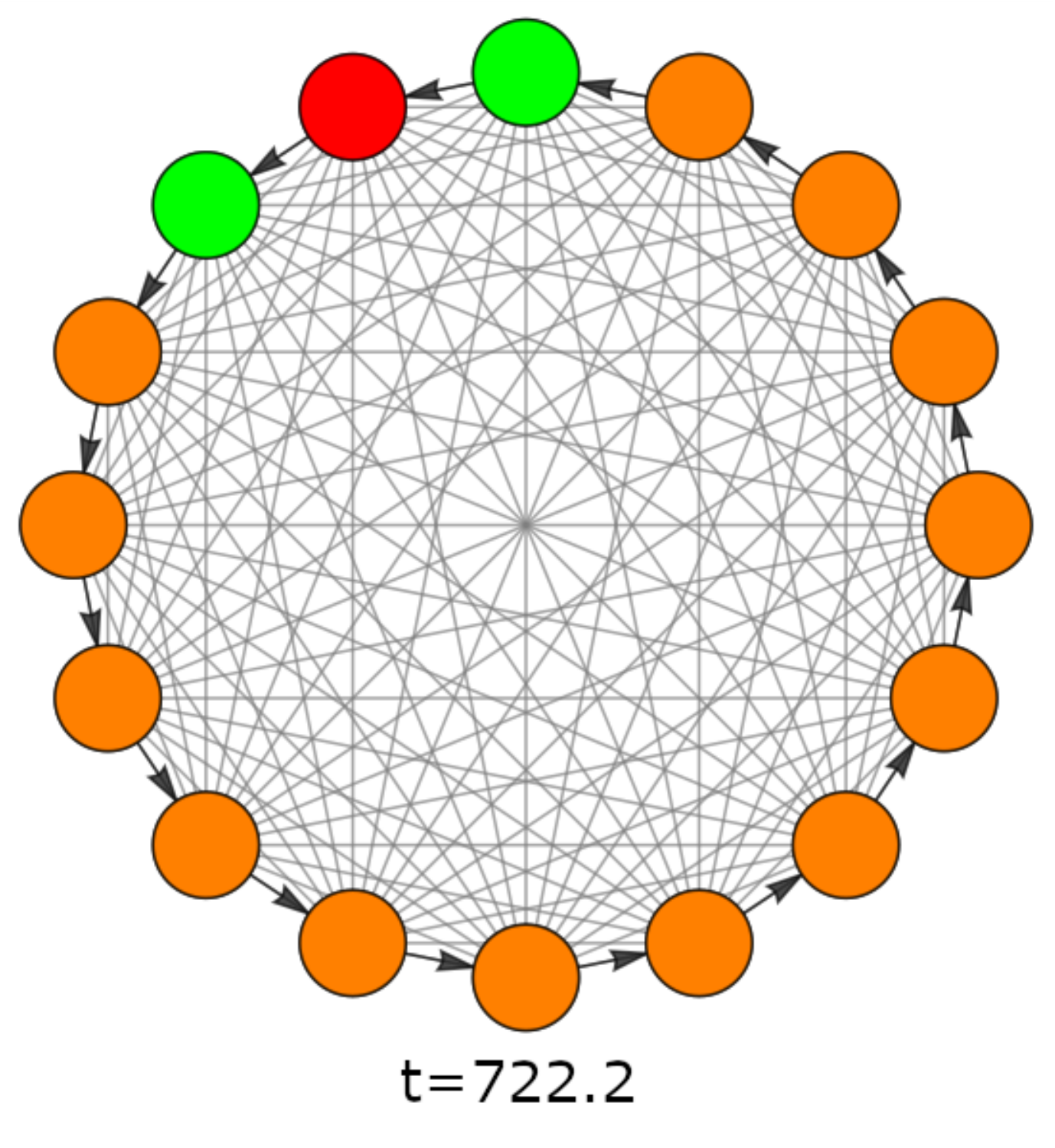}
		\includegraphics[width=0.23\linewidth]{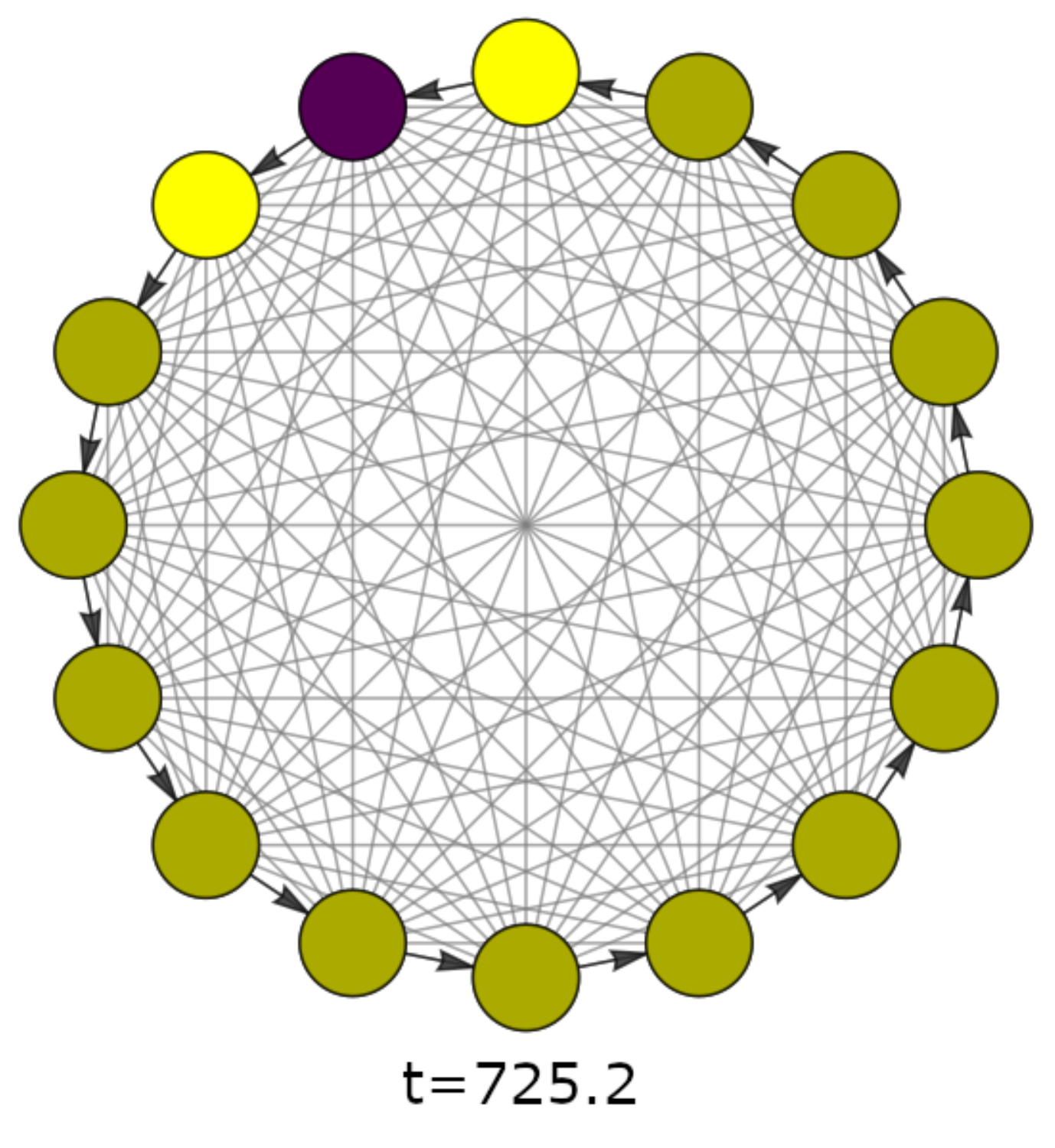}
		\includegraphics[width=0.23\linewidth]{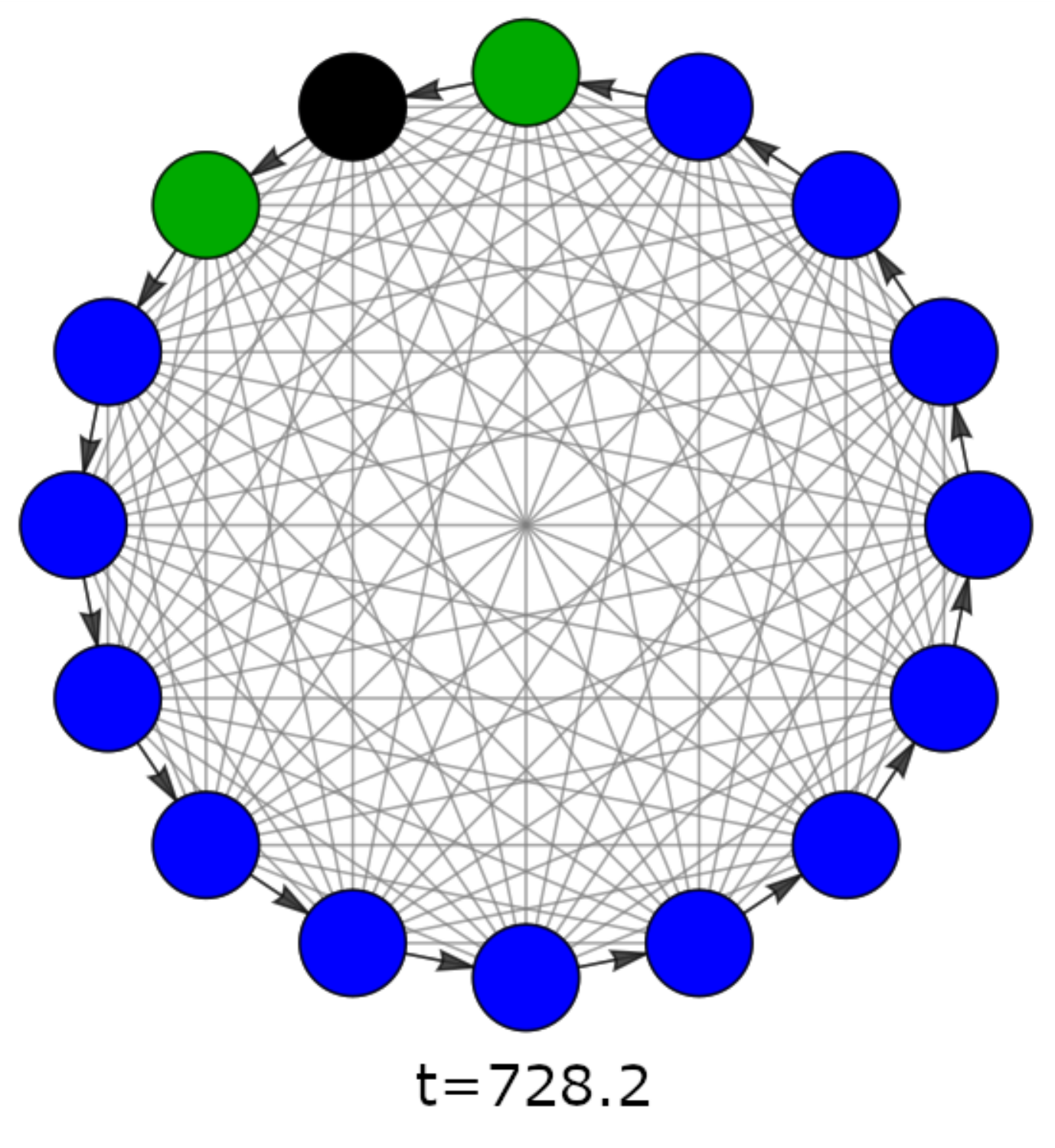}
		(d) \\
		\includegraphics[width=0.5\linewidth]{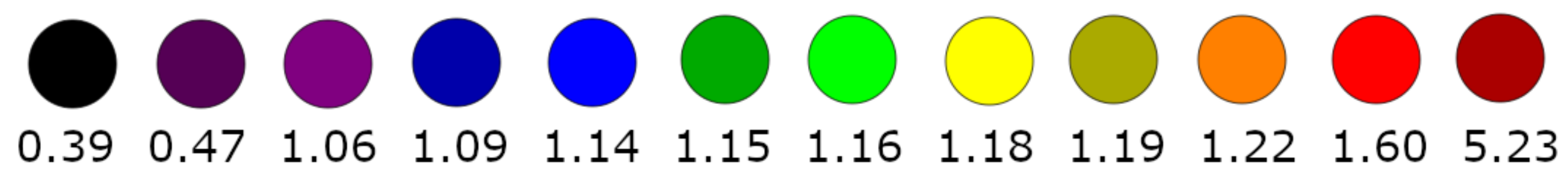}
		\caption{Patterning corresponding to the adjacency matrix \Eqref{Eq:Adjacency_1} with $a_{1}=0,a_{2}=1$, and where the constants $\beta_{i},i=1,\dots,16$ are given by \Eqref{Eq:Global_PieceWise}. In (a) we plot the function $u_1$, in (b) we plot the functions $u_2$ and $u_{16}$, and in (c) we plot the values of the function $u$ at $t=719.2,722.2,725.2,728.2$.}
		\label{fig:m_Limit}
	\end{figure}

	This behaviour is numerically confirmed in \Figref{fig:m_Limit}, where we see that the functions $u_i$, for $i=1,\dots,16$, approach limit-cycles. In \Figref{fig:m_Limit} we see that the values of $u_i$ at each node naturally separate into three groups. The first of these groups contains only $u_1$ and is shown in \subfig{fig:m_Limit}{(a)}. Here we see that $u_1$ takes the most extreme values but nevertheless approaches a stable limit-cycle at around $t\approx 300$. In \subfig{fig:m_Limit}{(b)} we show the second grouping, which consists of $u_2$ and $u_{16}$. Similarly, the third group is shown in \subfig{fig:m_Limit}{(c)}, and consists of $u_3$--$u_{15}$. Note that while group two and three appear to oscillate at the same frequency, they have different amplitudes. As with group one, both groups two and three approach a stable limit-cycle at around $t\approx 300$. We show one full period of the limit cycle in \subfig{fig:m_Limit}{(d)}, at four moments of time. It is interesting to compare this Turing-wave instability to the one shown in \Figref{fig:M}. In \Figref{fig:m_Limit} we clearly see that the unknown approaches a stable limit cycle. Conversely, in \Figref{fig:M} we see that although $u_i$ appears to remain bounded its behaviour is somewhat more chaotic, and as such it is unclear whether or not the resulting behaviour describes a stable limit cycle. 
	
	\begin{figure}[t!]
		\centering
		\includegraphics[width=0.3\linewidth]{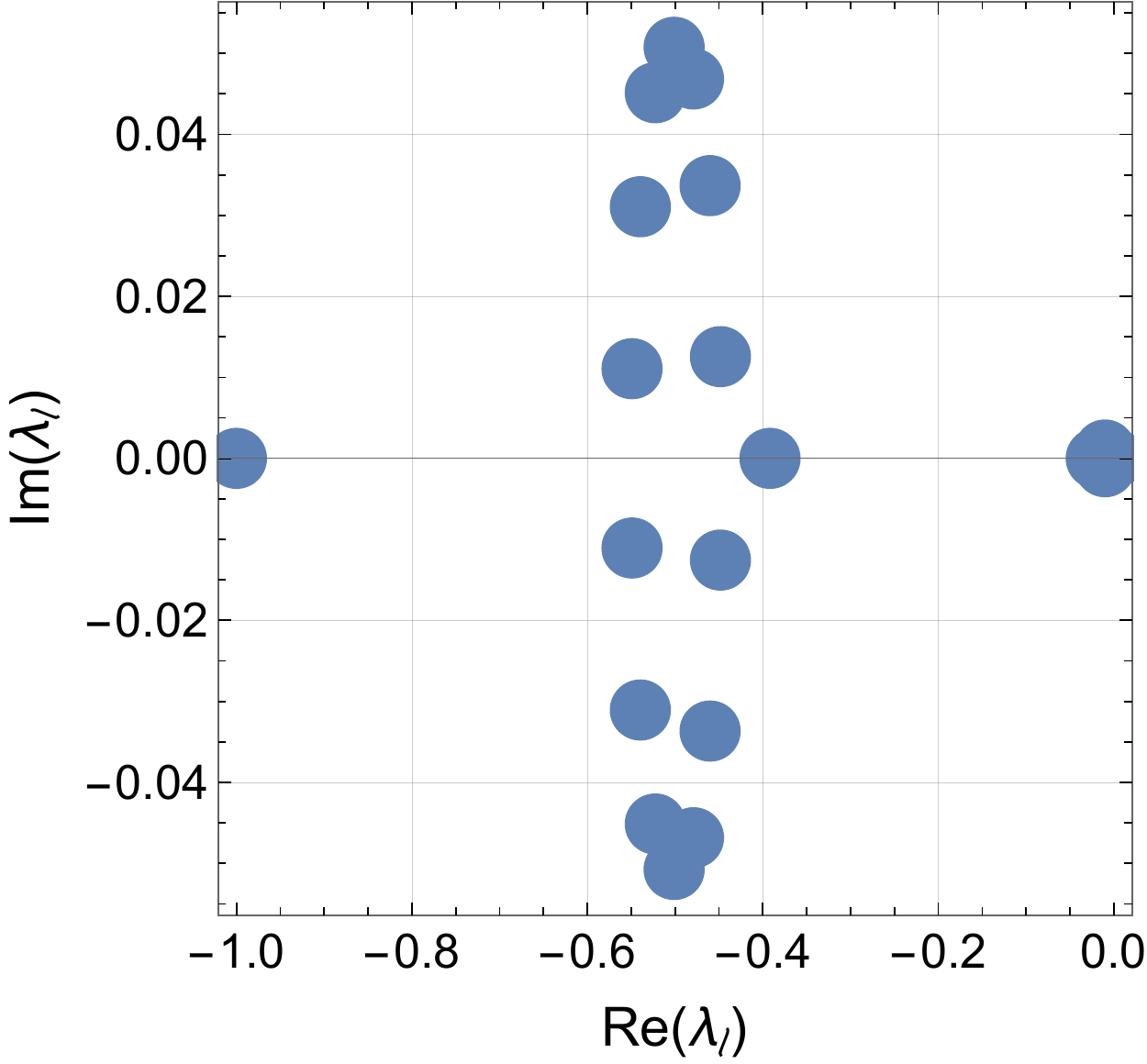}
		(a)
		\includegraphics[width=0.3\linewidth]{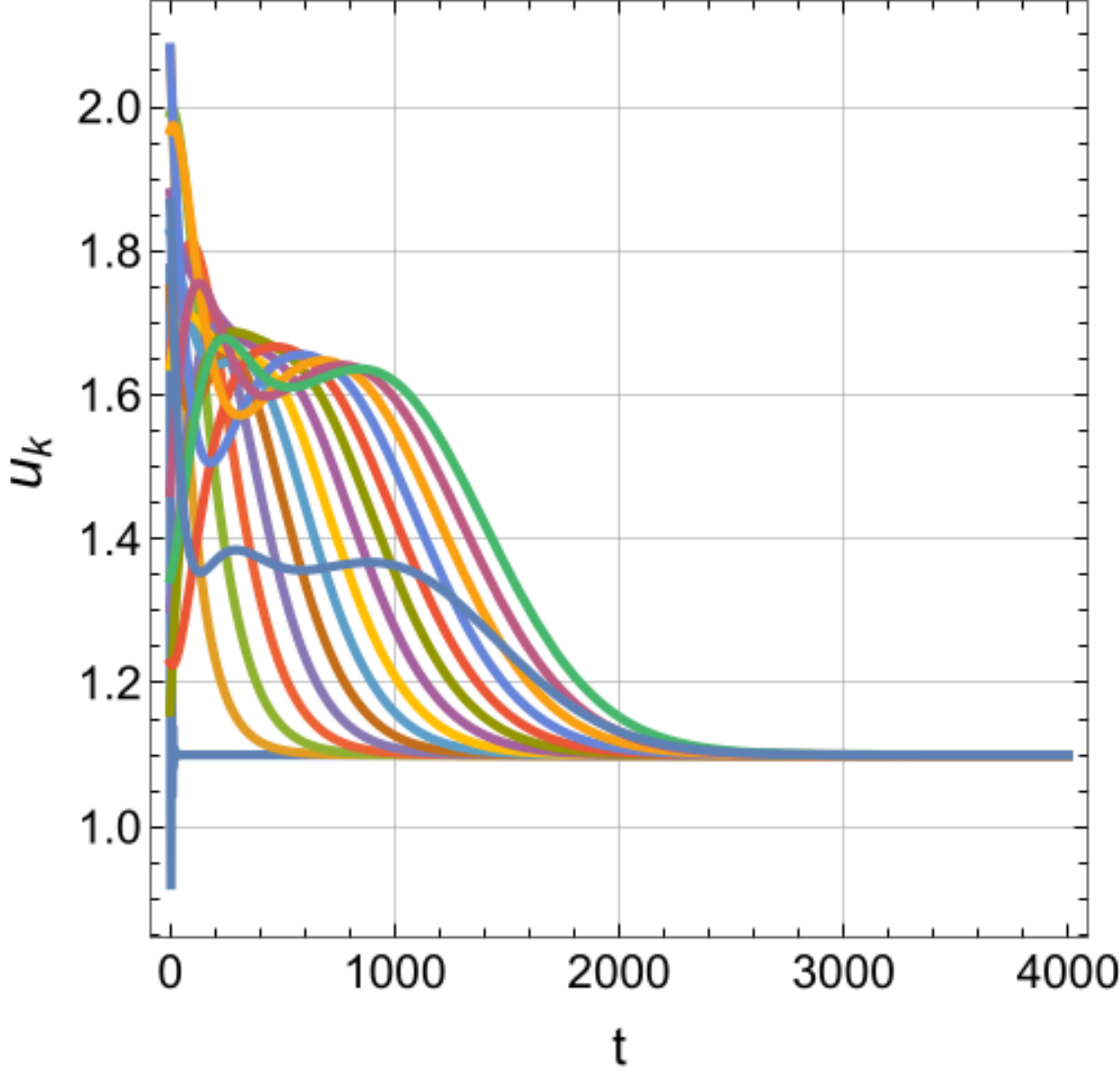}
		(b)
		\includegraphics[width=0.25\linewidth]{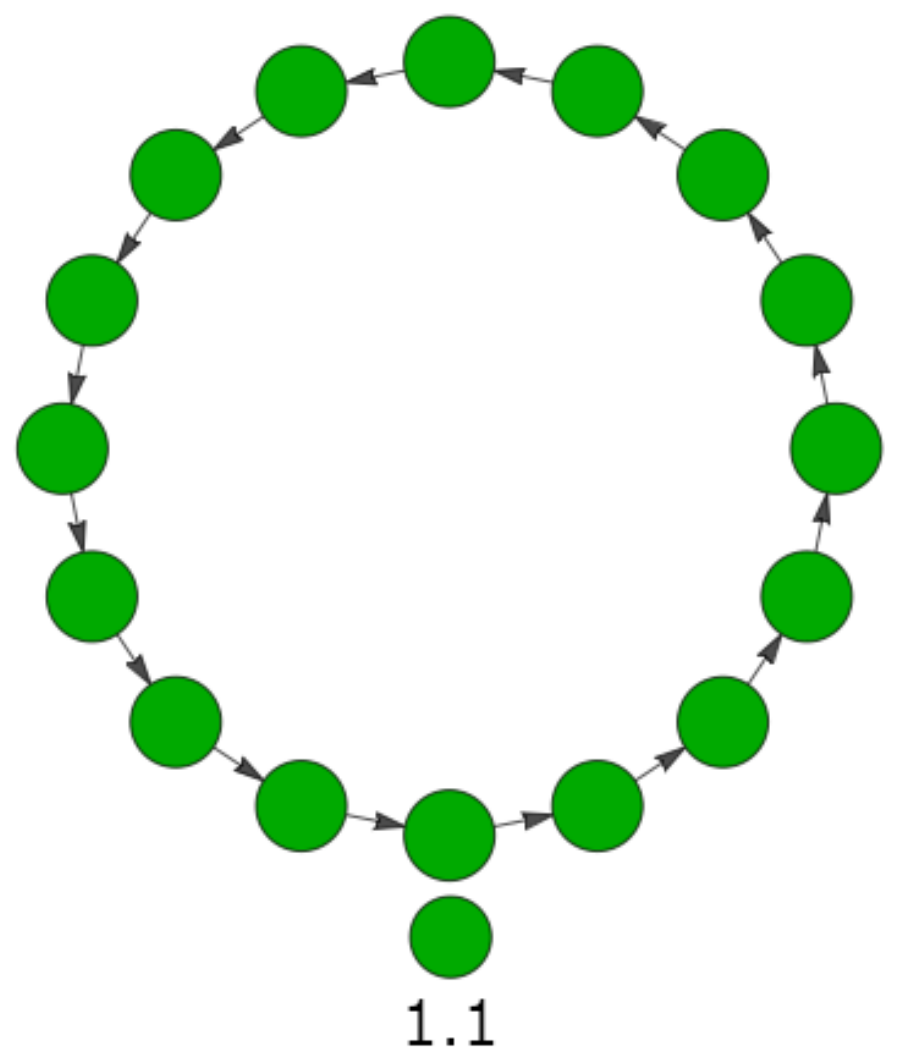}
		(c)
		\caption{Lack of patterning corresponding to the adjacency matrix \Eqref{Eq:Adjacency_1} with $a_{1}=a_{2}=0$, and where the constants $\alpha_{i},i=1,\dots,16$ are given by \Eqref{Eq:Global_PieceWise}. In (a) we give the eigenvalues of the matrix $M$. In (b) we show the behaviour of the function $u$, for each of the nodes. Finally, in (c) we give show the lack of a pattern for the function $u$ at $t=4000$.}
		\label{fig:glob_noPattern}
	\end{figure}
	Let us now end this subsection by discussing the effect of global reaction kinetics. One may be tempted to take the perspective that global reaction kinetics generically lead to patterning. From this perspective one argues that if the unknowns interact differently at each of the nodes then one would also expect their final values to vary from node to node. Although there is some merit to this argument is worth noting that the underlying topology of the graph can still have a significant effect on the formation of patterning. To demonstrate this we once again pick the constants $\beta_{i},i=1,\dots,n$ as in \Eqref{Eq:Global_PieceWise}. Furthermore, we suppose the adjacency matrix $A_{ij}$ describes an incomplete cycle graph (which corresponds to the choices $a_{1}=a_{2}=0$). The numerical results for this these choices are shown in \Figref{fig:glob_noPattern}. In \subfig{fig:glob_noPattern}{(a)} we show the eigenvalues of the matrix $M$. Here we see that our linear theory does \emph{not} predict patterning, with $\text{Re}(\lambda_\ell)=-0.01<0$. Consistent with this we see that $u_i$, shown in \subfig{fig:glob_noPattern}{(b)}, exhibits non-trivial behaviour for a short time only before approaching the steady state solution at around $t\approx 2200$. In \subfig{fig:glob_noPattern}{(c)} we show the values of $u$ on each node at $t=4000$. This example shows that even though global reaction kinetics can allow a large class of equations to generate patterns, the underlying topology nevertheless has a significant effect on the emergence of a pattern.

	\section{Numerical examples of pattern formation of temporal directed networks}
	\label{Sec:Numerical_examples_of_pattern_formation_of_temporal_directed_networks}
	We now consider examples of pattern formation on directed temporal networks. The goal of this section is to provide examples of the various types on non-autonomous reaction diffusion equations that can lead to pattern formation.  In this section, as in \Sectionref{Sec:Numerical_examples_of_pattern_formation_on_static_directed_networks}, we consider only two types of networks. We once again emphasise that many of the results we present here can be applied to other types of networks. However, we find that the ones considered here are sufficient for our purposes.  
	\subsection{Numerical methods}
	The numerical methods we employ here, when solving equations of the form \Eqref{Eq:Gen_Temp}, are largely the same as in \Sectionref{Sec:Numerical_methods}. In addition to numerically solving systems of the form \Eqref{Eq:Gen_Temp}, we also solve the corresponding linearized system \Eqref{Eq:Gen_Temp_Lin}. Recall that $U_{\alpha,i}$ is the unknown of the linear system \Eqref{Eq:Gen_Temp_Lin}. In regards to initial data we set $U_{\alpha,i}(0)=\xi_{\alpha,i}$, where $\xi_{\alpha,i}$ is a random real number which we calculate using the Mathematica function \emph{RandomReal[1]}. In order to present our numerical results it is useful to now introduce the following quantity: 
	\begin{align}
		w = \max_{i}\left\{ \|\mathbf{U}_{1}\|,\dots,\|\mathbf{U}_{n} \| \right\},
		\quad
		\|\mathbf{U}_{k}\| = \left( \sum_{\beta=1}^{m}U_{\beta,k}^2 \right)^{1/2}.
	\end{align}
	If there is an instability, then one expects that $w\rightarrow \infty$ as $t\rightarrow \infty$. Conversely, if the steady state is linearly stable then $w\rightarrow 0$ as $t\rightarrow \infty$.	Finally, for later convenience, we define the function $\Omega(t,\eta):\mathbb{R}^2\rightarrow \mathbb{R}$ as
	\begin{align}
		\Omega(t,\eta)=\frac{1}{2}\tanh\left( \frac{1}{10}( t- \eta ) \right).
	\end{align}
	In this section here we use $\Omega(t,\eta)$ as a smooth switch function, where $t=\eta$ is the ``switch point''.

	\subsection{Pattern formation on directed temporal networks}
\begin{figure}[t!]
		\centering
		\includegraphics[width=0.28\linewidth]{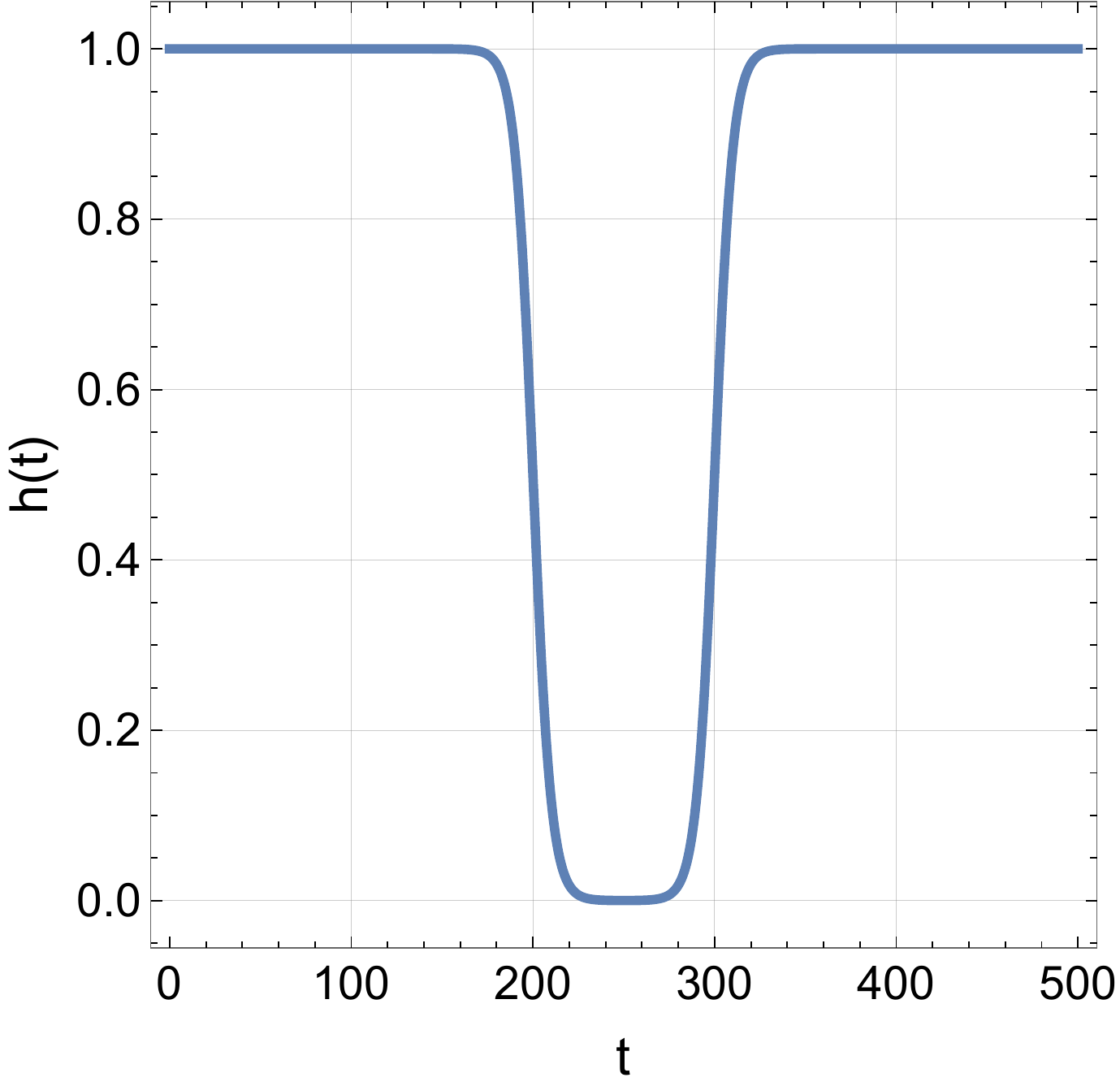}
		(a)
		\includegraphics[width=0.28\linewidth]{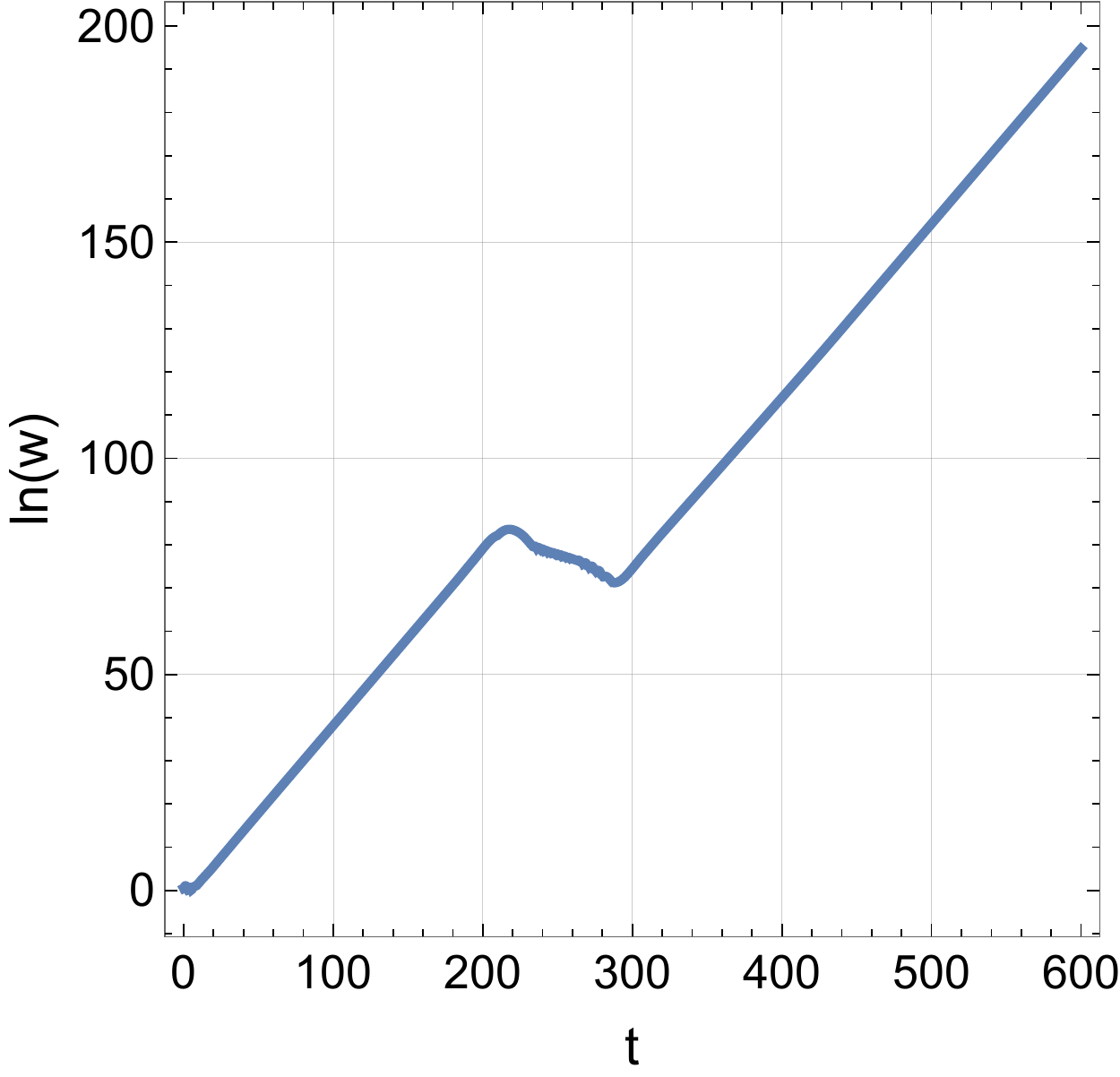}
		(b)
		\includegraphics[width=0.27\linewidth]{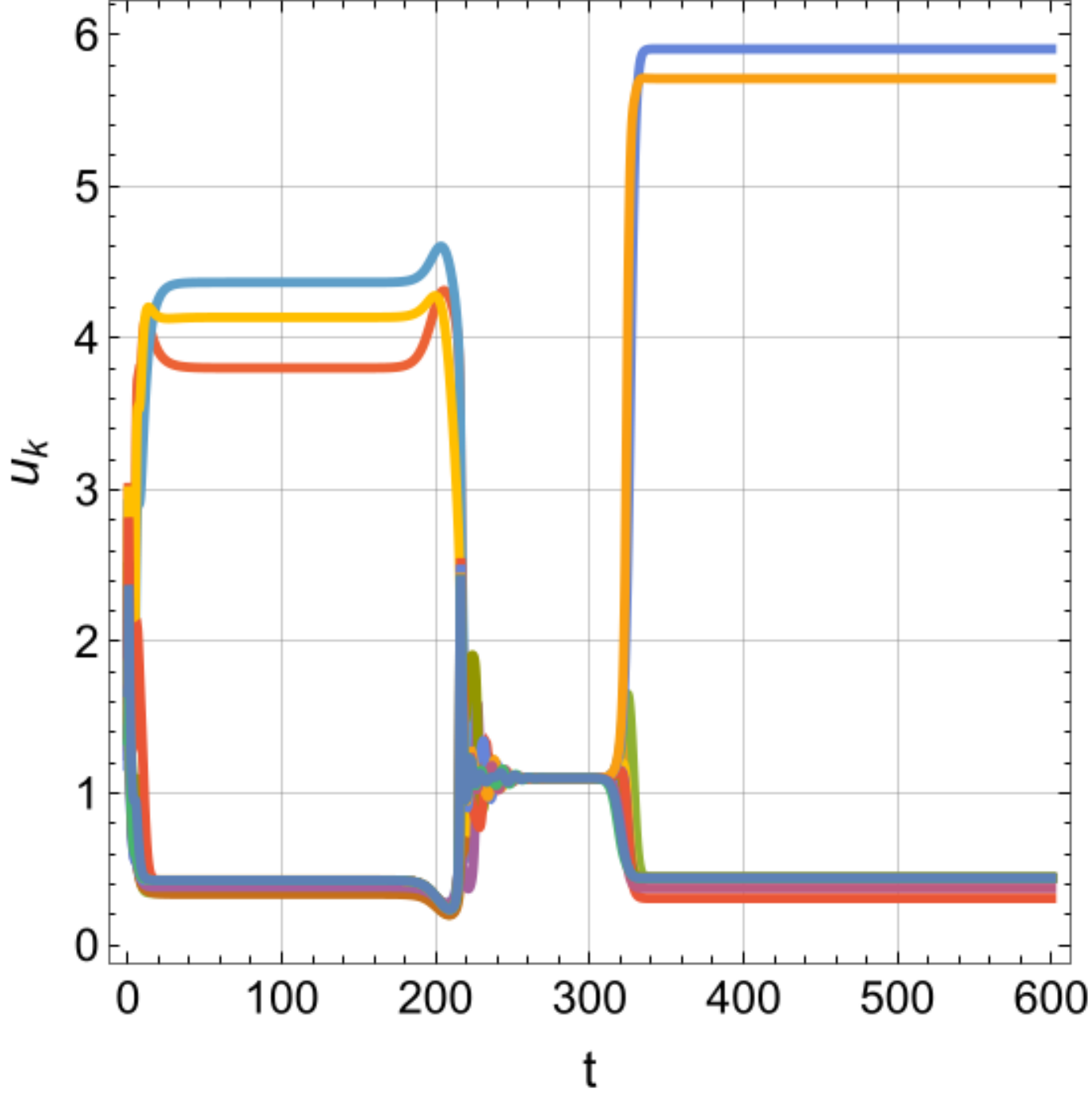}
		(c) \\
		\includegraphics[width=0.23\linewidth]{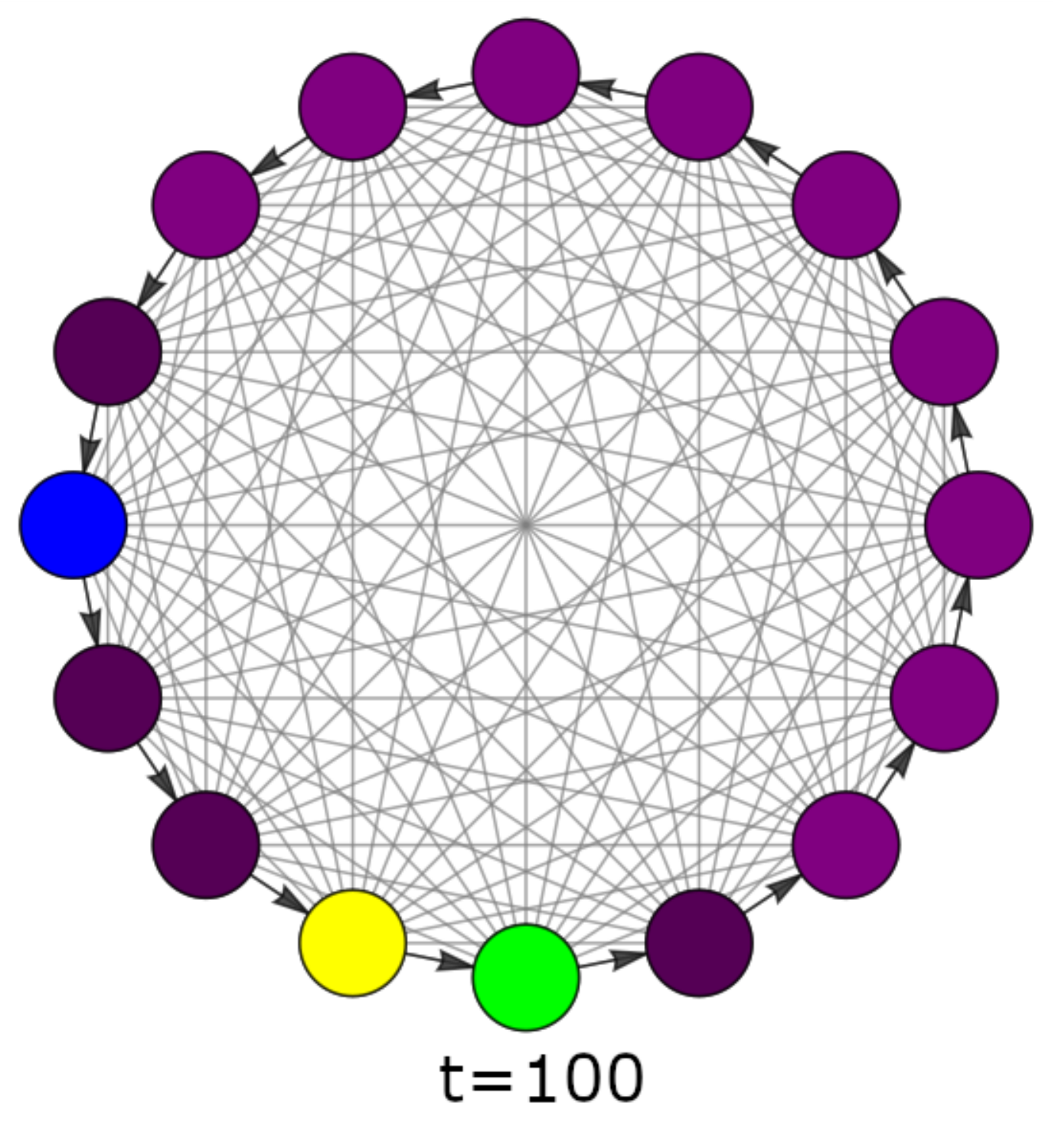}
		\includegraphics[width=0.23\linewidth]{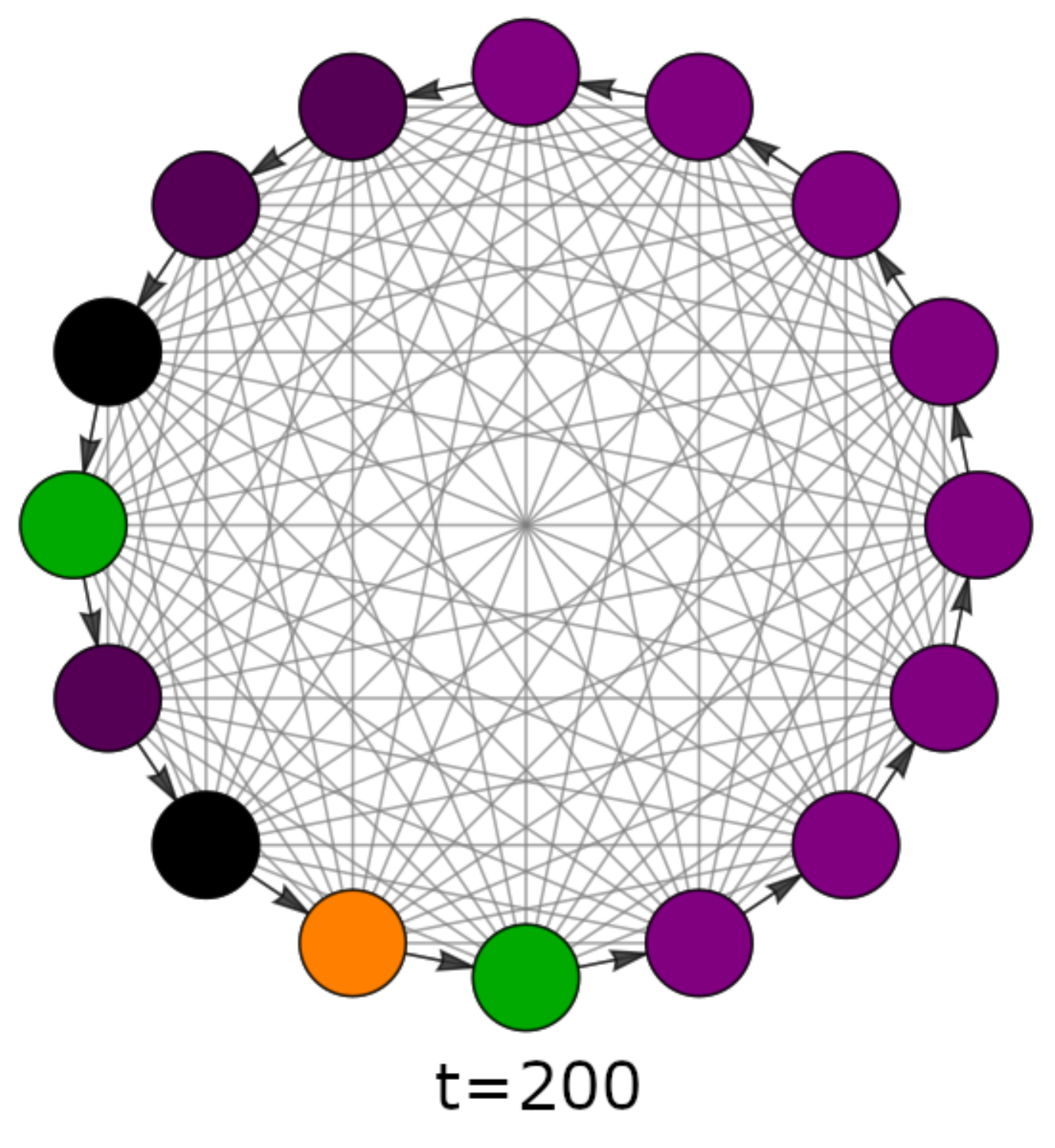}
		\includegraphics[width=0.23\linewidth]{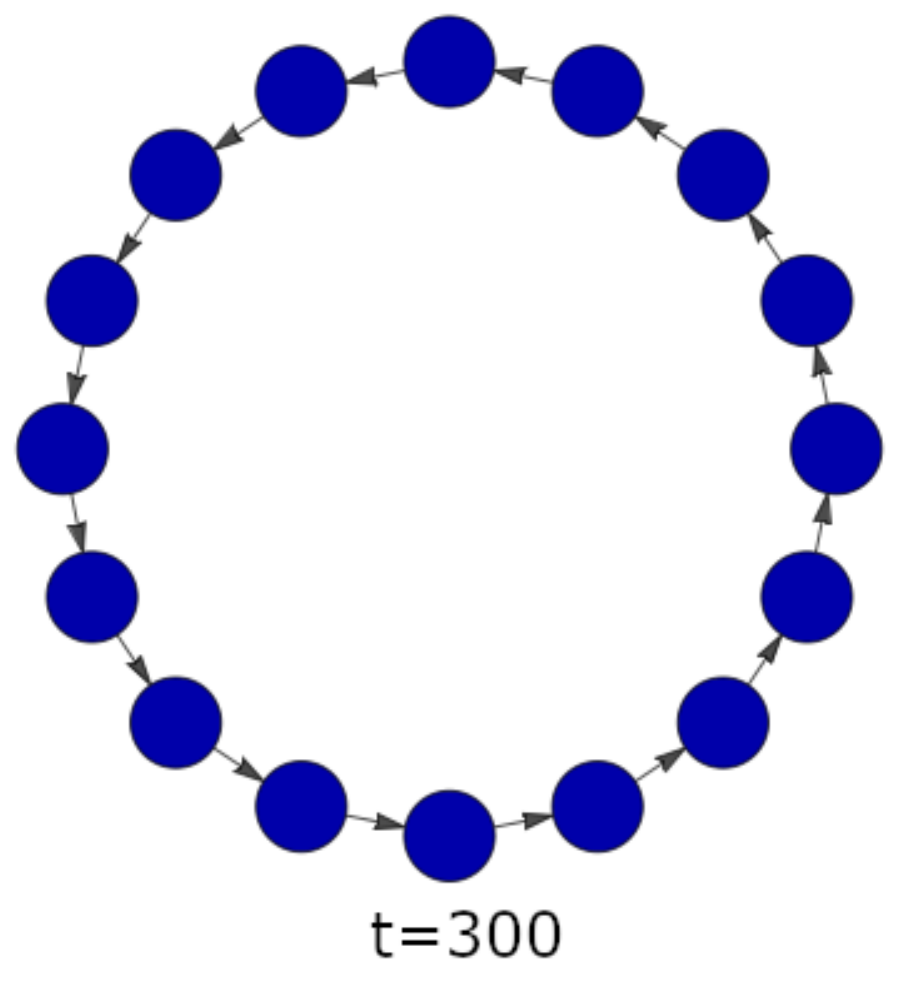}
		\includegraphics[width=0.23\linewidth]{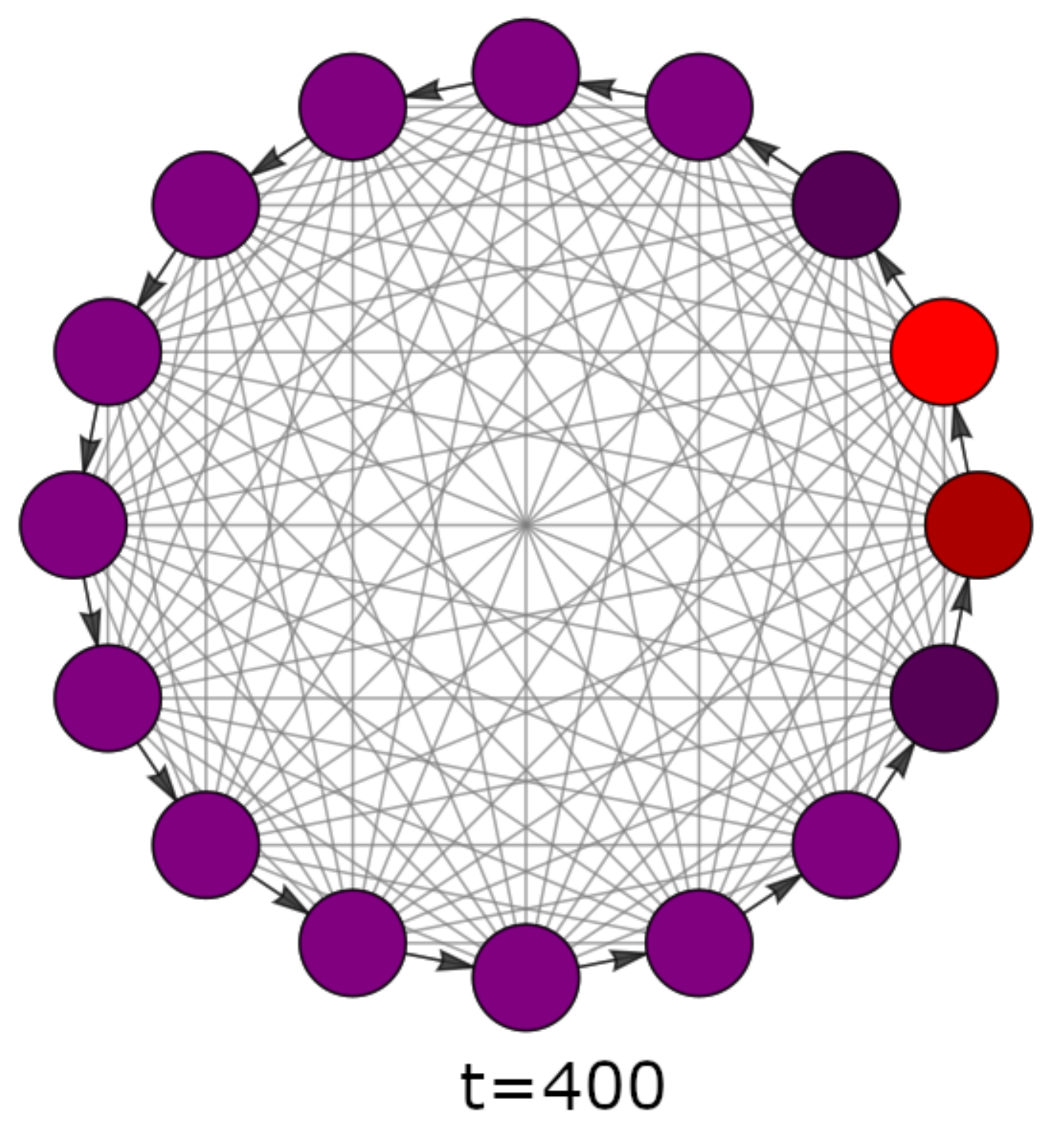}
		(d) \\
		\includegraphics[width=0.5\linewidth]{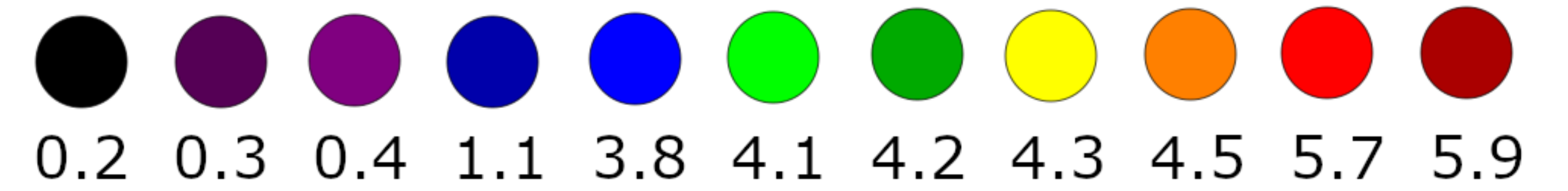}
		\caption{Patterning corresponding to \Eqsref{Eq:u_t}--\eqref{Eq:Adjacency_t} with $h(t)=1-\Omega(t,200)+\Omega(t,300)$, and $\beta_{i}(t)=1$. In (a) we plot the function $h(t)$. In (b) we show the quantity $\ln(w)$. In (c) we plot the functions $u_i$ for $i=1,\dots,16$. Finally, in (d) we show the values of $u_{i}$ on the graph at $t=100,200,300,400$.}
		\label{fig:Cycle_2_t}
	\end{figure}

	For our first example, of pattern formation on a directed temporal networks, we consider the following system:
	\begin{align}
		\frac{d u_{i}}{d t} =\, & \frac{1}{50}\sum_{j=1}^{n}A_{ij}(t)\left( u_{j} - u_{i} \right) + \beta_{i}(t)\left(  \frac{1}{10} - u_{i} + u_{i}^{2}v_{i} \right),
		\label{Eq:u_t}
		\\
		\frac{d v_{i}}{d t} =\, & \sum_{j=1}^{n}A_{ij}(t)\left( v_{j} - v_{i} \right) + \beta_{i}(t)\left( 1 - u_{i}^{2}v_{i} \right),
		\label{Eq:v_t}
	\end{align} 
	for freely specifiable functions $\beta_{i}:\mathbb{R}\rightarrow \mathbb{R}$, with $i=1,\dots,n$. The adjacency matrix $A_{ij}$ is 
	\begin{align}
		A_{ij}=
		\begin{cases}
			1/2 \quad &\text{if}\quad i-j=1\quad \text{or}\quad i-j=n-1,
			\\
			0 \quad &\text{if}\quad i-j=0
			\\
			h(t) \quad &\text{otherwise},
		\end{cases}
		\label{Eq:Adjacency_t}
	\end{align}
	where $h(t)$ is also a freely specifiable function. In this example here we set $h(t)=1-\Omega(t,200)+\Omega(t,300)$, and $\beta_{i}(t)=1$ for all $i=1,\dots,n$ with $n=16$. Notice that, for $\beta_{i}(t)=1$, \Eqsref{Eq:u_t} and \eqref{Eq:v_t} have local reaction kinetics that do not depend on time. However, the network itself \emph{does} change in time. The function $h(t)$ is shown in \subfig{fig:Cycle_2_t}{(a)}. In \subfig{fig:Cycle_2_t}{(a)} we see that $h(t)\approx 1$ if $t<200$ or $t>300$, and therefore describes a complete cycle graph in these regions. However, for $t\in(200,300)$ we have that $h(t)\approx 0$, in which case the adjacency matrix \Eqref{Eq:Adjacency_t} corresponds to an \emph{incomplete} cycle graph.

	In \Figref{fig:Cycle_2_t} we show the numerical results corresponding to this network. In \subfig{fig:Cycle_2_t}{(b)} we show the numerically calculated quantity $\ln(w)$. Here, we see that $w$ clearly grows for $t<200$ and $t>300$, indicating an instability in these regions. Conversely, for $t\in(200,300)$, we see that $w$ is \emph{deceasing}. We therefore conclude that the steady state solution $(u_i,v_i)=(11/10,100/121)$ is linearly stable in this region (and therefore does not produce a pattern). This behaviour is verified in \subfig{fig:Cycle_2_t}{(c)}, where we see that a pattern occurs for $t<200$ and $t>300$, but is suppressed for $t\in(200,300)$.  This result is not surprising as network topology is known to significantly effect patterning (see, for example, \cite{Mimar:2019,Asllani:2016,VanGorder:2020}).

	\subsection{Pattern formation with time dependent reaction kinetics}
	In addition to being able to control the topology, the theory presented in \Sectionref{Sec:Linear_instability_analysis_on_temporal_networks} also allows one to investigate non-autonomous reaction diffusion equations on \emph{static} networks. In this subsection here we first consider systems with time dependent reaction kinetics. To this end, we again consider \Eqsref{Eq:u_t}--\eqref{Eq:v_t}. Here, we consider two different choices of the adjacency matrix $A_{ij}$. 
	
\begin{figure}[t!]
		\centering
		\includegraphics[width=0.28\linewidth]{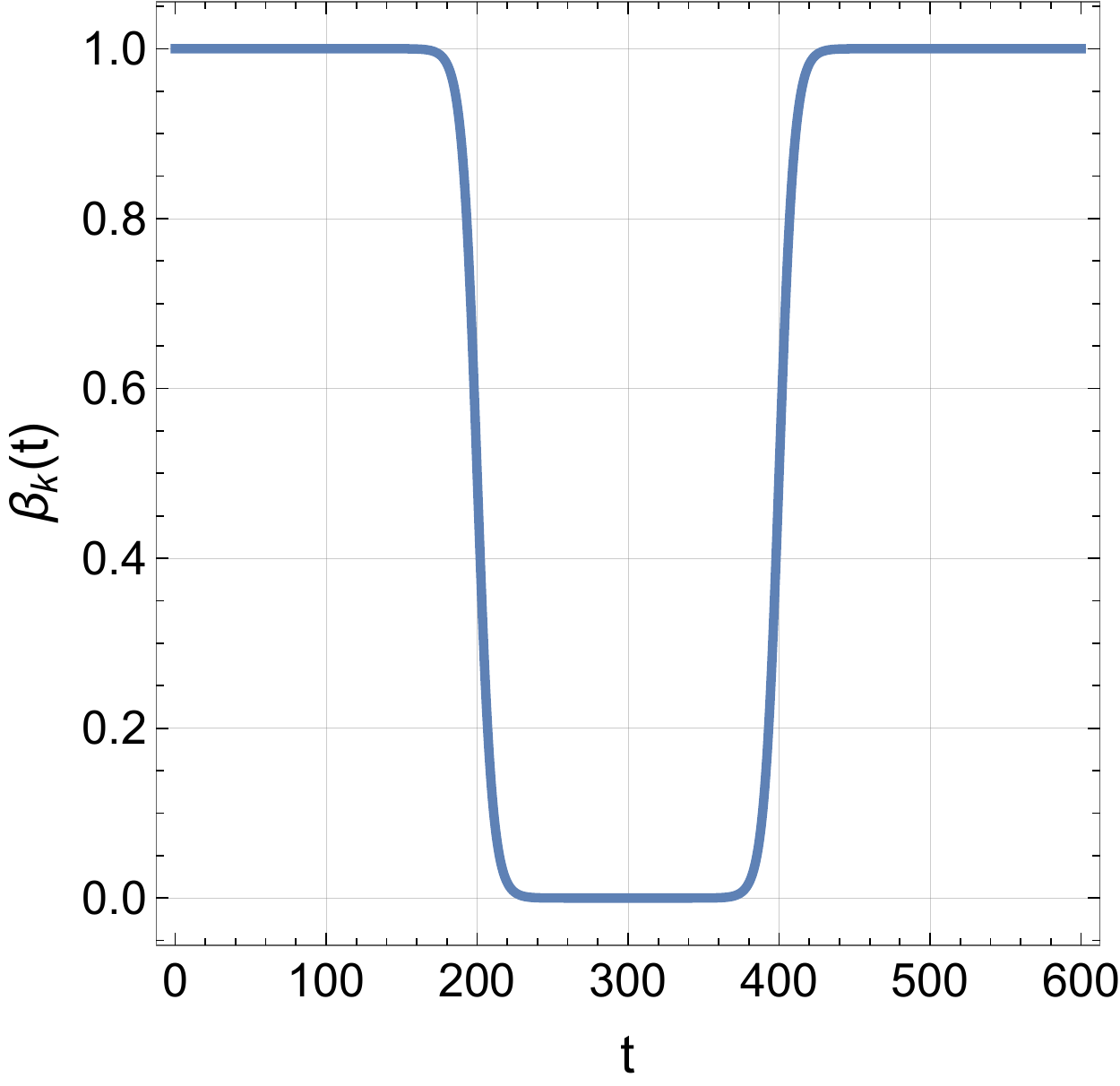}
		(a)
		\includegraphics[width=0.28\linewidth]{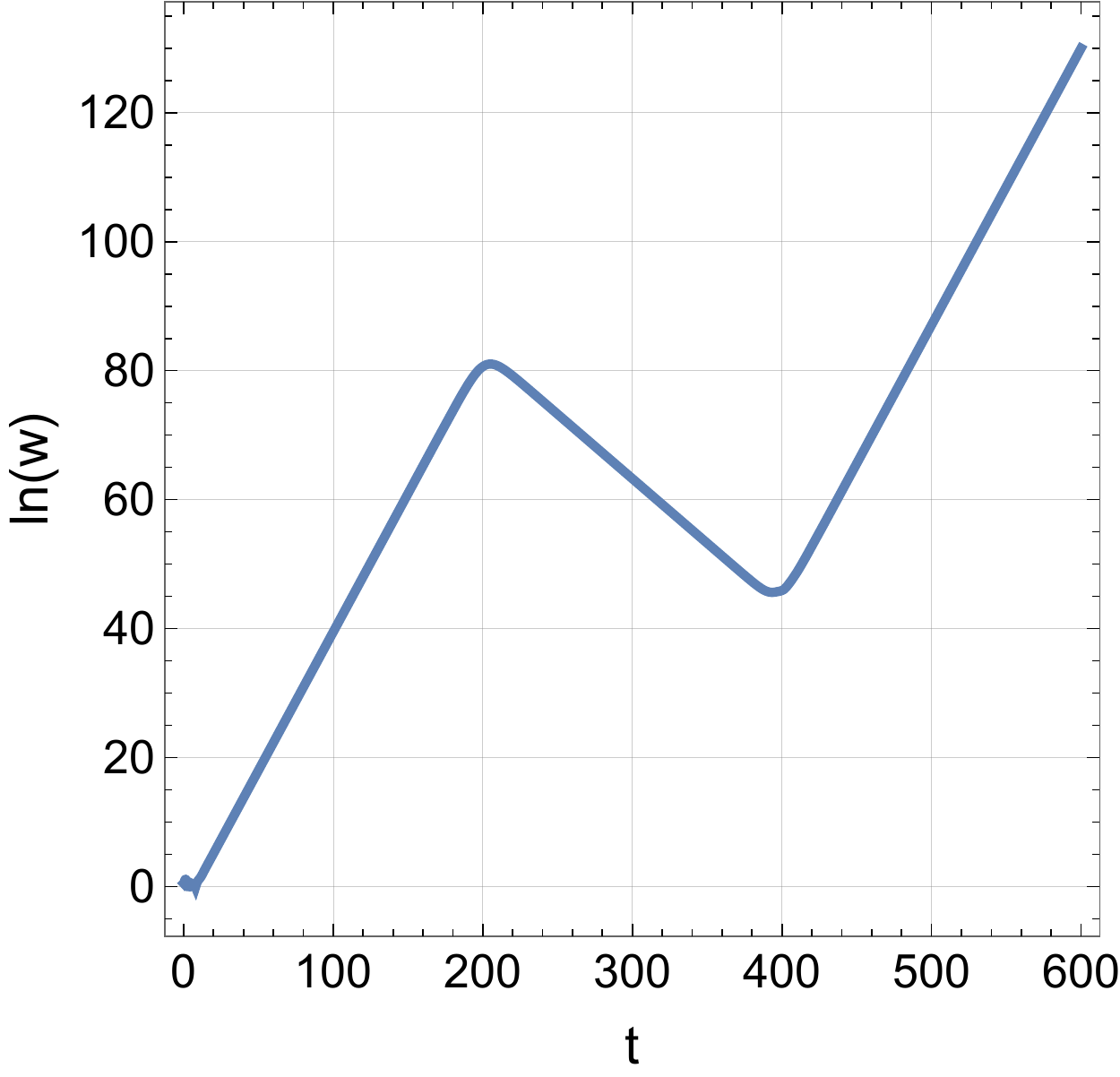}
		(b)
		\includegraphics[width=0.27\linewidth]{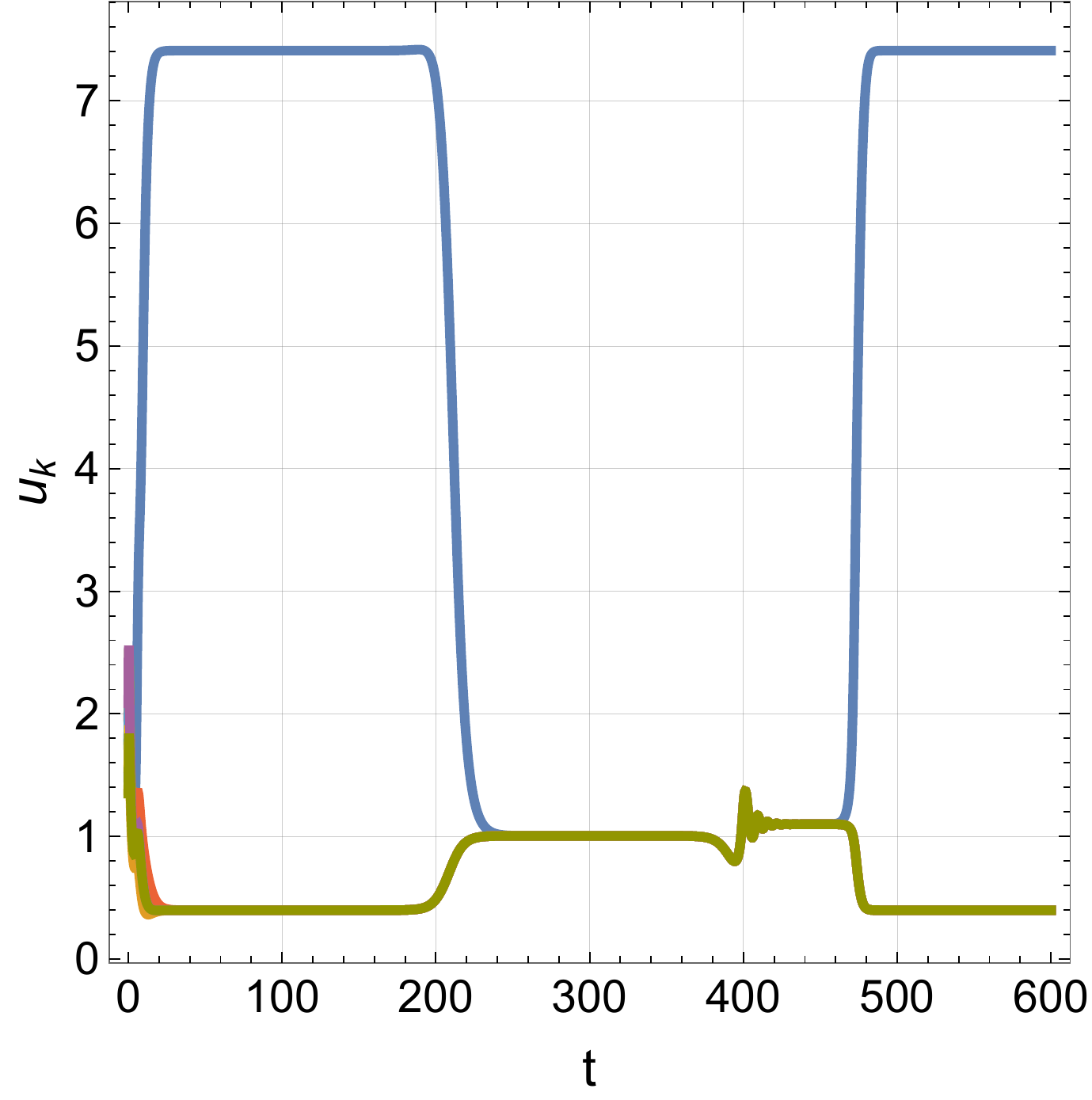}
		(c) \\
		\includegraphics[width=0.23\linewidth]{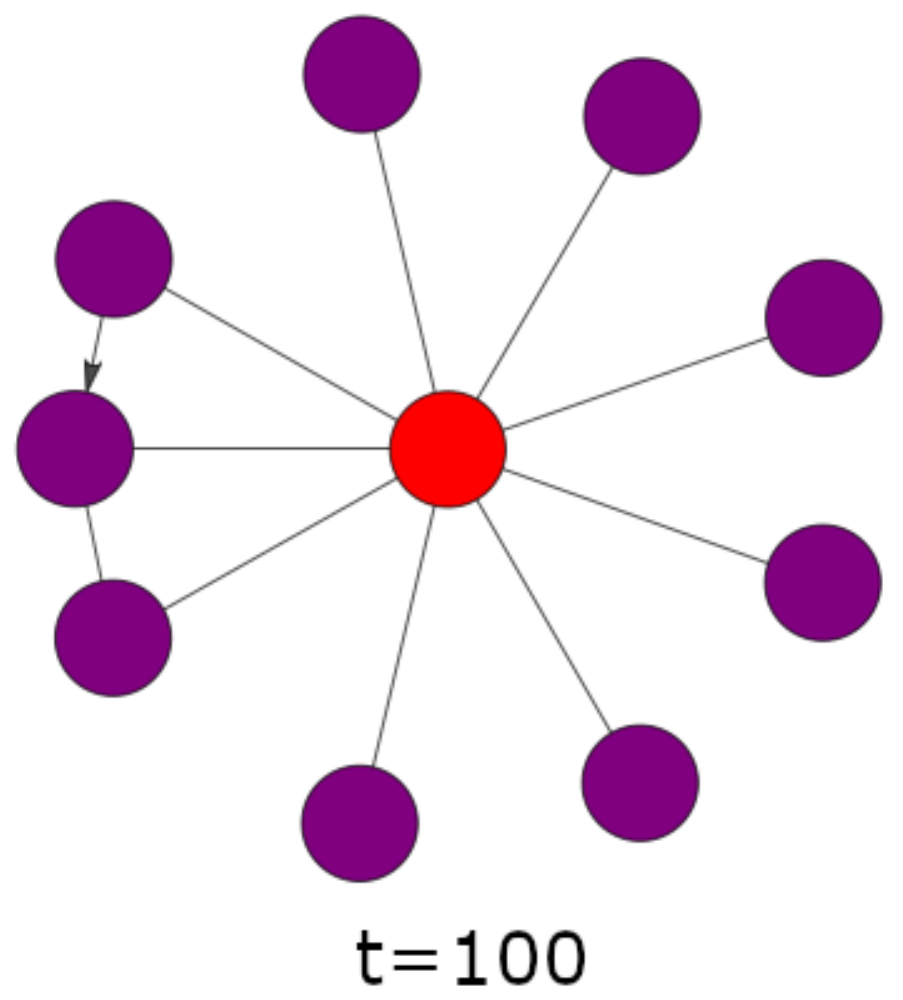}
		\includegraphics[width=0.23\linewidth]{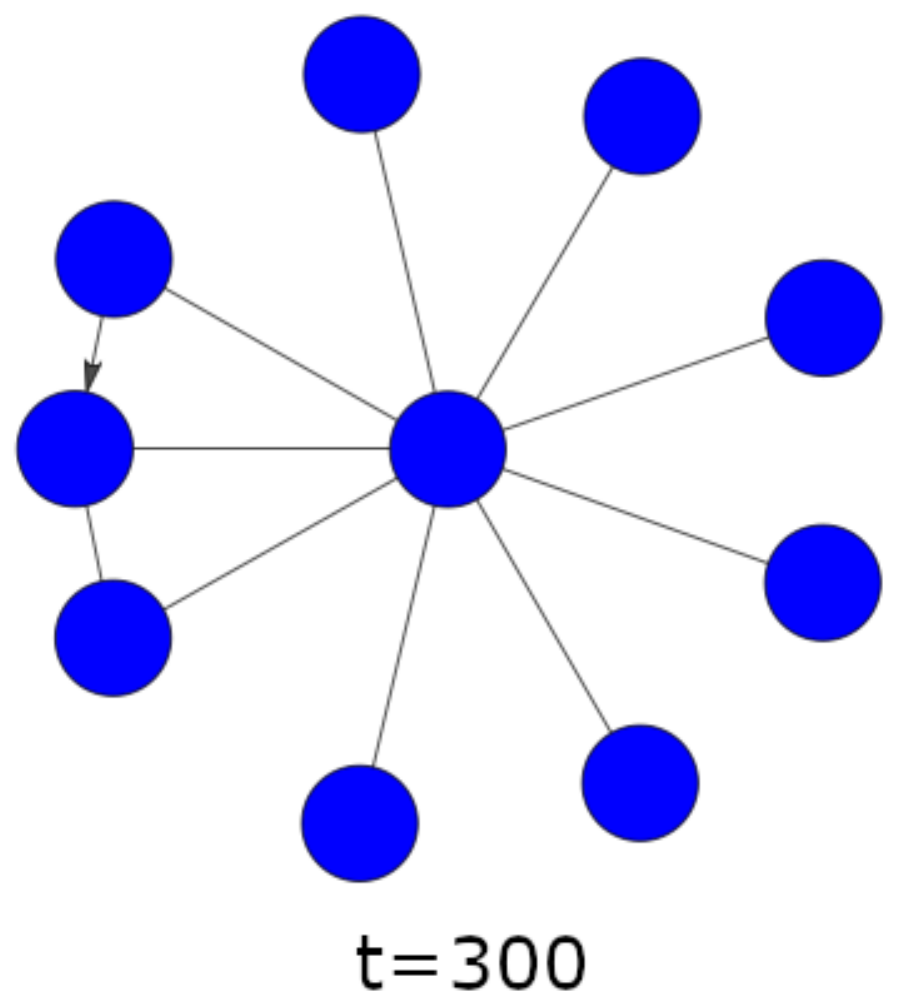}
		\includegraphics[width=0.23\linewidth]{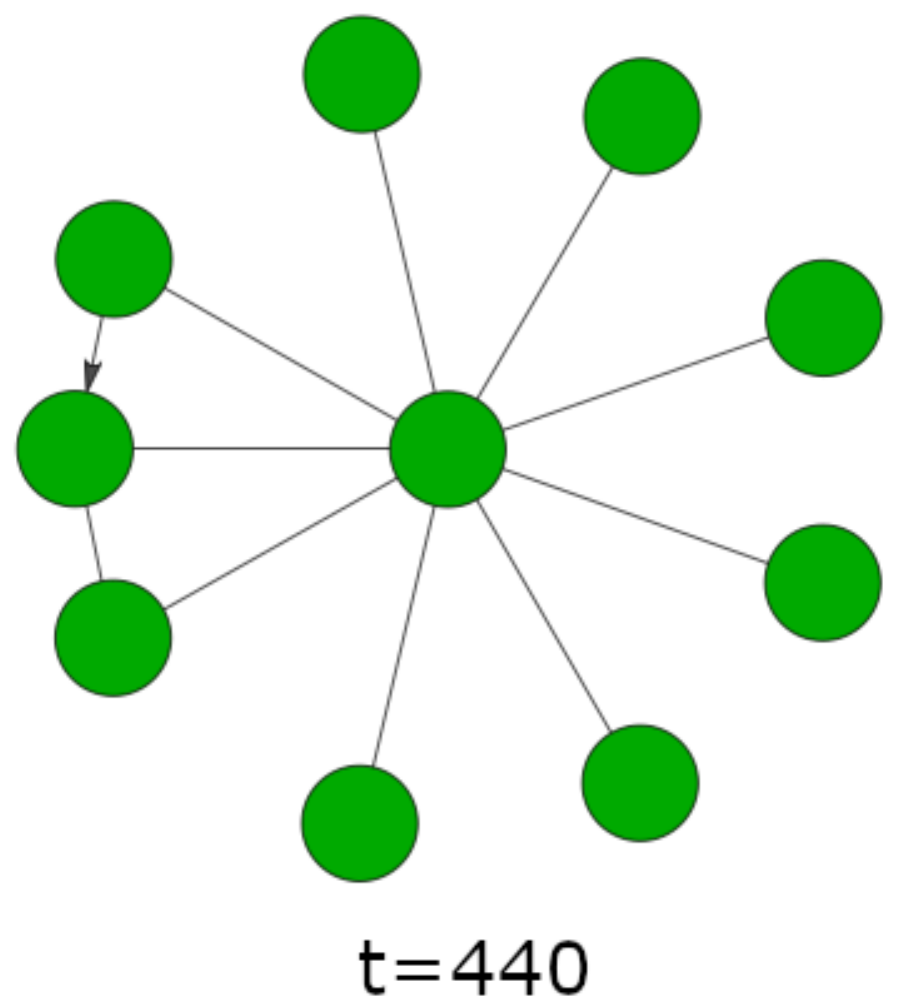}
		\includegraphics[width=0.23\linewidth]{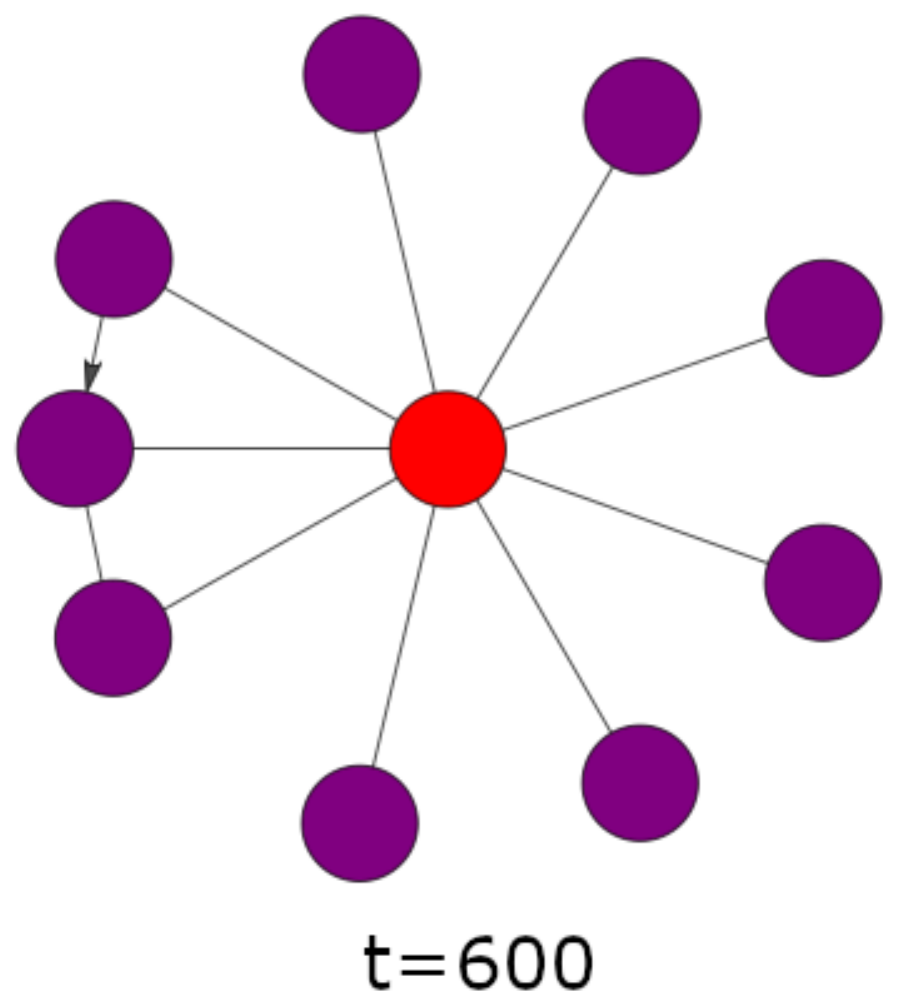}
		(d) \\
		\includegraphics[width=0.17\linewidth]{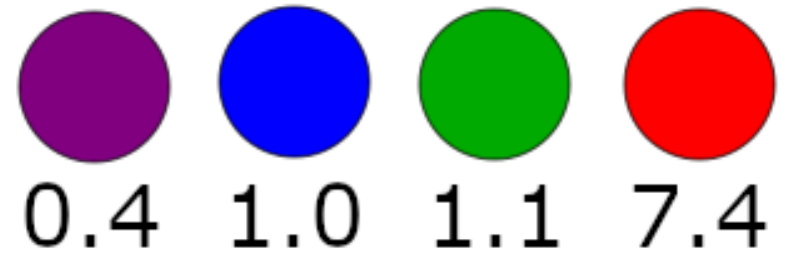}
		\caption{Patterning corresponding to \Eqsref{Eq:u_t}--\eqref{Eq:v_t} with $A_{ij}$ given by \Eqref{Eq:Adjacency_Hyp_1} and $\beta_{i}(t)=1-\Omega(t,200)+\Omega(t,400)$ for all $i=1,\dots,10$. In (a) we plot the function $\beta_k(t)$. In (b) we show the quantity $\ln(w)$. In (c) we plot the functions $u_i$ for $i=1,\dots,10$. Finally, in (d) we show the values of $u_{i}$ on the graph at $t=100,300,440,600$.}
		\label{fig:w}
	\end{figure}

	We first suppose that $n=10$ and that the adjacency matrix $A_{ij}$ is given by \Eqref{Eq:Adjacency_Hyp_1}. Recall that this choice of adjacency matrix describes a ``modified star graph''. Moreover, we set $\beta_{i}(t)=1-\Omega(t,200)+\Omega(t,400)$ for $i=1,\dots,10$, so that \Eqsref{Eq:u_t}--\eqref{Eq:v_t} have local reaction kinetics. A plot of $\beta_{k}(t)$ is shown in \subfig{fig:w}{(a)}. In \subfig{fig:w}{(a)} we see that $\beta_k(t)\approx 1$ if $t<200$ or $t>400$. Conversely, for $t\in(200,400)$ we see that $\beta_{k}(t)\approx 0$. This has the effect of turning the reaction kinetics ``off'' for $t\in(200,400)$ and we would therefore not expect a pattern to occur in this region.  
	The corresponding numerical results are shown in \Figref{fig:w}. In \subfig{fig:w}{(b)} we show the numerically calculated quantity $\ln(w)$. Here, we see that $w$ clearly grows for $t<200$ and $t>400$, indicating an instability in these regions. Conversely, for $t\in(200,400)$, we see that $w$ is \emph{deceasing}. We therefore conclude that the steady state solution $(u_i,v_i)=(11/10,100/121)$ is linearly stable in this region (and therefore does not produce a pattern). This behaviour is verified in \subfig{fig:w}{(c)}, where we see that a pattern occurs for $t<200$ and $t>400$, but is suppressed for $t\in(200,400)$. The corresponding pattern is shown in \subfig{fig:w}{(d)} at four time points.
	
	We now consider examples of pattern formation with global reaction kinetics. For this we again consider \Eqsref{Eq:u_t}--\eqref{Eq:v_t}, where the adjacency matrix $A_{ij}$ is given by \Eqref{Eq:Adjacency_t} with $n=2$. We now consider two choices of the functions $h(t)$ and $\beta_{i}(t)$.
	
\begin{figure}[t!]
			\centering
			\includegraphics[width=0.28\linewidth]{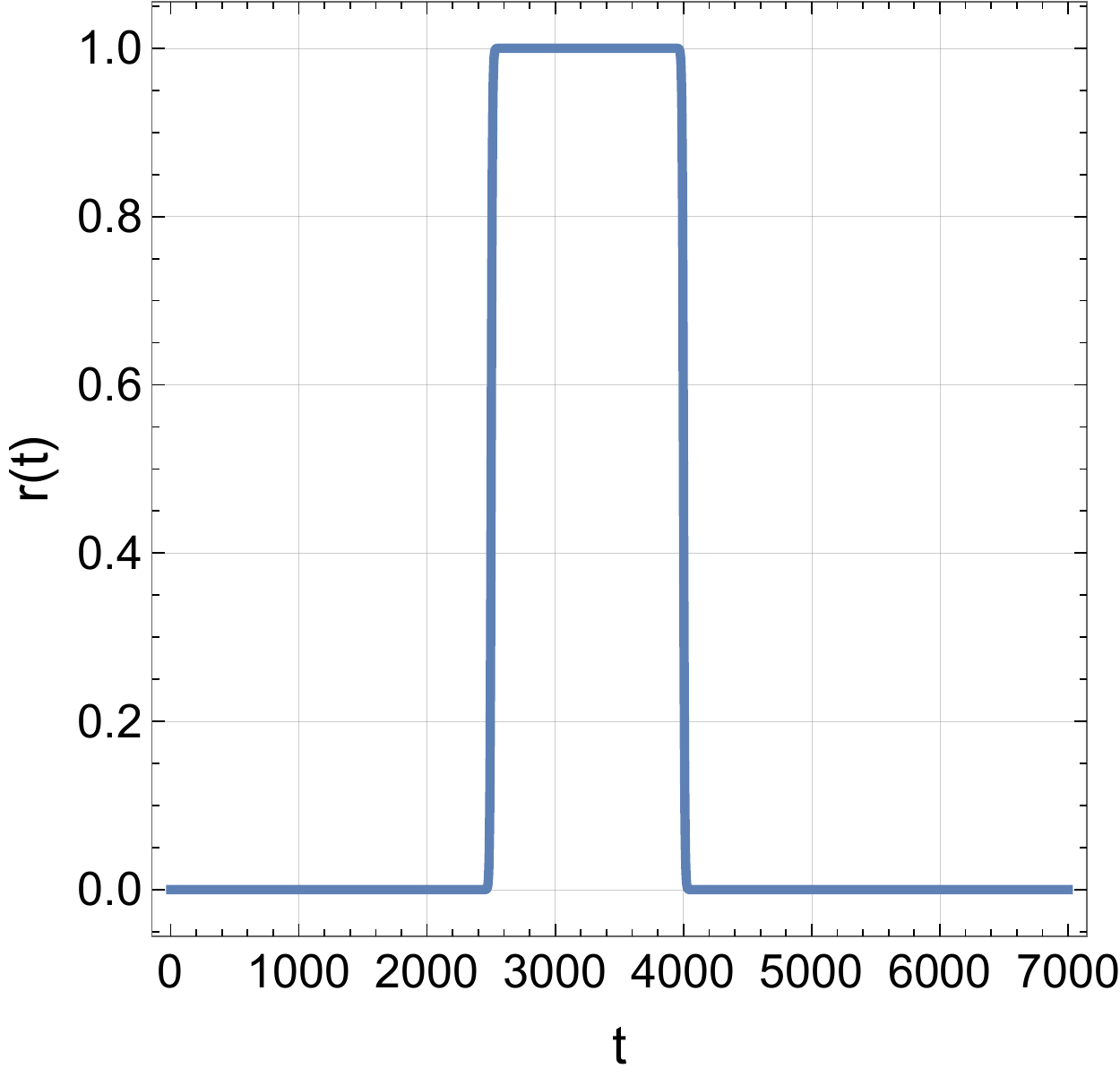}
			(a)
			\includegraphics[width=0.28\linewidth]{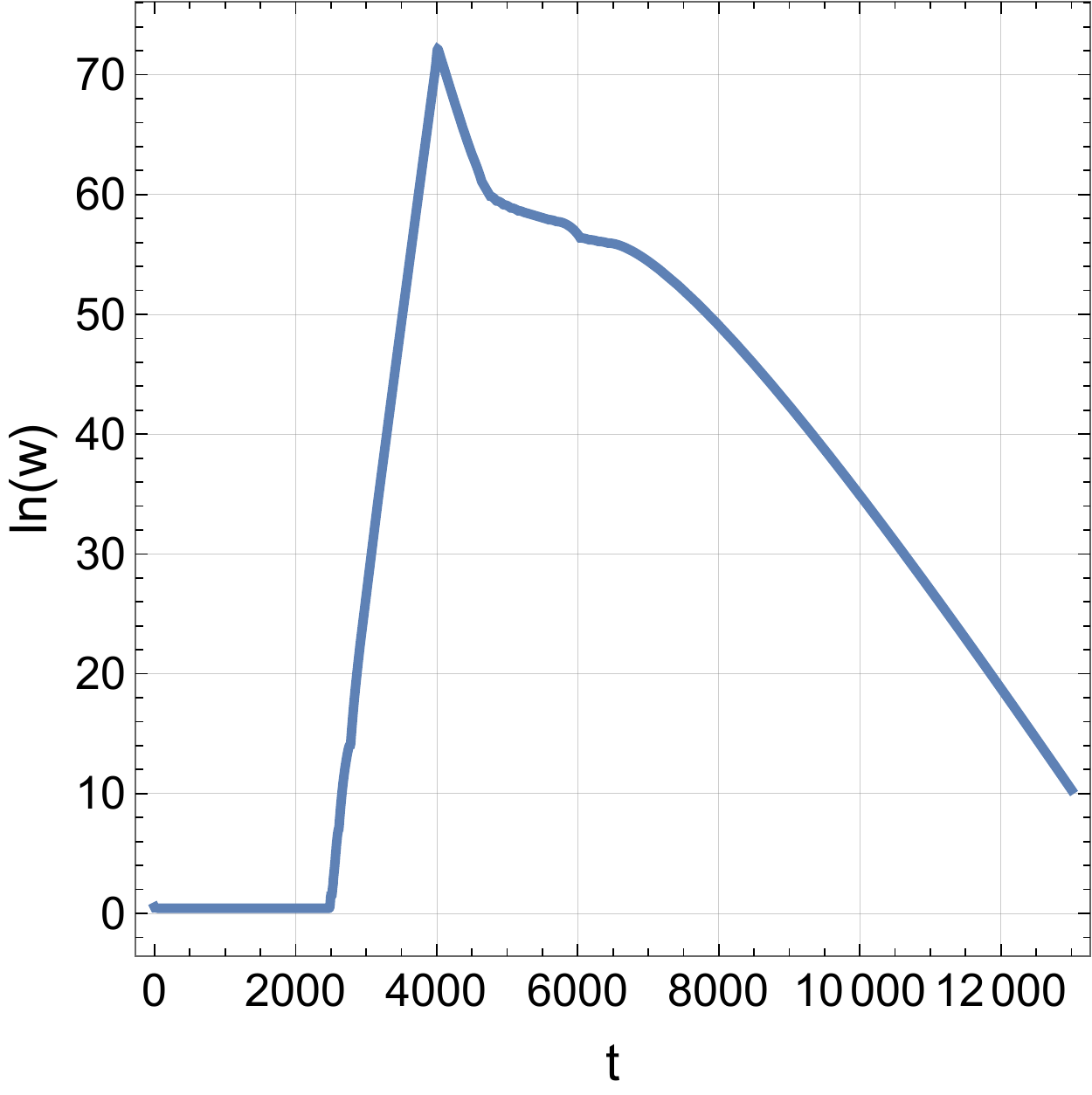}
			(b)
			\includegraphics[width=0.27\linewidth]{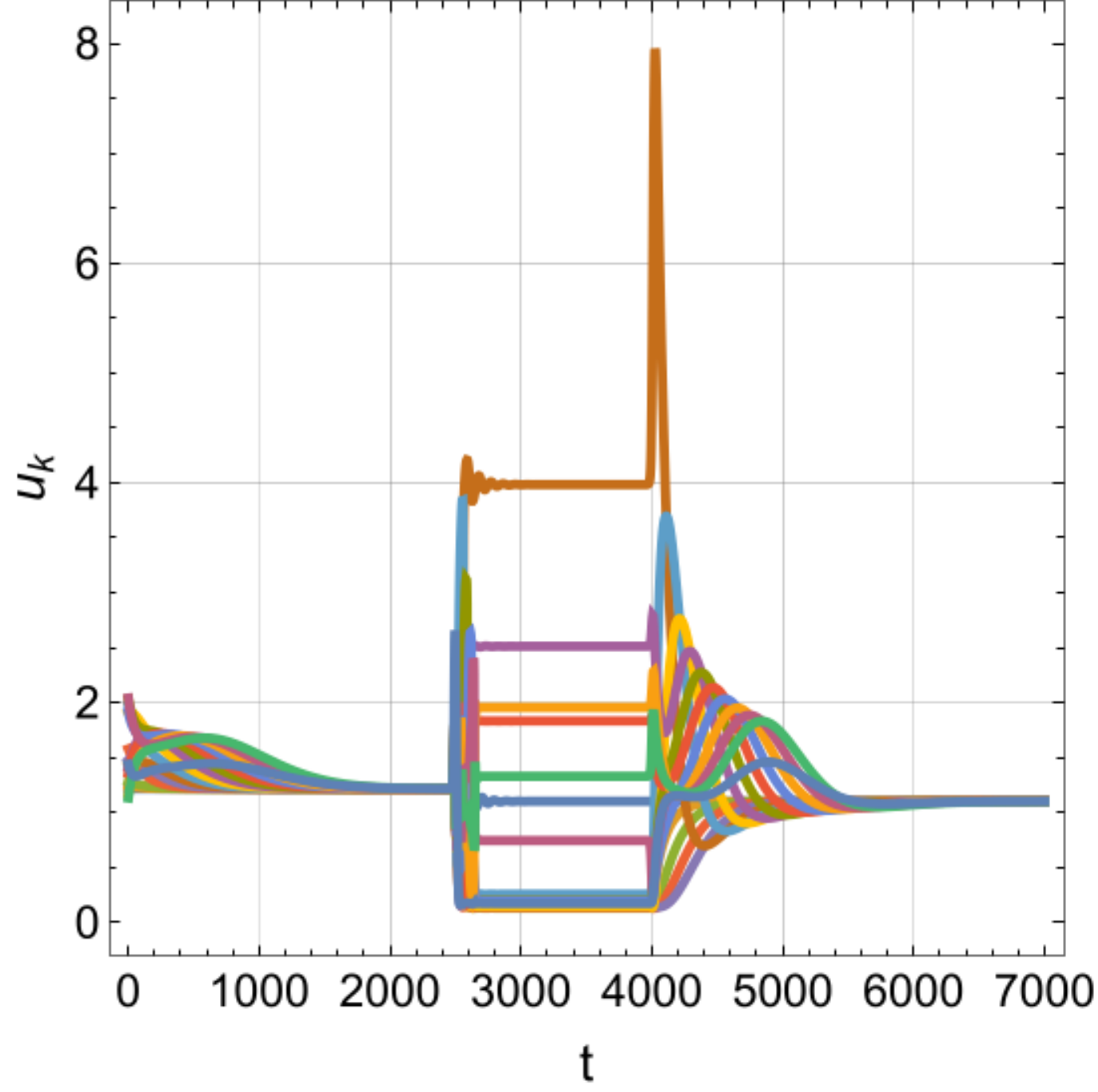}
			(c) \\
			\includegraphics[width=0.23\linewidth]{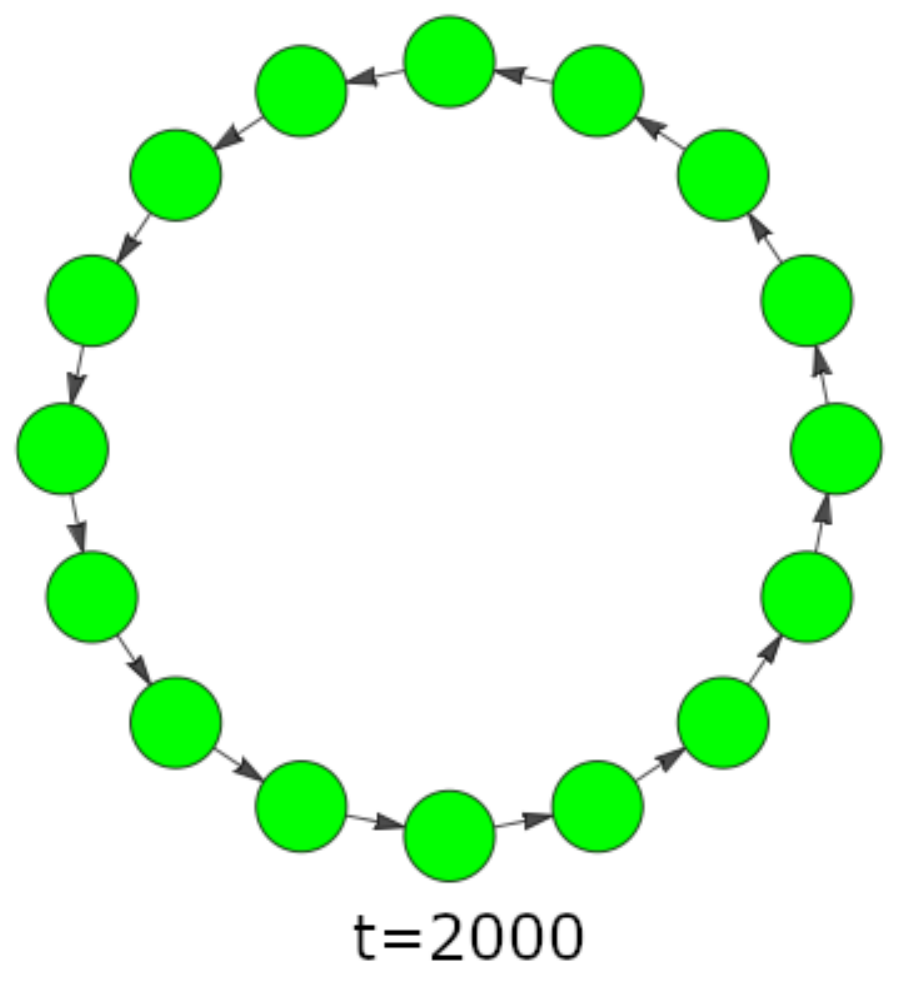}
			\includegraphics[width=0.23\linewidth]{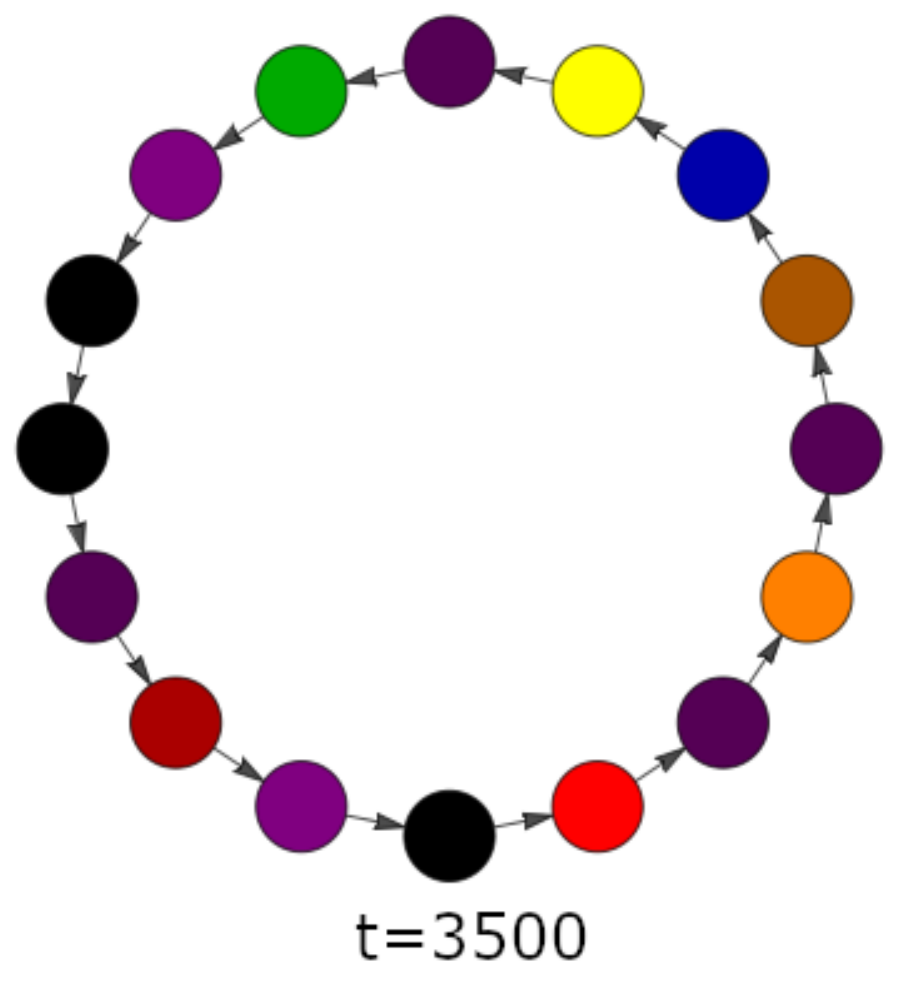}
			\includegraphics[width=0.23\linewidth]{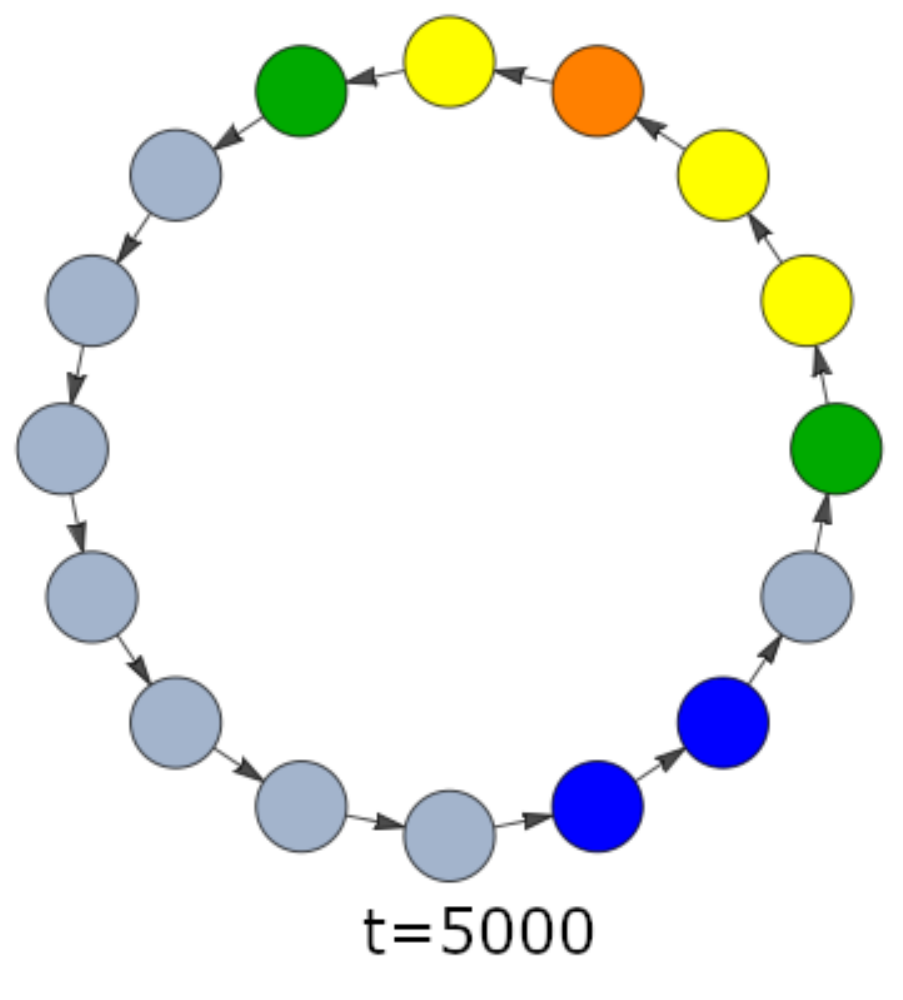}
			\includegraphics[width=0.23\linewidth]{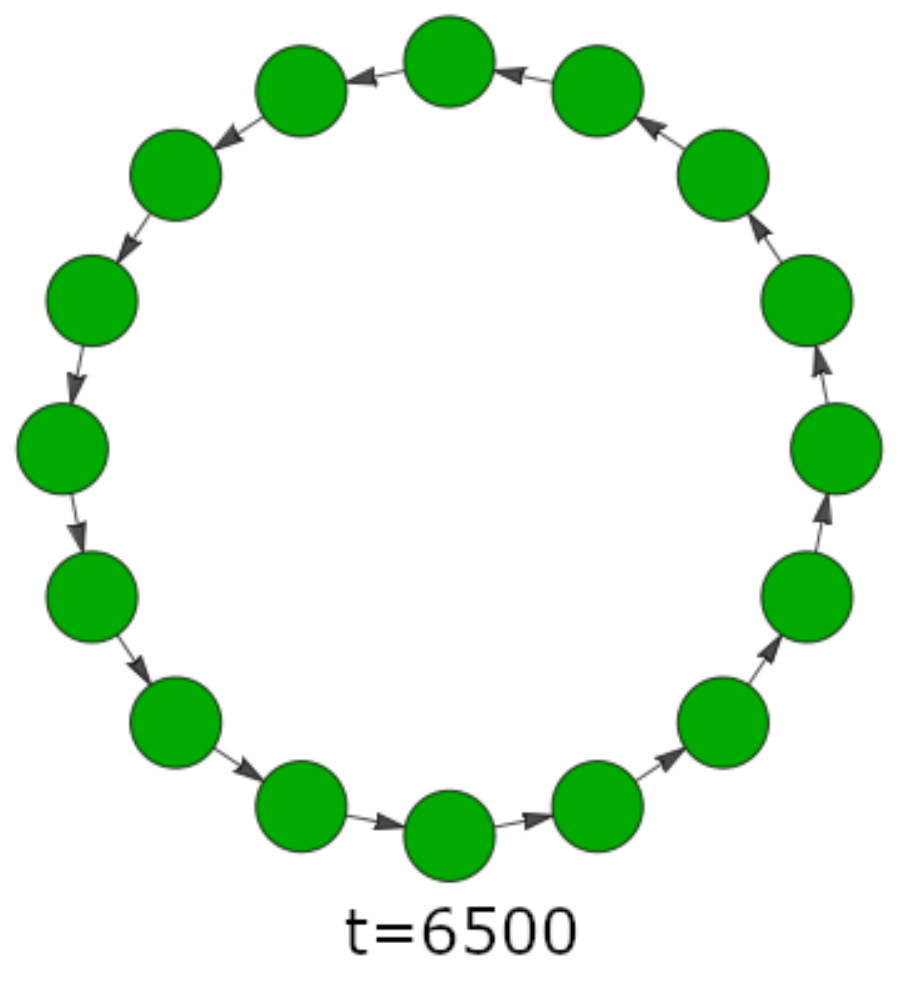}
			(d) \\
			\includegraphics[width=0.6\linewidth]{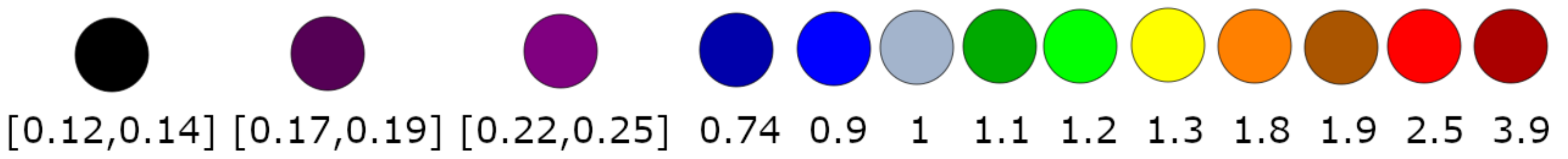}
			\caption{Patterning corresponding to \Eqsref{Eq:u_t}--\eqref{Eq:Adjacency_t} with $h(t)=0$, and where $\beta_{i}(t)$ is given by \Eqref{Eq:beta_t_1} for $i=1,\dots,n=16$. In (a) we plot the function $h(t)$. In (b) we show the numerically calculated quantity $\ln(w)$. In (c) we plot the functions $u_i$ for $i=1,\dots,16$. Finally, in (d) we show the values of $u_{i}$ on the graph at $t=297,3500,5000,6500$.}
		\label{fig:Glob_Cycle_2}
	\end{figure}

	We first set $h(t)=0$. Recall that, for this particular choice of $h(t)$, the underlying network is an incomplete cycle graph. Moreover, we pick 
	\begin{align}
		\beta_{i}(t)=\frac{i}{16}r(t),\quad r(t)= \Omega(t,2500)-\Omega(t,4000).
		\label{Eq:beta_t_1}
	\end{align}
	A plot of the function $r(t)$ is shown in \subfig{fig:Glob_Cycle_2}{(a)}. In \subfig{fig:Glob_Cycle_2}{(a)} we see that $r(t)\approx 1$ for $t\in (2500,4000)$, and $r(t)\approx 0$ for $t\in(0,2500)\cup (4000,\infty)$. This has the effect of ``switching on'' the reaction kinetics for a short while.
	The numerical results corresponding to these choices are shown in \Figref{fig:Glob_Cycle_2}. In \subfig{fig:Glob_Cycle_2}{(b)} we show the numerically calculated quantity $\ln(w)$. Here we see that $w\approx 0$ for $t\in (0,2500)$, and we therefore do not expect a pattern to form in this region. However, when the kinetics are ``switched on'' at $t=2500$, $w$ quickly grows. When the reaction kinetics are again ``switched off'' we see that $w$ begins to decrease toward $w\approx 0$. This behaviour is confirmed in \subfig{fig:Glob_Cycle_2}{(c)}. In \subfig{fig:Glob_Cycle_2}{(d)} we show the patterning resulting from $u_{i}$ with $i=1,\dots,n$ at different times. 
	
\begin{figure}[t!]
		\centering
		\includegraphics[width=0.28\linewidth]{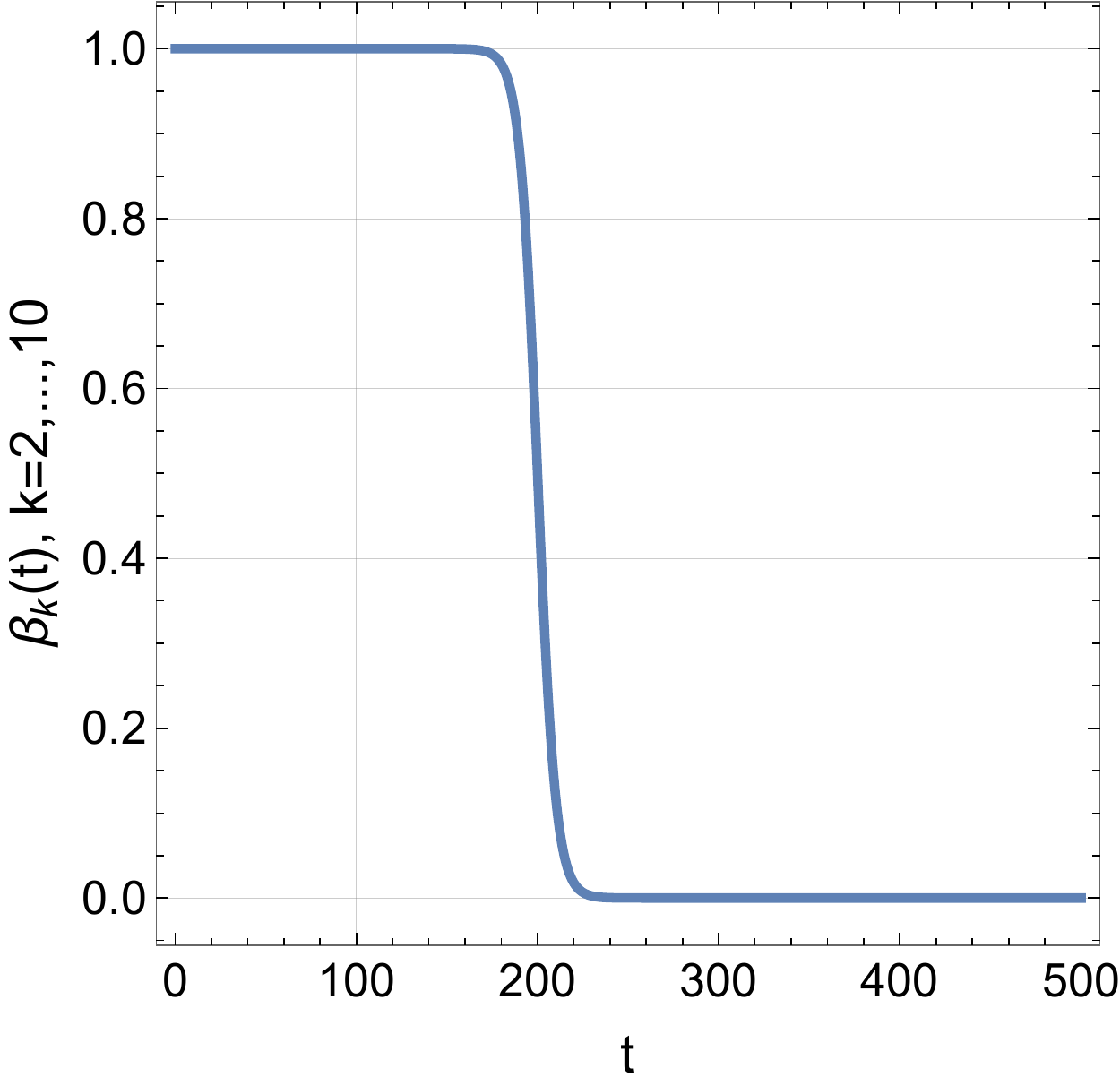}
		(a) 
		\includegraphics[width=0.28\linewidth]{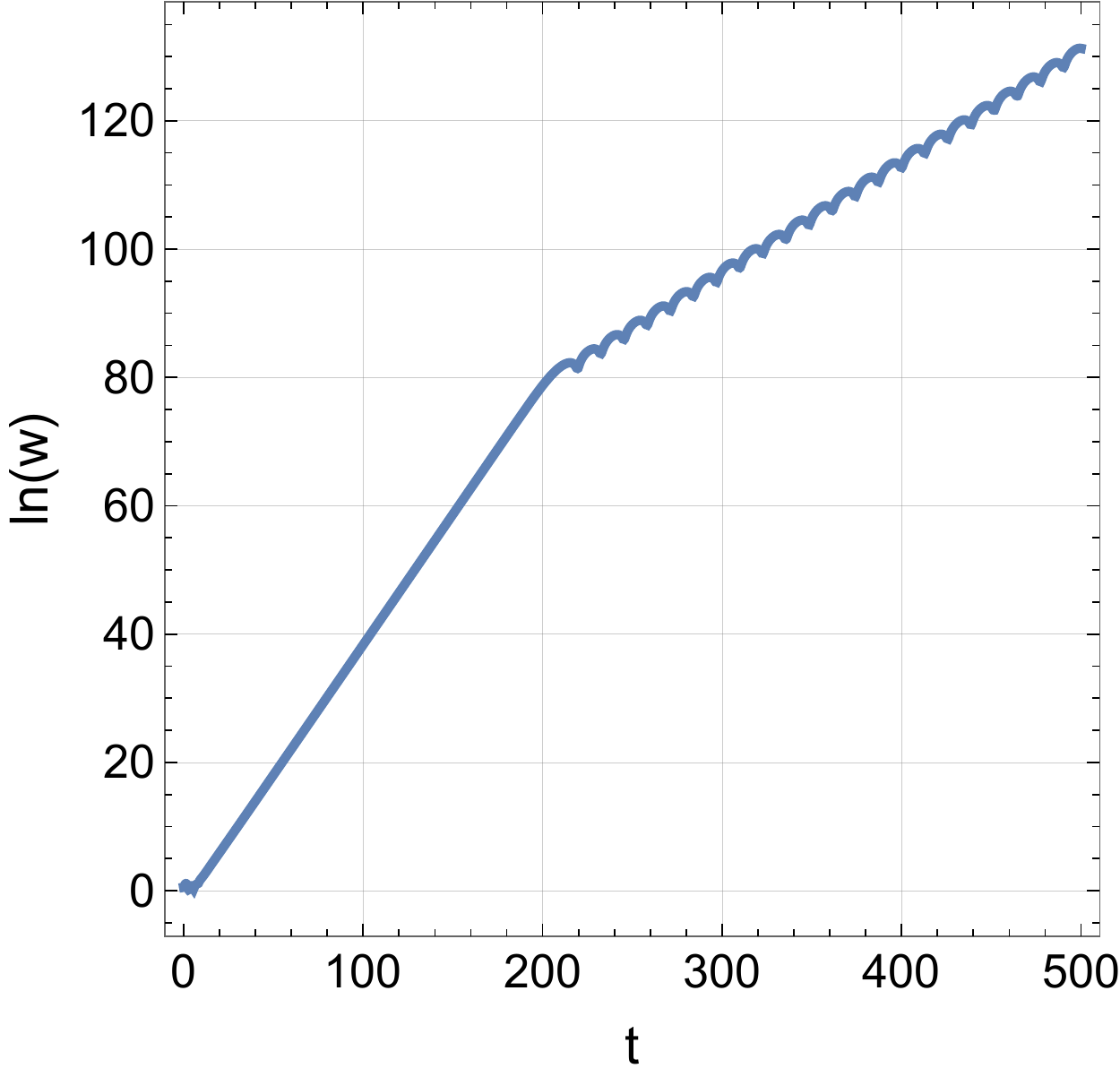}
		(b)
		\includegraphics[width=0.28\linewidth]{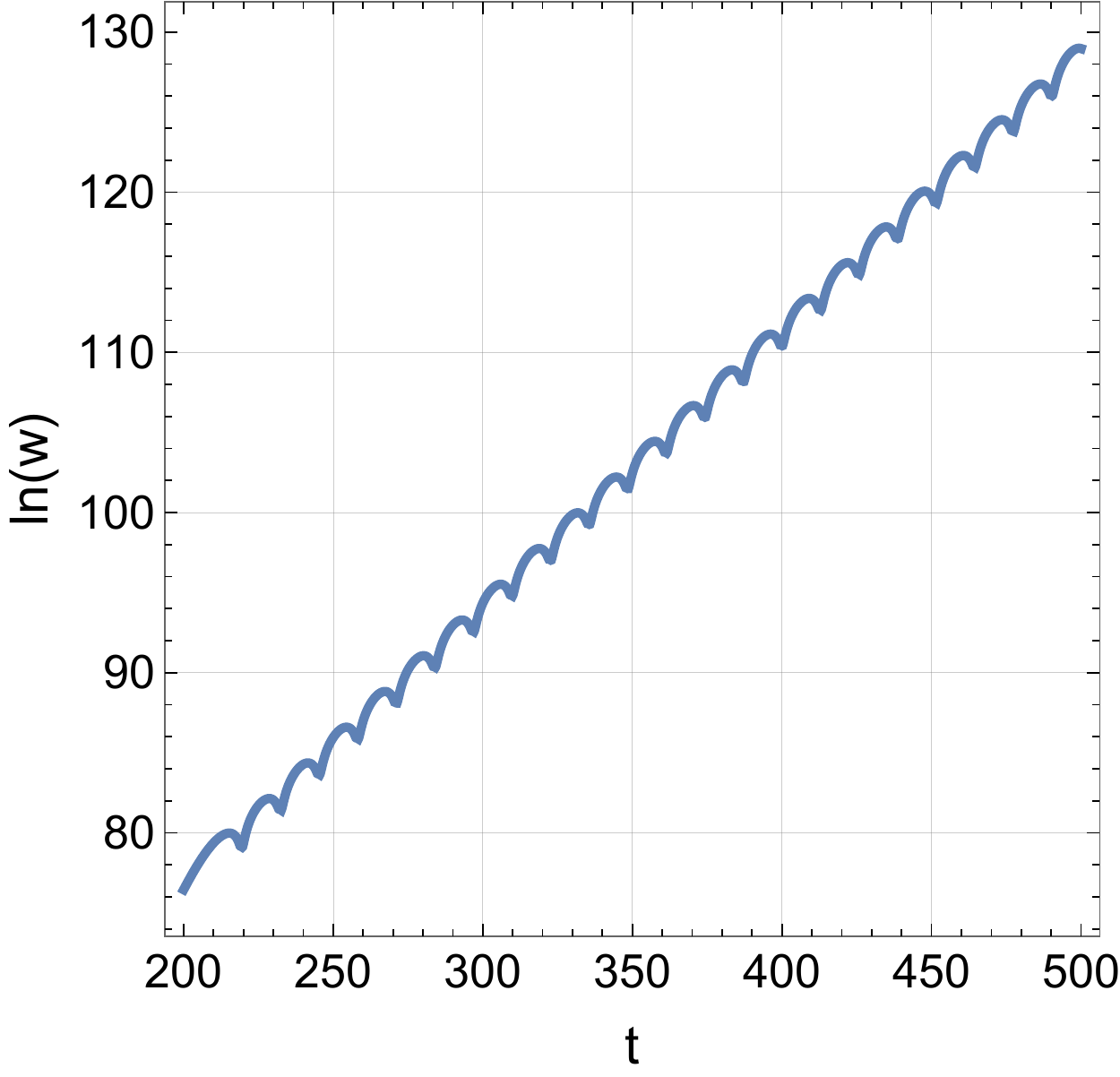}
		(c) 
		\includegraphics[width=0.28\linewidth]{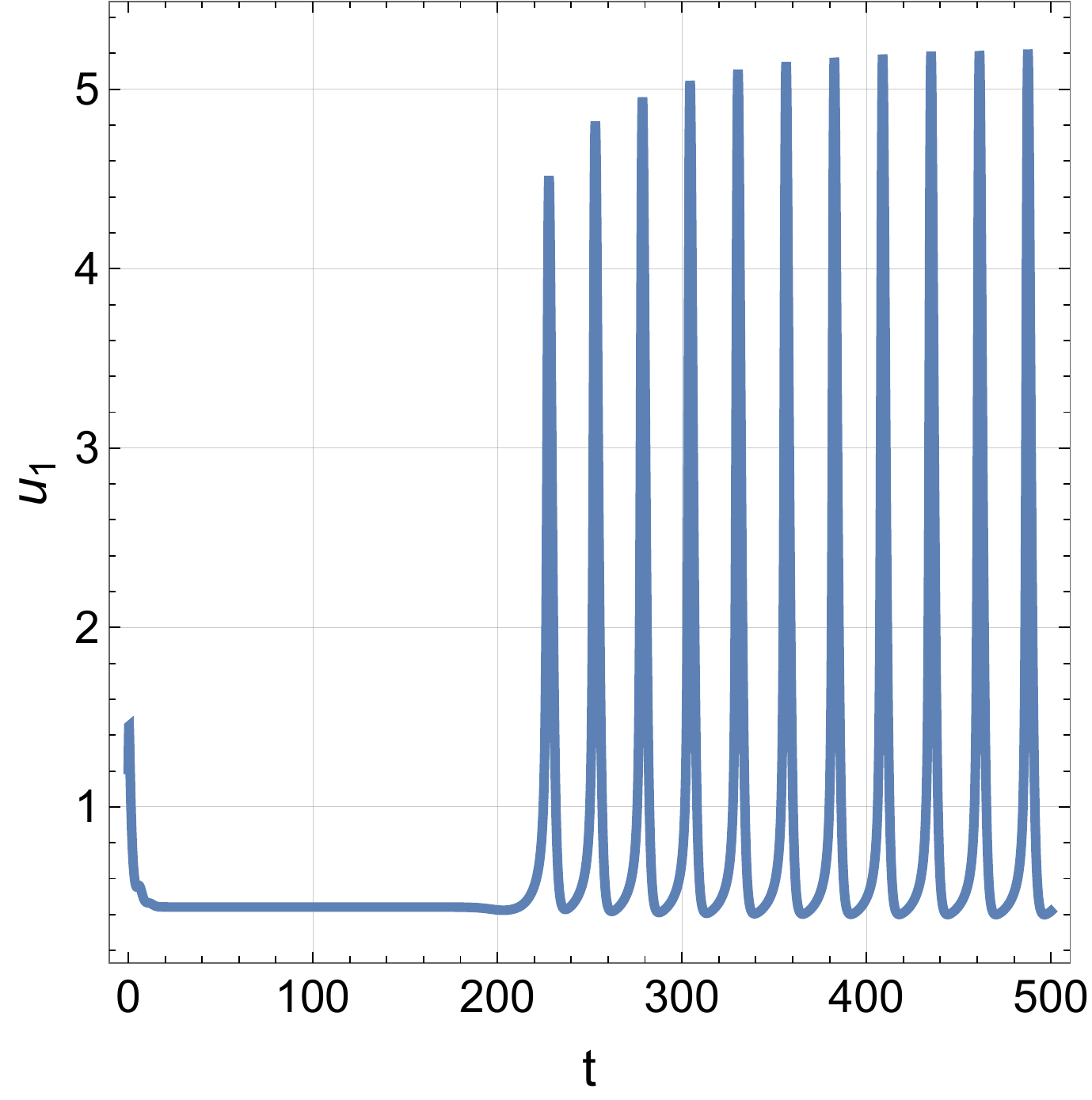}
		(d) 
		\includegraphics[width=0.28\linewidth]{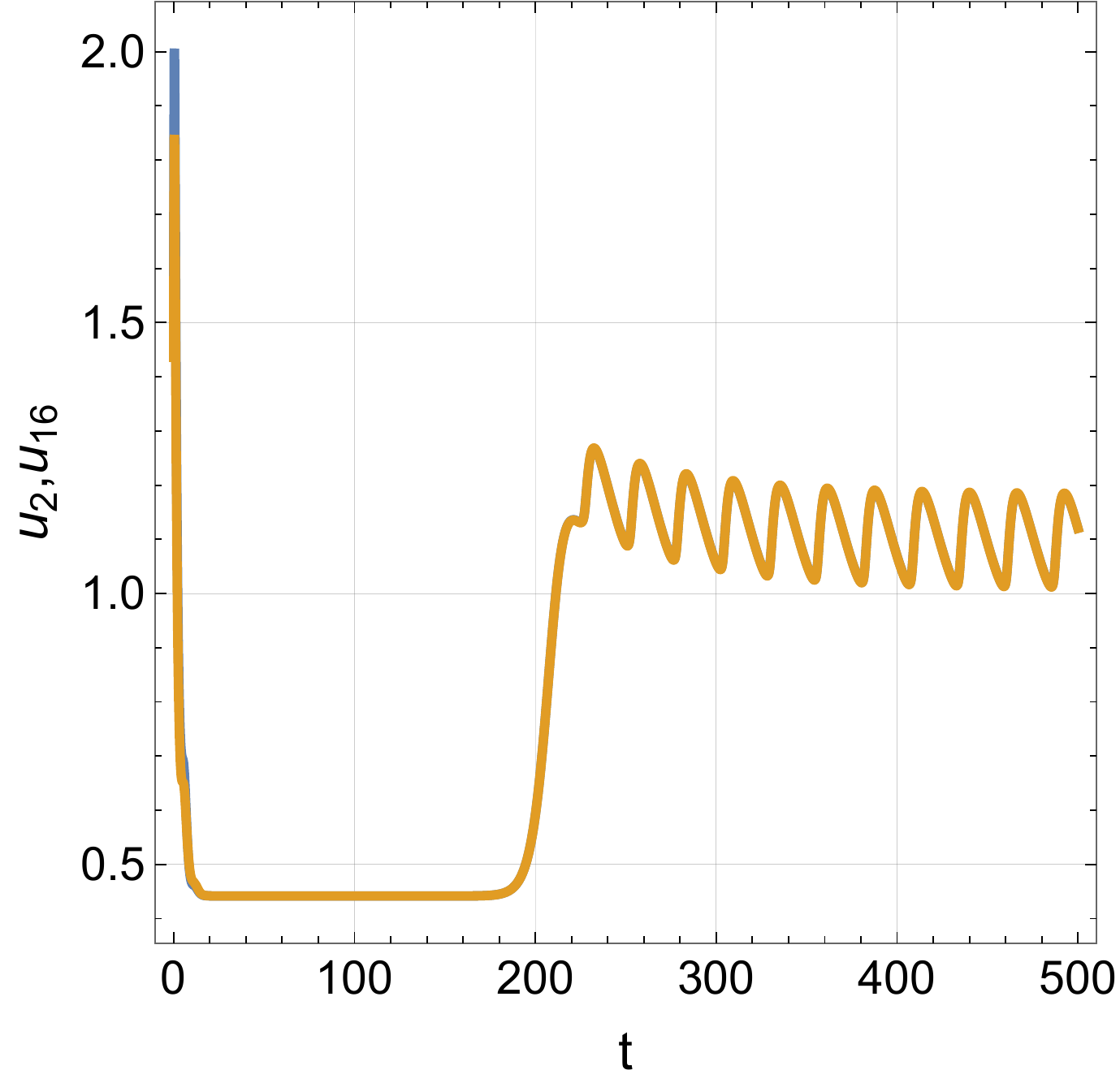}
		(e) 
		\includegraphics[width=0.28\linewidth]{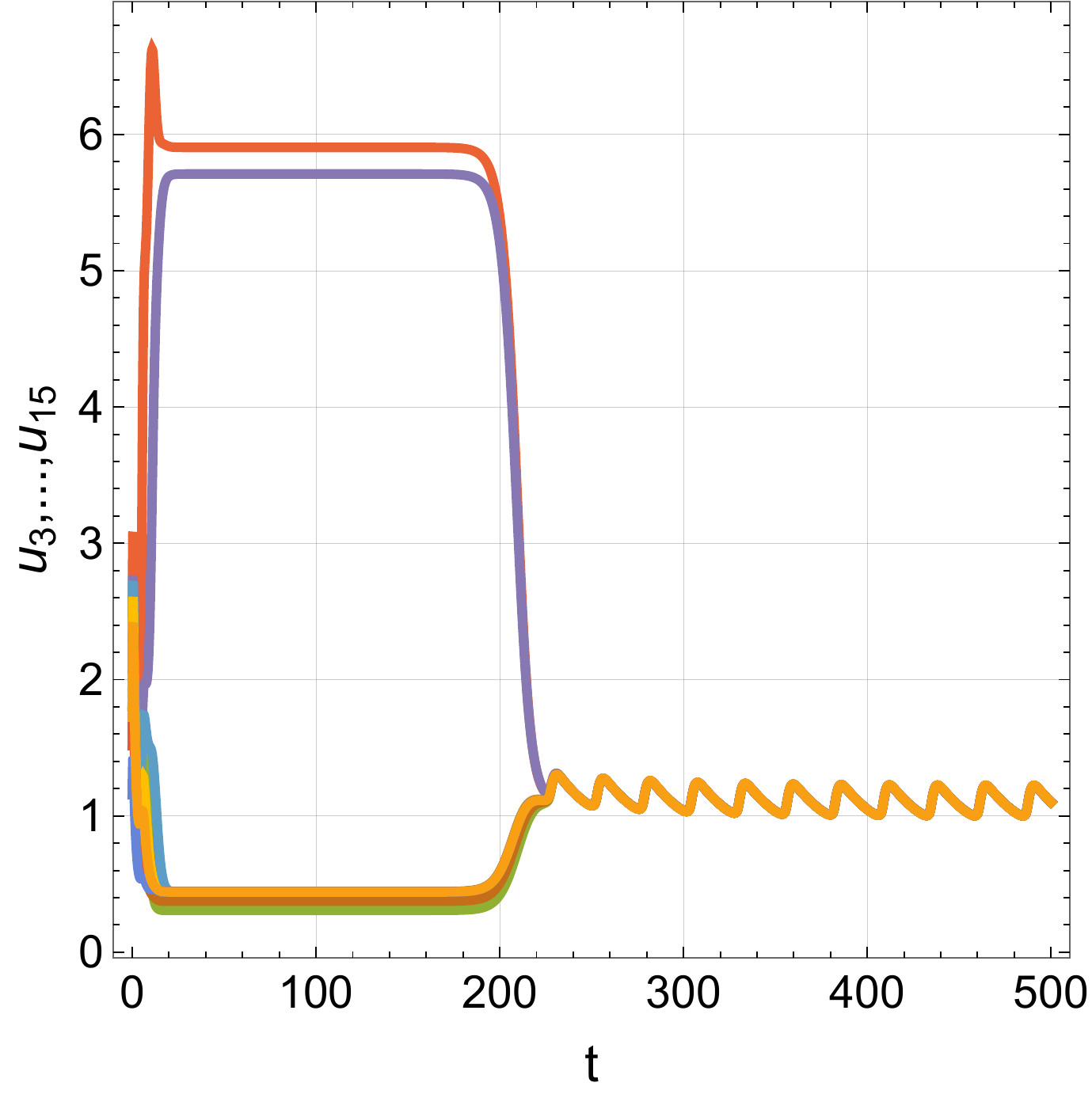}
		(f) \\
		\includegraphics[width=0.23\linewidth]{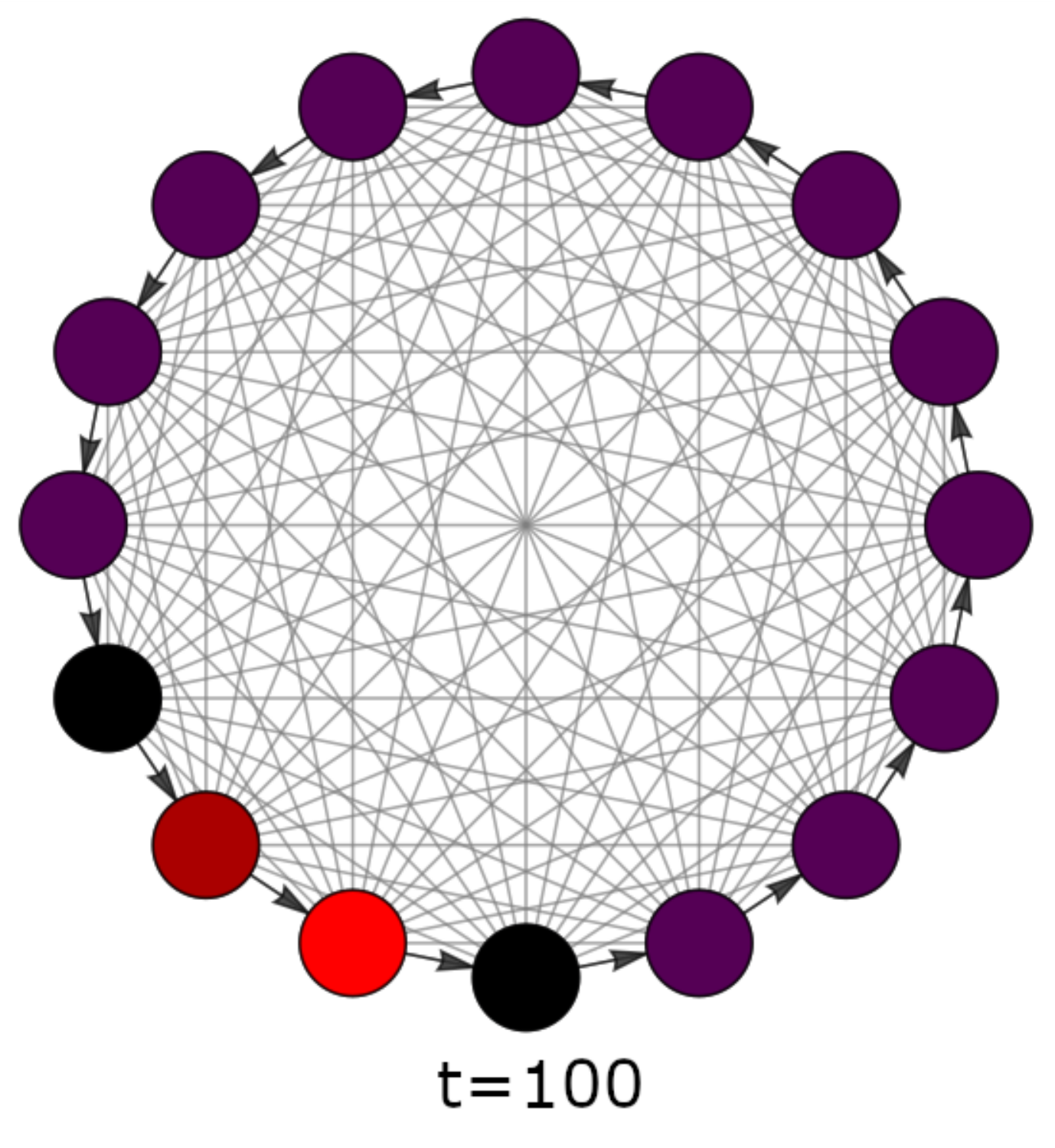}
		\includegraphics[width=0.23\linewidth]{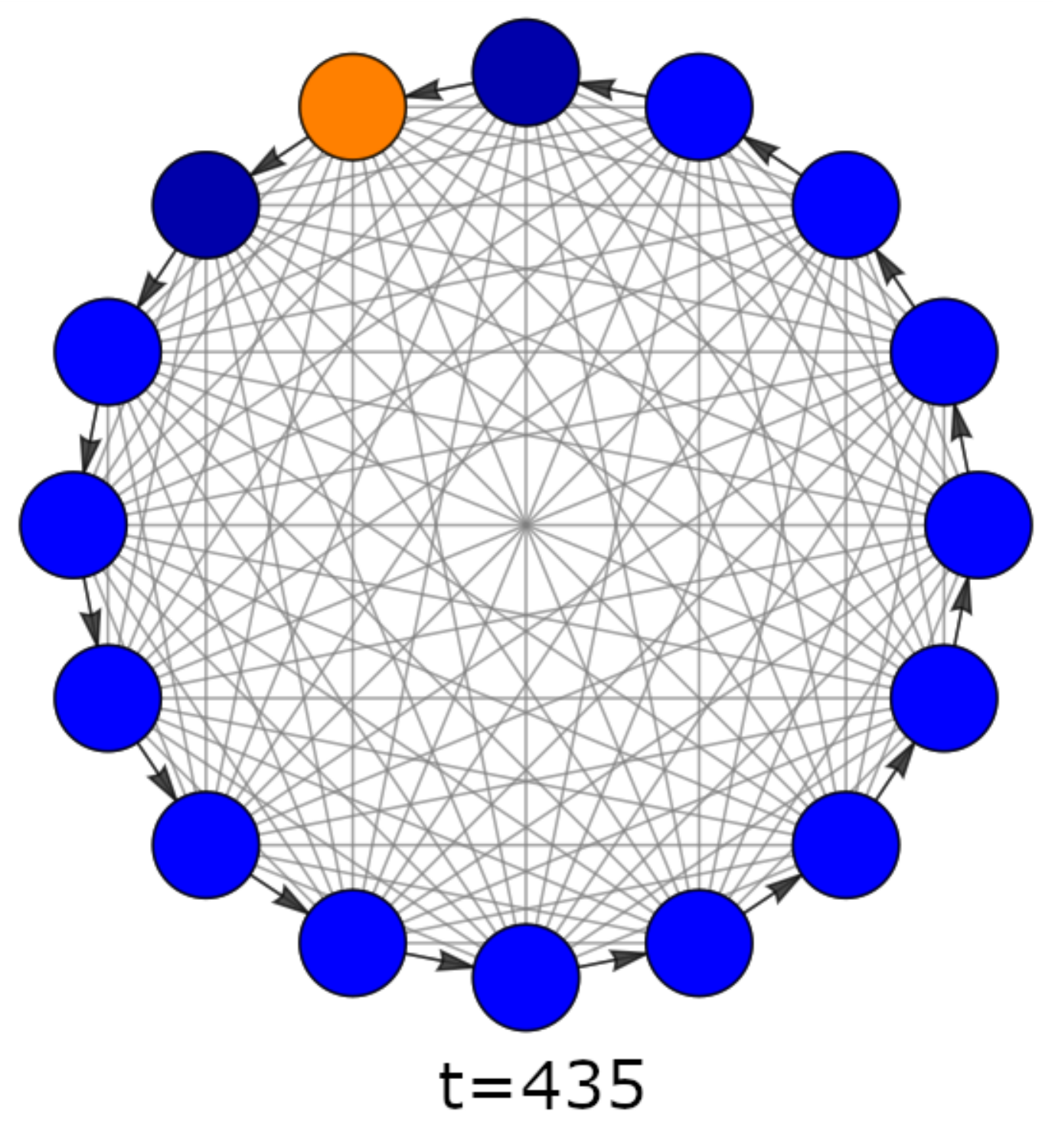}
		\includegraphics[width=0.23\linewidth]{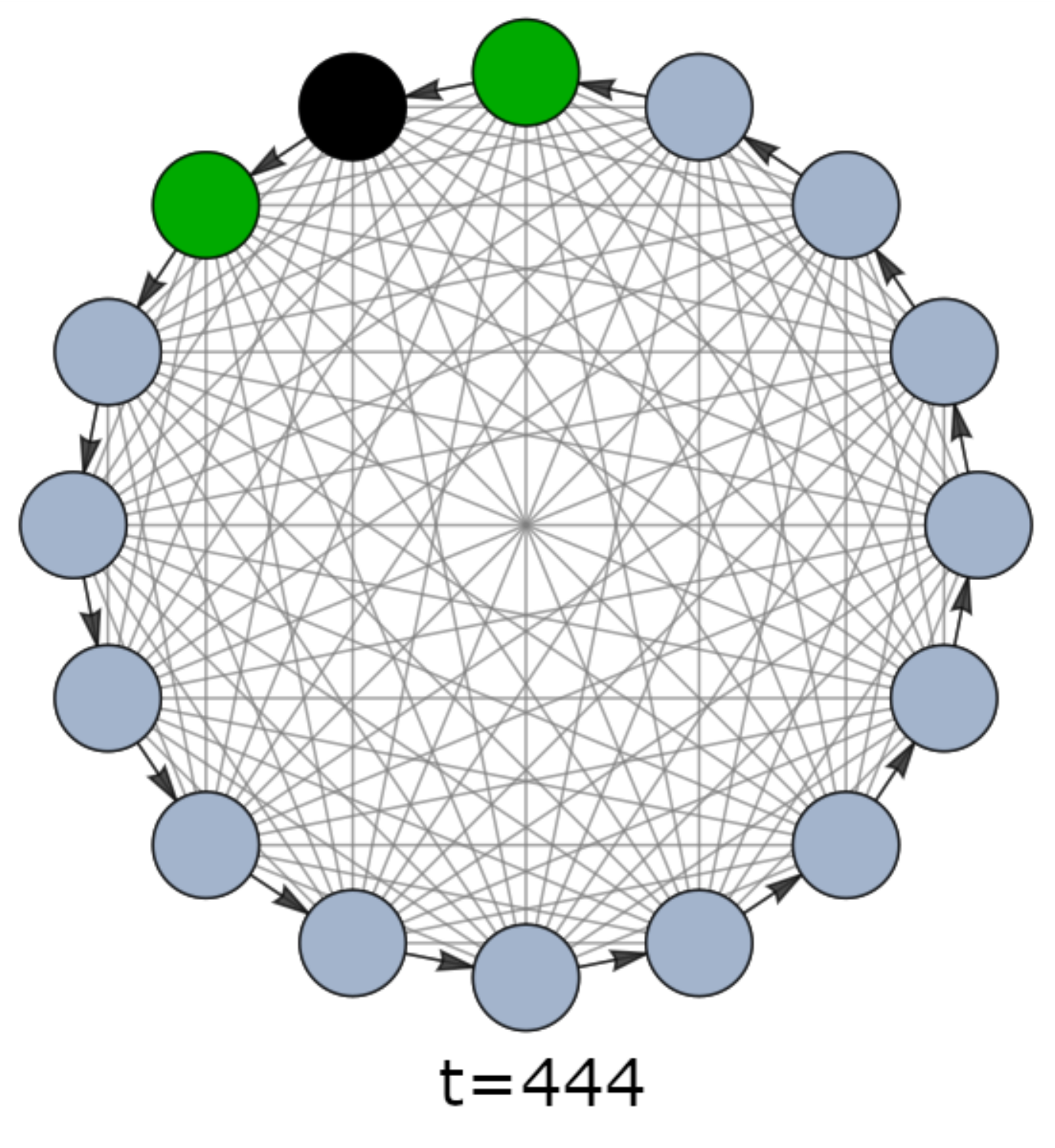}
		\includegraphics[width=0.23\linewidth]{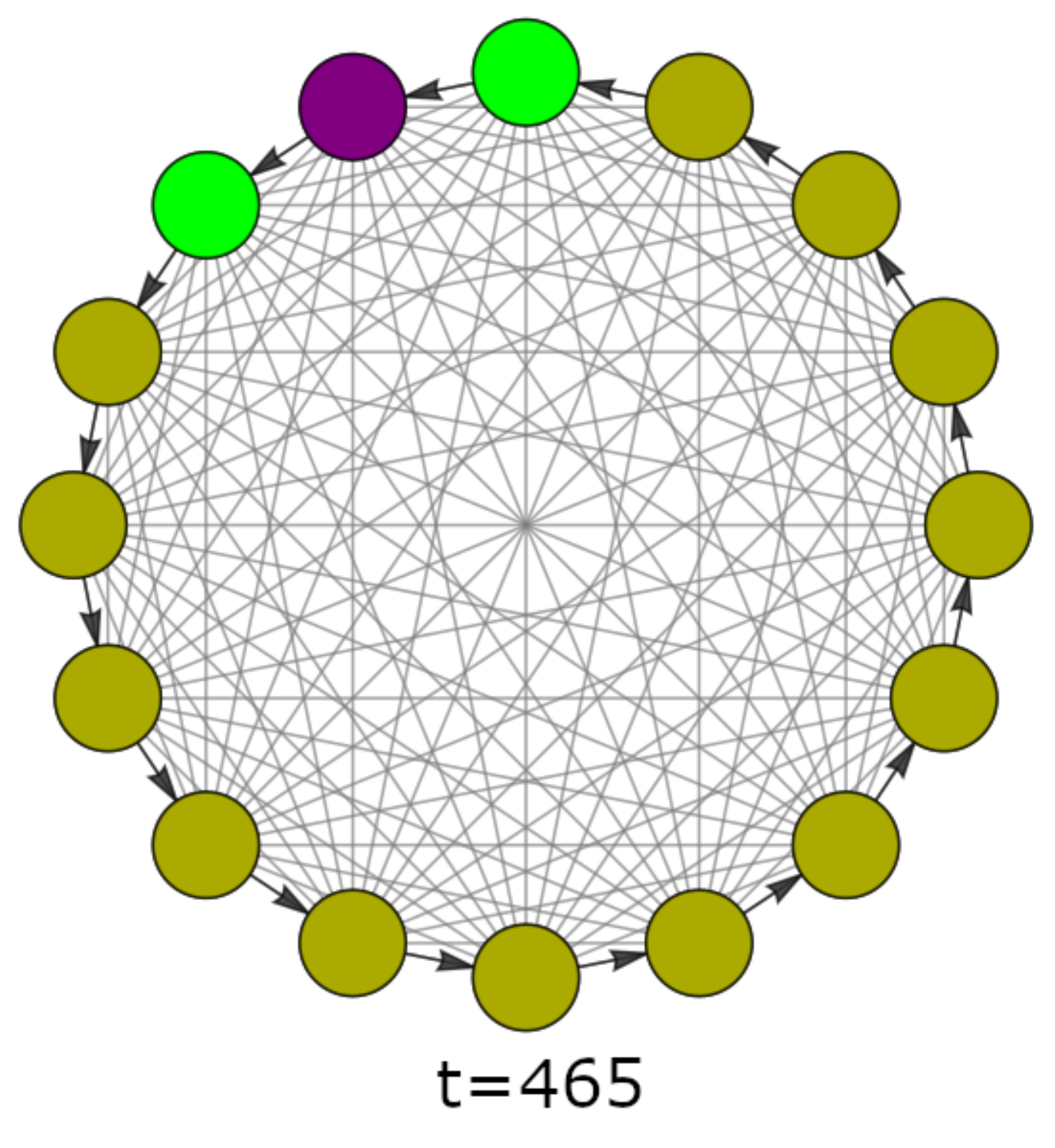}
		(g) \\
		\includegraphics[width=0.55\linewidth]{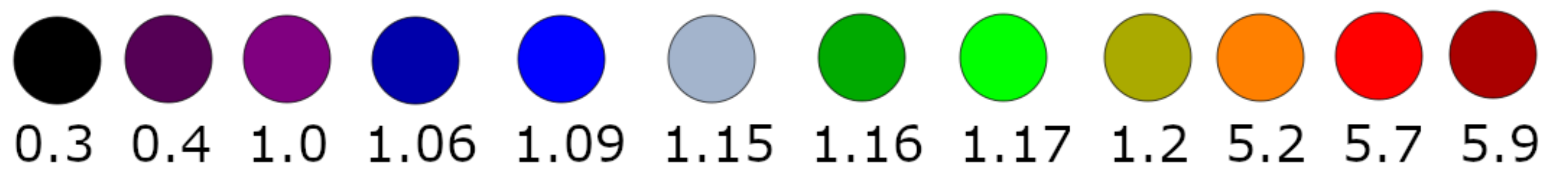}
		\caption{Patterning corresponding to \Eqsref{Eq:u_t}--\eqref{Eq:Adjacency_t} with $h(t)=1$, and where $\beta_{i}(t)$ is given by \Eqref{Eq:beta_t} for $i=1,\dots,n=16$. In (a) we plot the function $1/2 - \Omega(t,200)$. In (b) we show the numerically calculated quantity $\ln(w)$. In (d) we plot the function $u_1$, in (e) we plot the functions $u_2$ and $u_{16}$, and in (f) we show the functions $u_{3}$--$u_{15}$. Finally, in (g) we show the values of $u_{i}$ on the graph at $t=100,435,444,465$.}
		\label{fig:h_t}
	\end{figure}

	For our second example of global reaction kinetics, we set $h(t)=1$. Recall that, for this particular choice of $h(t)$, the underlying network is a complete cycle graph. Moreover, we set 
	\begin{align}
		\beta_{i}(t)=
		\begin{cases}
			1, &\text{if}\quad i=1, \\
			1/2 - \Omega(t,200), &\text{otherwise}.
		\end{cases}
		\label{Eq:beta_t}
	\end{align}
	A plot of the function $1/2 - \Omega(t,200)$ is given in \subfig{fig:h_t}{(a)}. In \subfig{fig:h_t}{(a)} we see that $1/2 - \Omega(t,200)\approx 1$ for $t\in (0,200)$, and $1/2 - \Omega(t,200)\approx 0$ for $t>200$. When $t<200$ the unknowns are allowed to interact at all of the nodes. In fact, for this region, the reaction kinetics are \emph{local}, not global. However, for $t>200$ the reactions become zero at all nodes, excluding node $1$. In this region the reaction kinetics are global. This is therefore an example that begins with local reaction kinetics that evolve into global ones.

	The numerical results corresponding to these choices are shown in \Figref{fig:h_t}.	In \subfig{fig:h_t}{(b)} we show the numerically calculated quantity $\ln(w)$. For $t<200$ we see that $w$ is an increasing function. This suggests that a stable pattern forms, at least while $t<200$. This behaviour changes when the reaction kinetics become global, which occurs for $t>200$. Here, we find that the quantity $w$ continues to grow. However, in addition to growth we can also clearly see small oscillations. These oscillations are also shown in \subfig{fig:h_t}{(c)}. We interpret this as follows: For $t>200$ the system experiences a \emph{Turning-wave} instability. Exactly this behaviour is shown in \subfig{fig:h_t}{(d)--(f)}. Here, as in \Sectionref{Sec:Pattern_formation_with_global_reaction_kinetics}, we find that the values of $u_i$ naturally split into three groups based on how they oscillate. The first group contains $u_1$ only and is shown in \subfig{fig:h_t}{(d)}. In \subfig{fig:h_t}{(e)} we show the second group, which consist of $u_2$ and $u_{16}$. In \subfig{fig:h_t}{(f)} we show the third group, which contains $u_{3}$--$u_{15}$. Finally, in (g) we show the values of $u_{i}$ on the graph at $t=100,435,444,465$.

	\subsection{Pattern formation with time-dependent diffusion coefficients}
	Another way in which non-autonomous reaction diffusion equations can on static networks can occur, is if the diffusion coefficients themselves depend on time. The purpose of this subsection here it provide one such example. To this end we consider the \emph{hyperbolic reaction-diffusion} system
	\begin{align}
		\frac{d^2 u_{i}}{dt^2}+\frac{du_{i}}{dt} &= d(t)\sum_{j=1}^{10}A_{ij}\left( u_{j}-u_{i} \right) + 5 - 10u_{i} + u_{i}^{2}v_{i},
		\label{Eq:HypEqs_u_t}
		\\
		\frac{d^2 v_{i}}{dt^2}+\frac{dv_{i}}{dt} &=  \sum_{j=1}^{10}A_{ij}\left( v_{j} - v_{i} \right) + 9u_{i} - u_{i}^{2}v_{i},
		\label{Eq:HypEqs_v_t}
	\end{align}
	\begin{figure}
		\centering
		\includegraphics[width=0.275\linewidth]{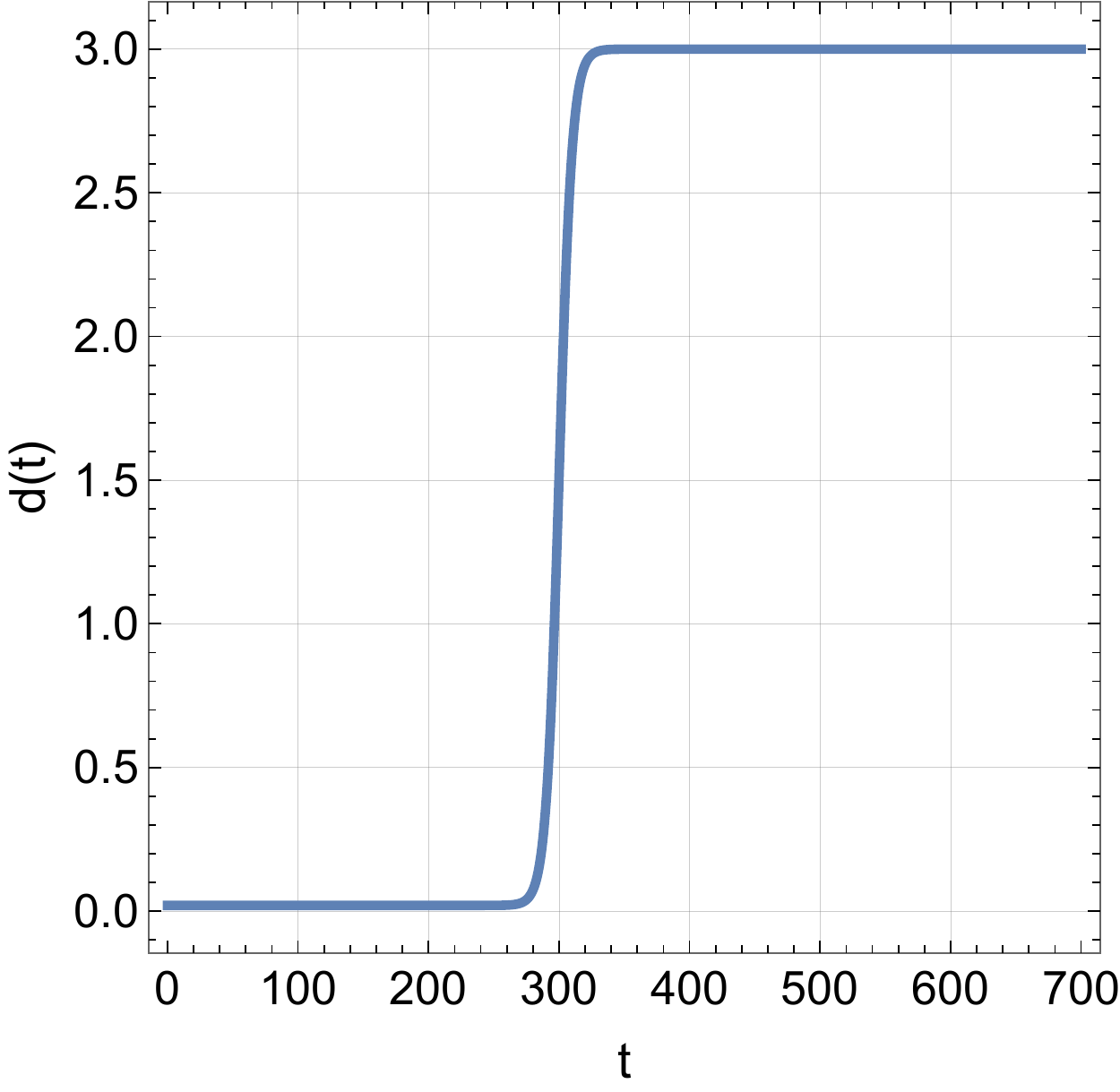}
		(a)
		\includegraphics[width=0.28\linewidth]{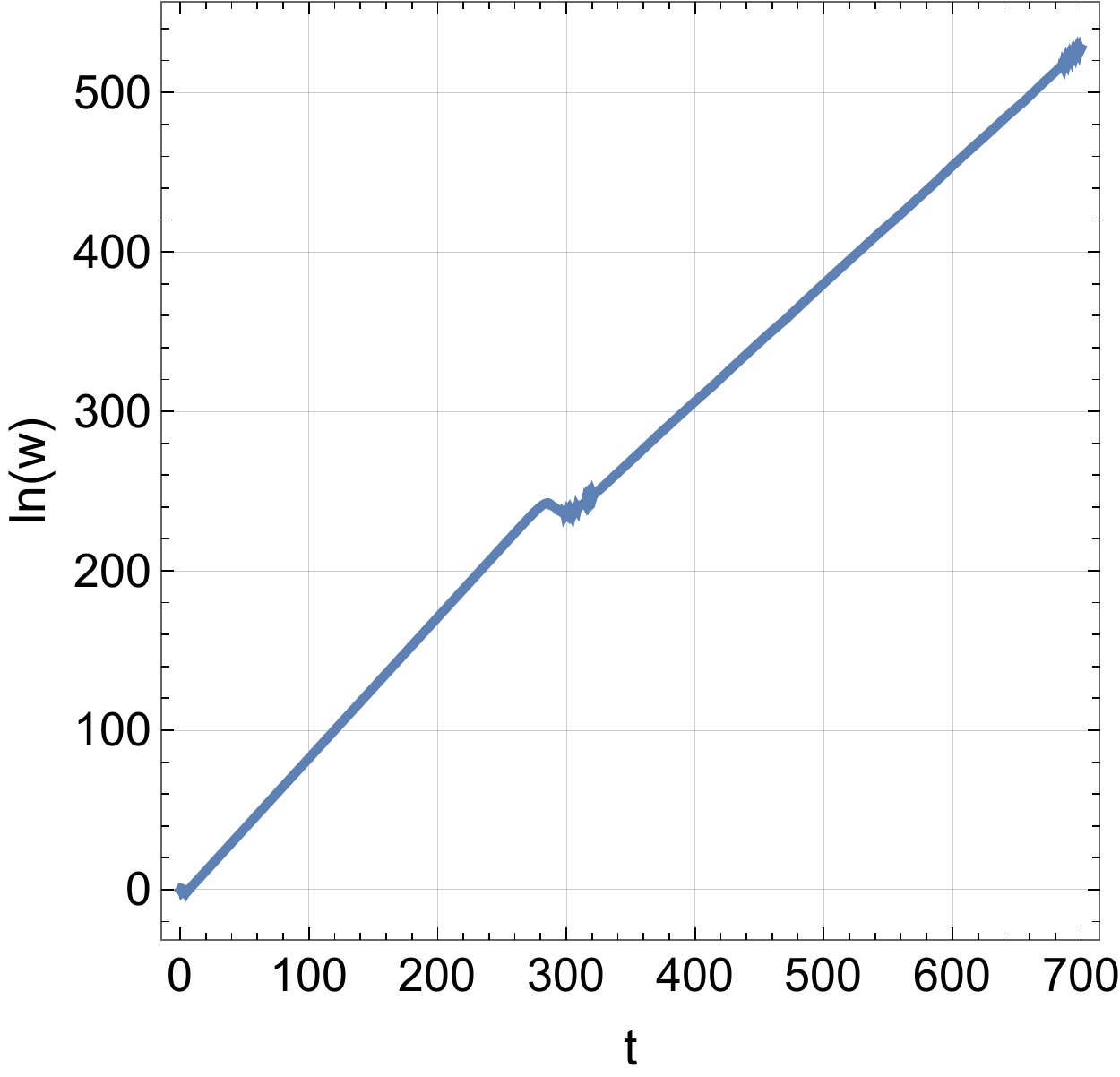}
		(b)
		\includegraphics[width=0.28\linewidth]{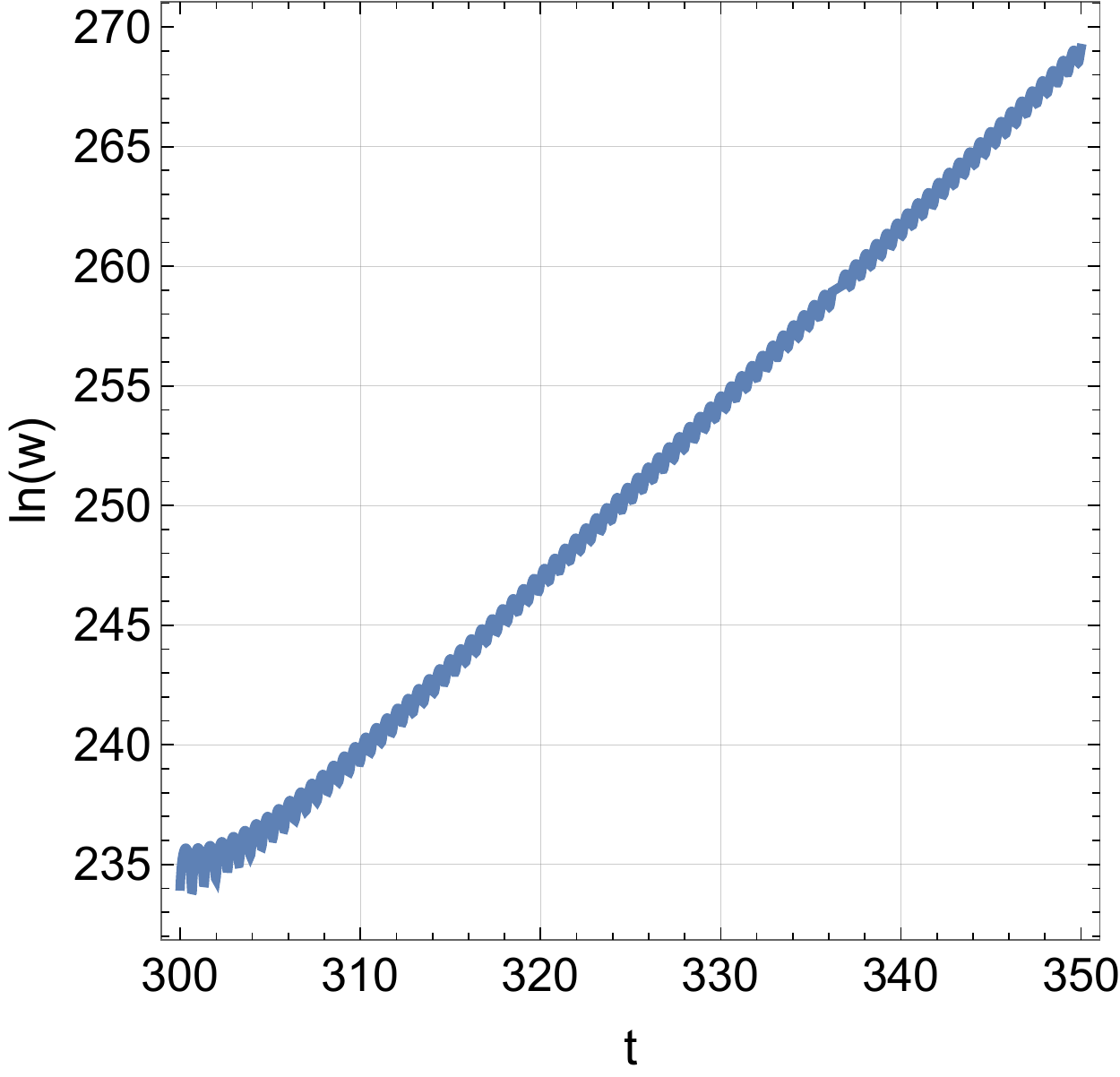}
		(c) \\
		\includegraphics[width=0.28\linewidth]{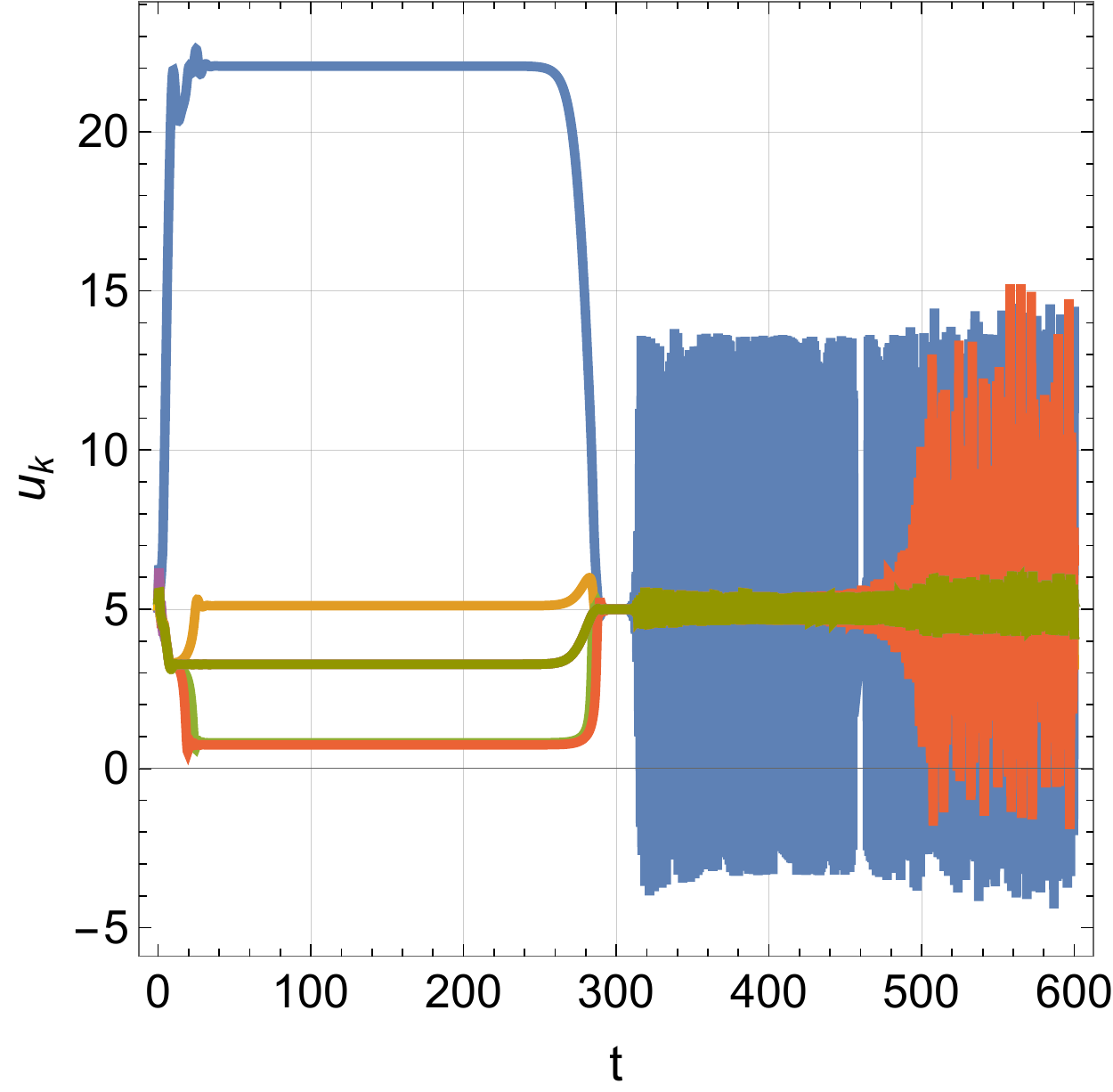}
		(d)
		\includegraphics[width=0.28\linewidth]{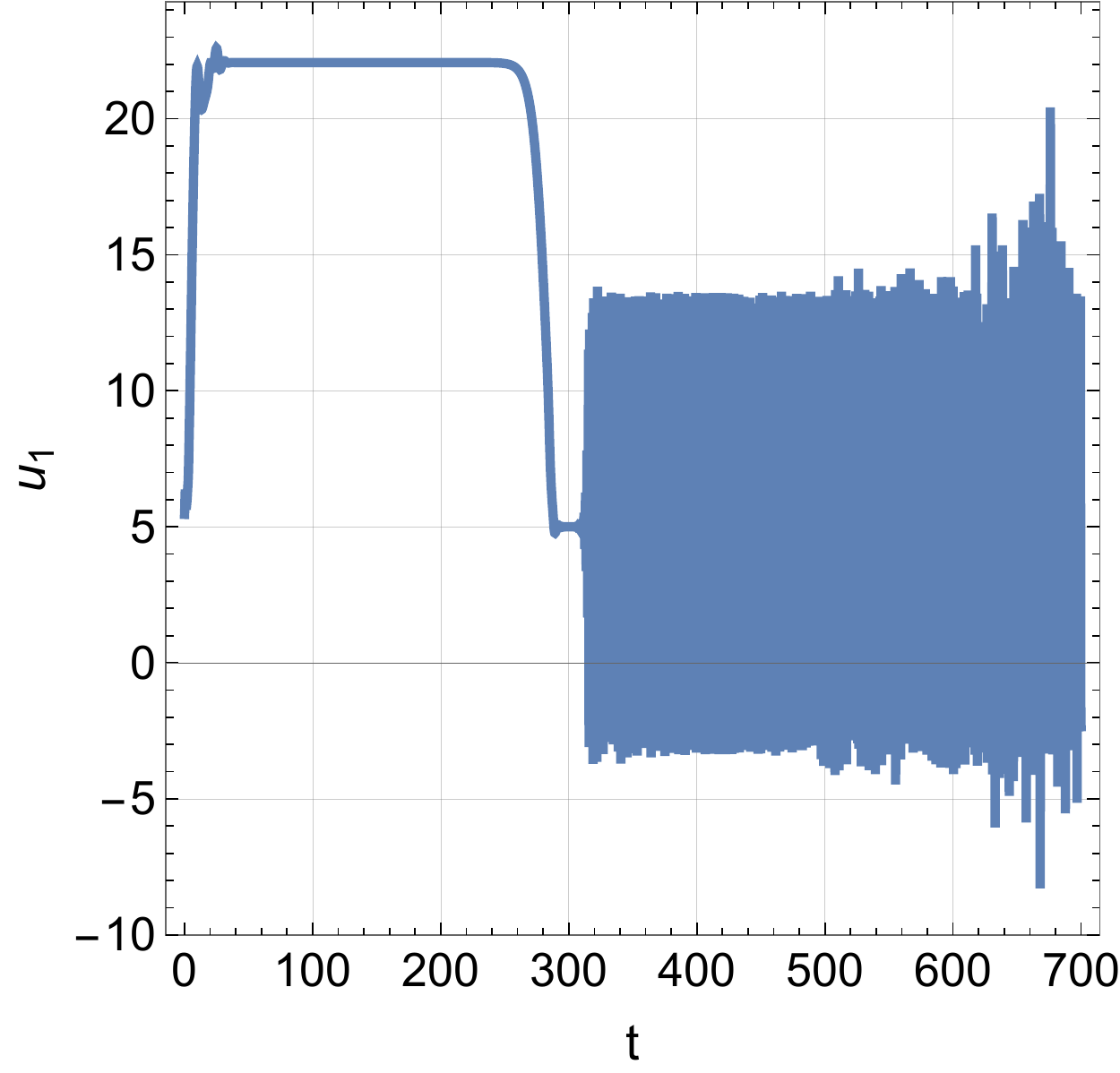}
		(e)
		\includegraphics[width=0.28\linewidth]{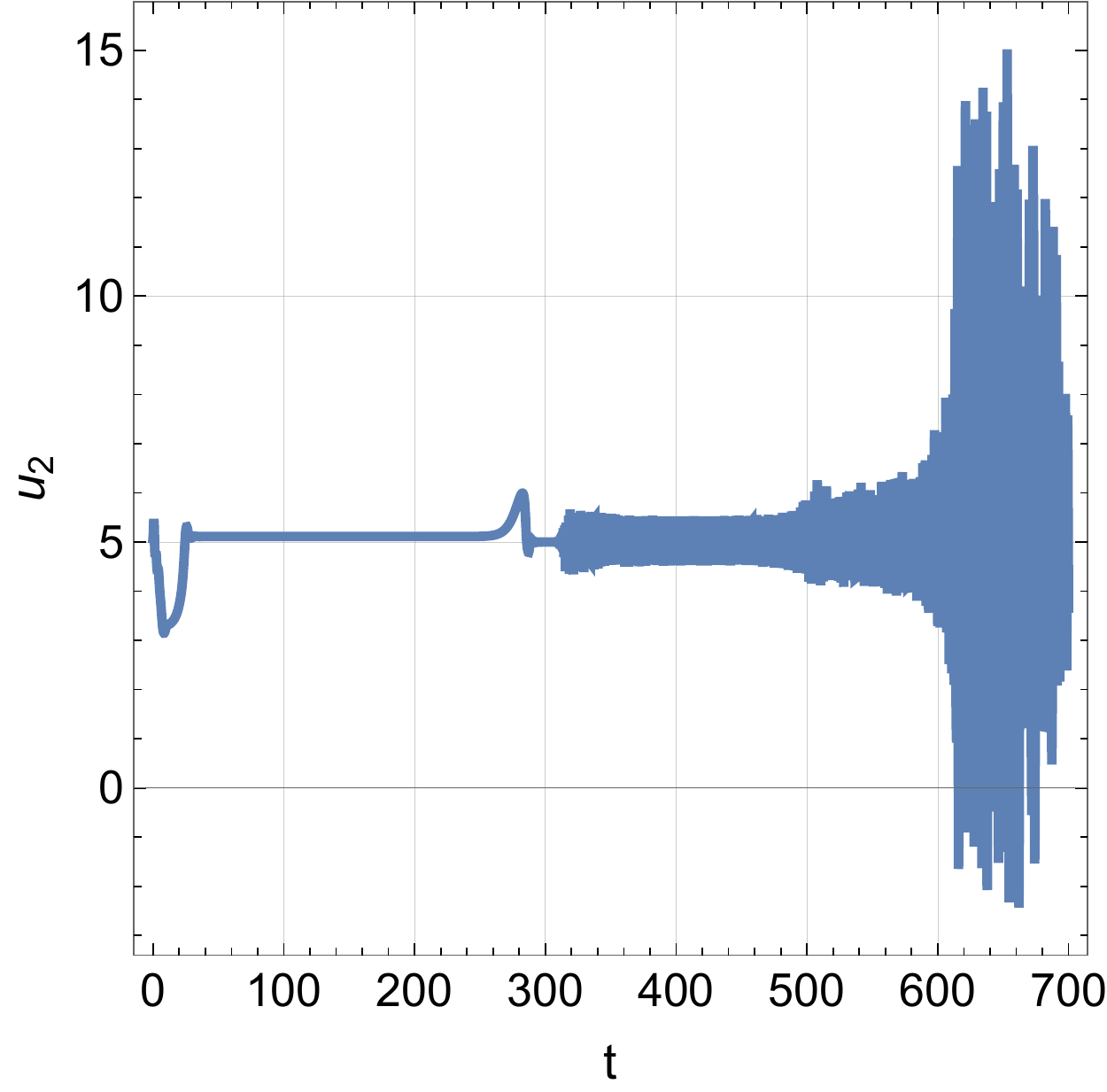}
		(f) \\
		\includegraphics[width=0.28\linewidth]{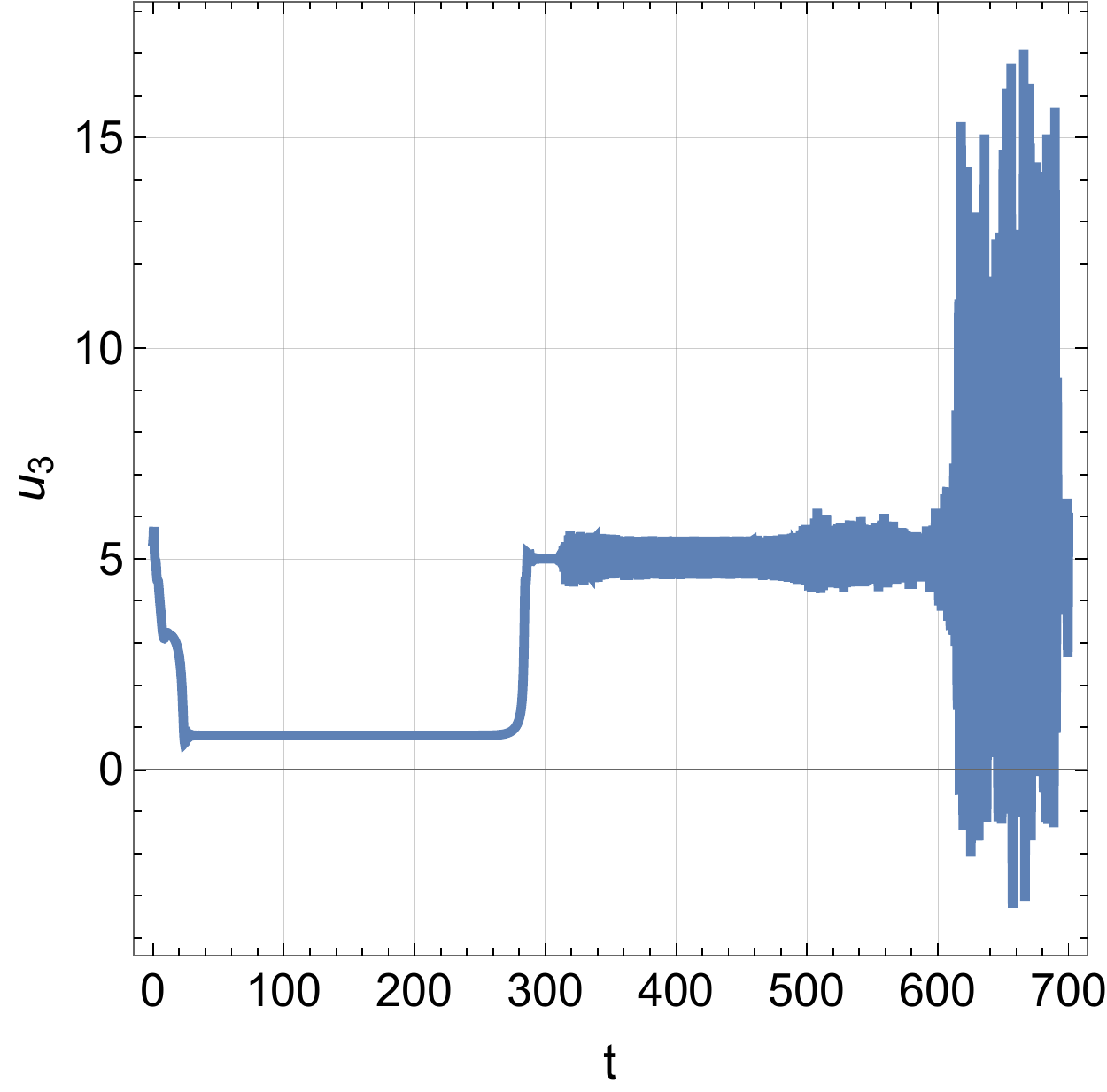}
		(g)
		\includegraphics[width=0.28\linewidth]{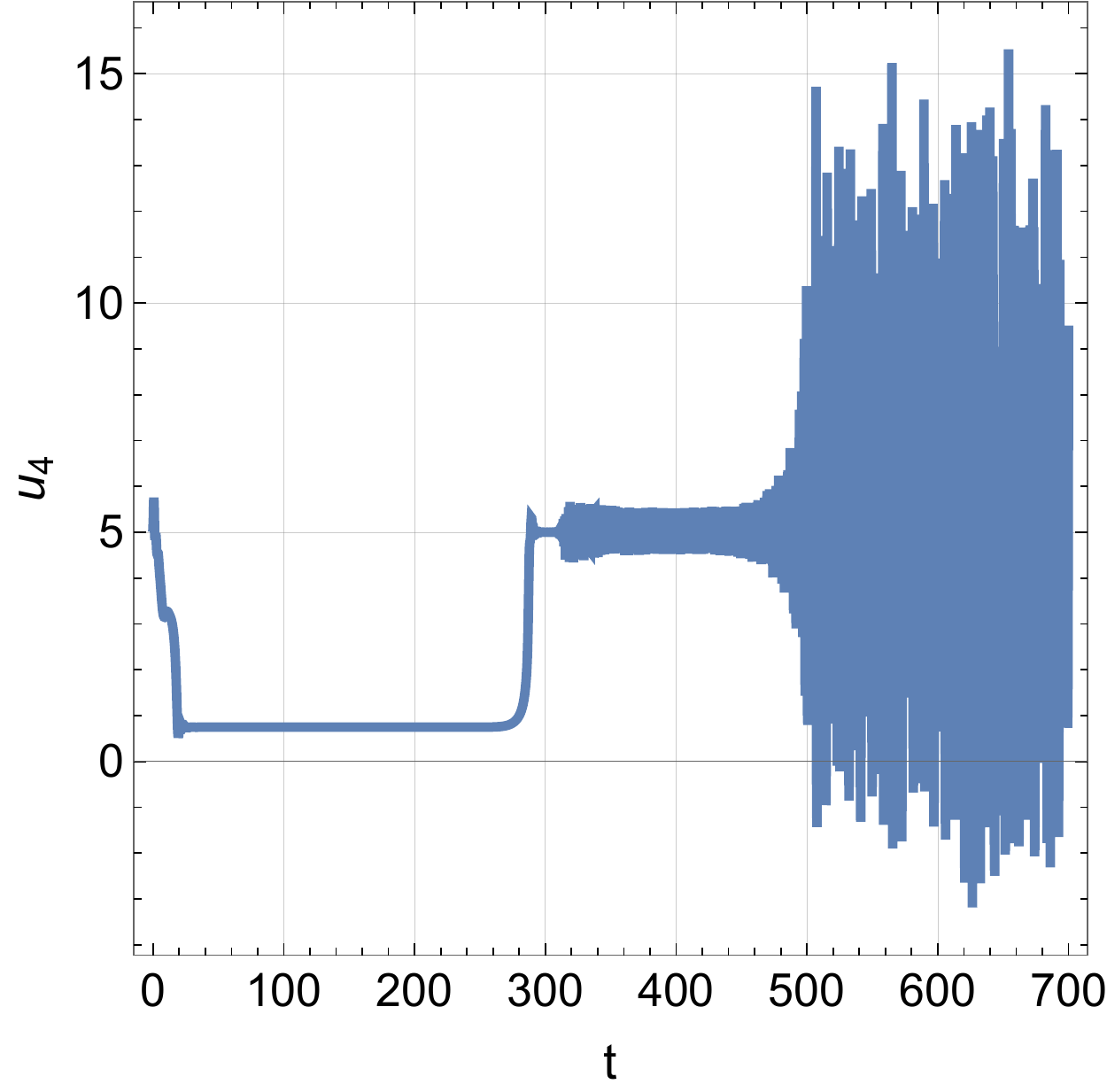}
		(h)
		\includegraphics[width=0.28\linewidth]{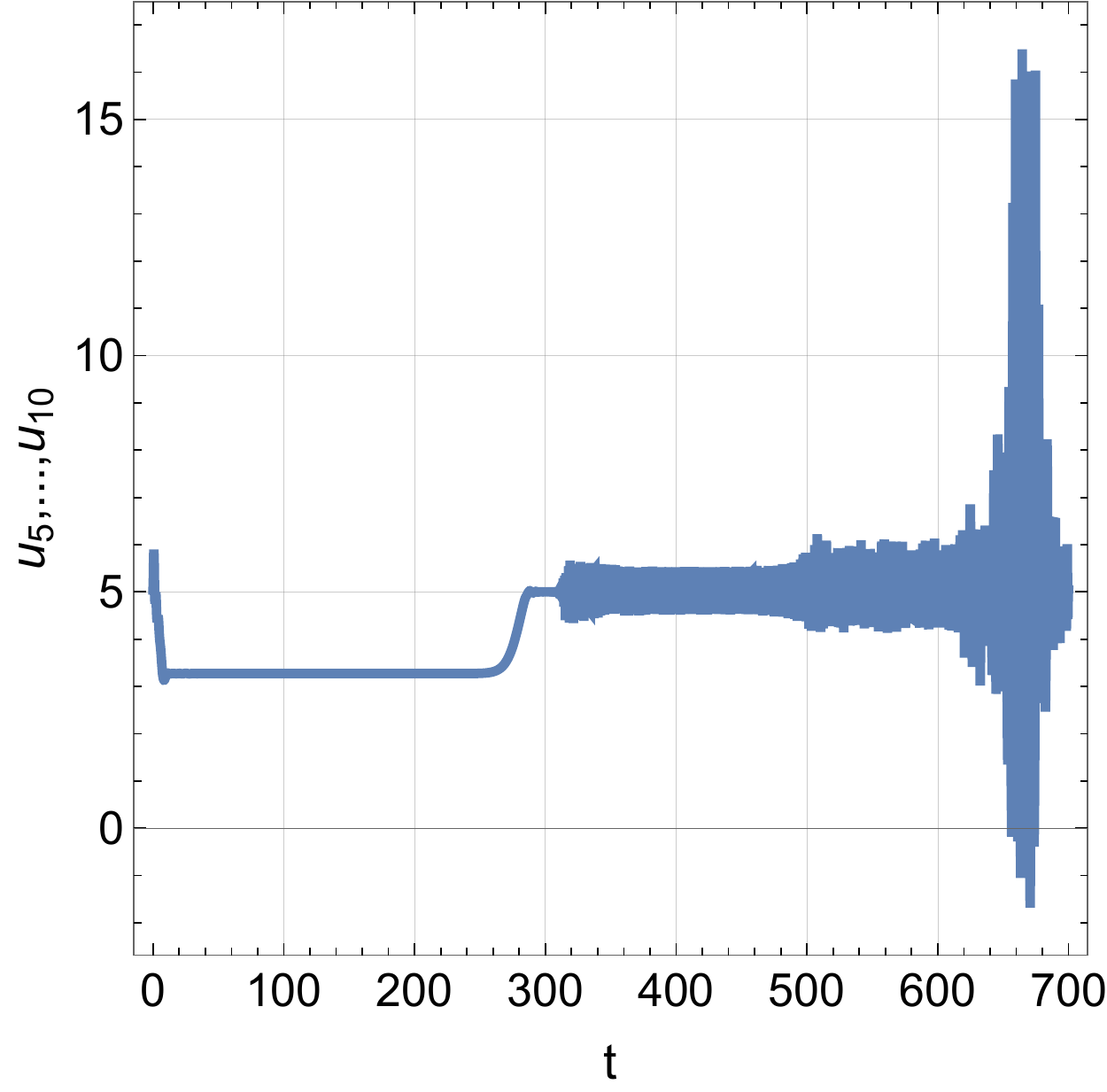}
		(i) \\
		\includegraphics[width=0.23\linewidth]{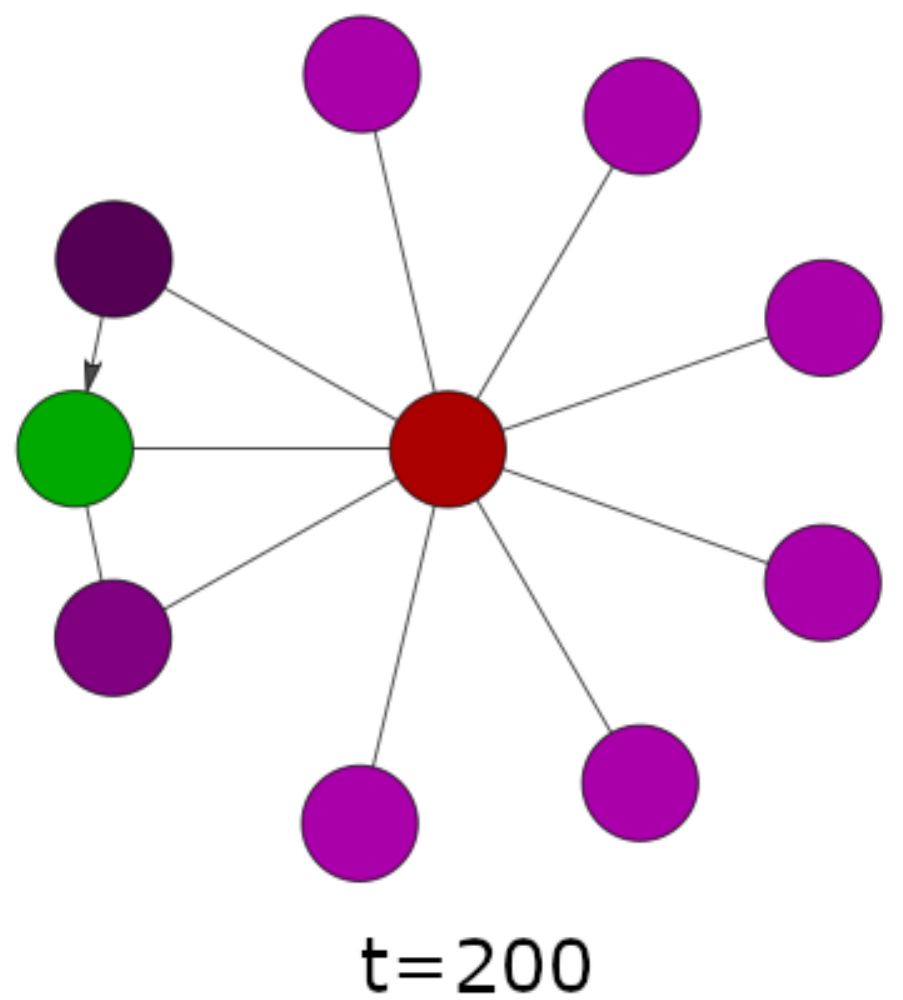}
		\includegraphics[width=0.23\linewidth]{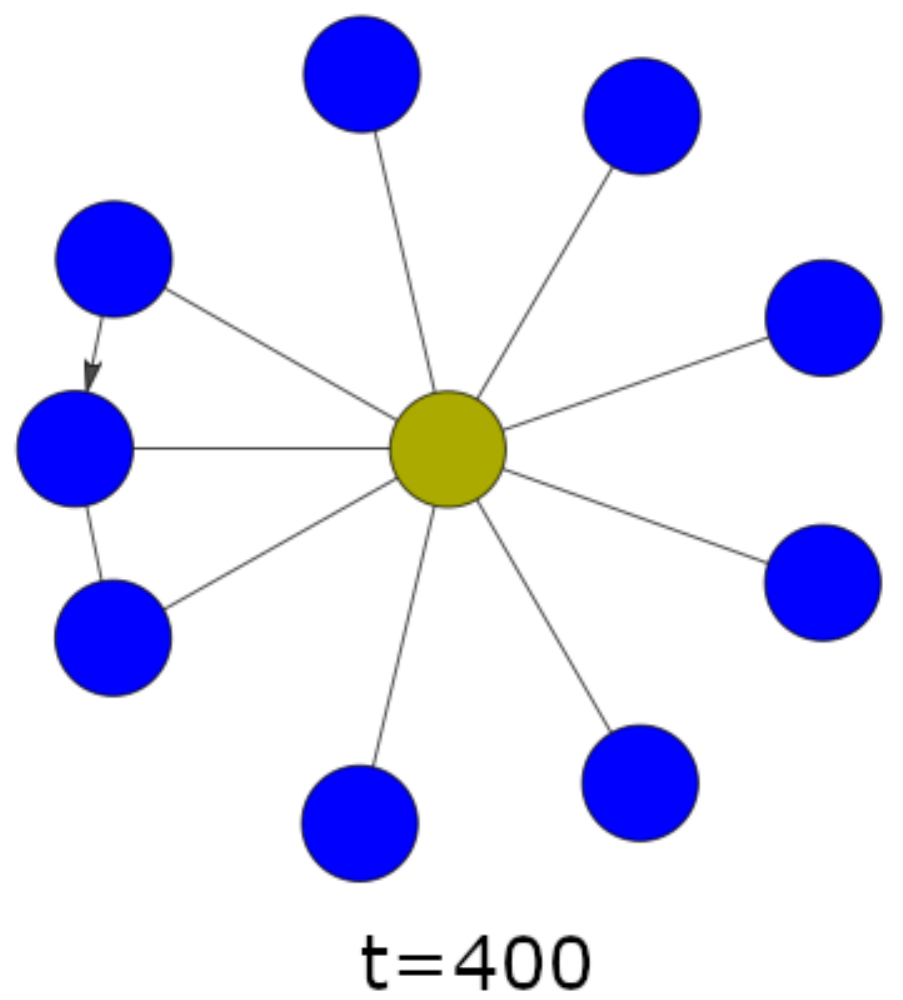}
		\includegraphics[width=0.23\linewidth]{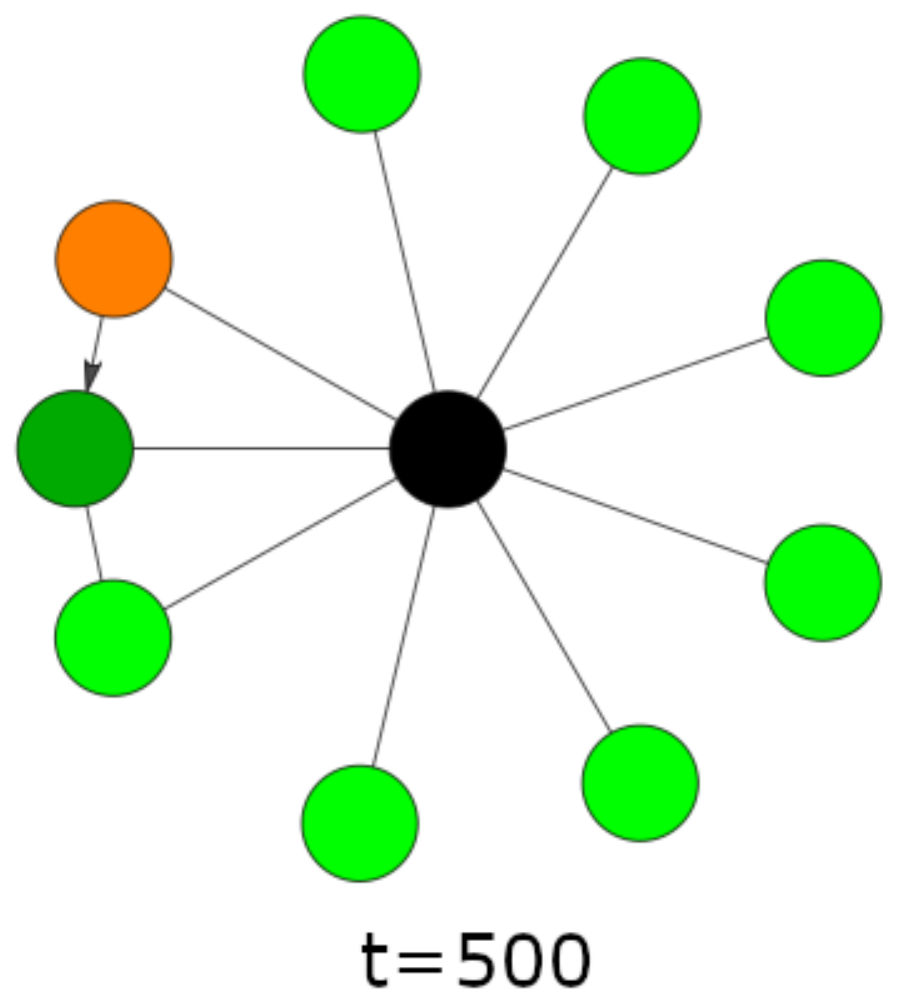}
		\includegraphics[width=0.23\linewidth]{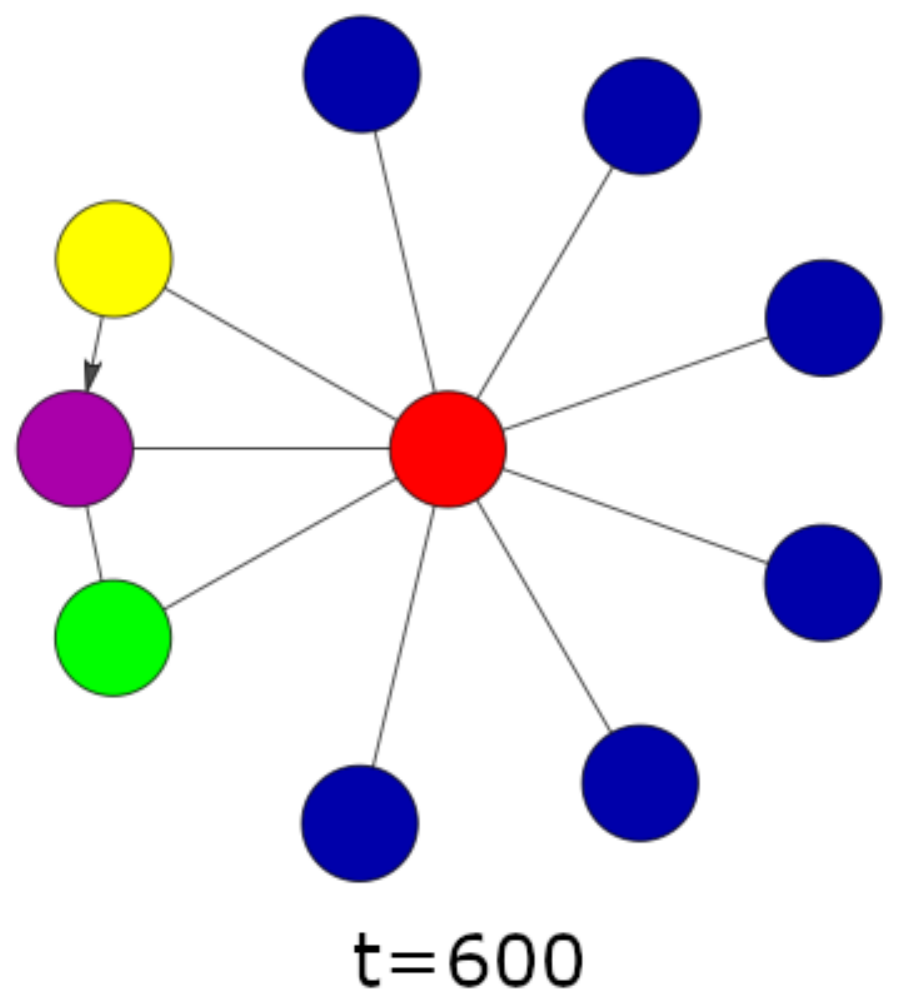}
		(j)	
		\\
		\includegraphics[width=0.55\linewidth]{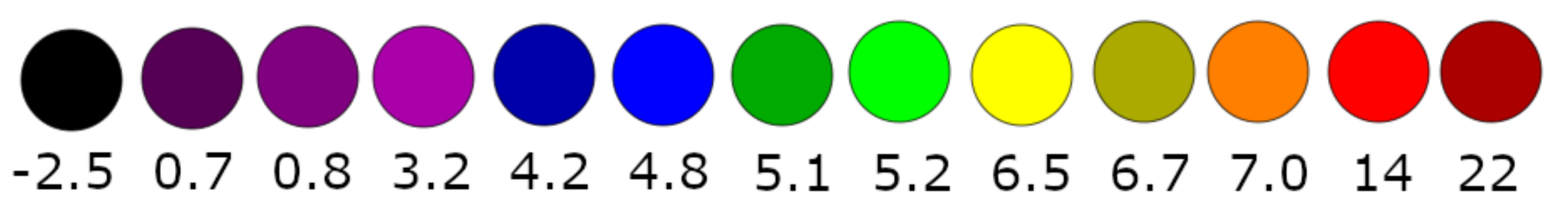}
		\caption{Patterning corresponding \Eqsref{Eq:HypEqs_u_t} and \eqref{Eq:HypEqs_v_t} with $A$ given by \Eqref{Eq:Adjacency_Hyp_1}. In (a) we plot the diffusion coefficient $d(t)$. In (b) and (c) we show $\ln(w)$. In, (d) we plot the the functions $u_{k}$ and in plots (e)--(h) we plot $u_1,u_2,u_3$, and $u_4$ respectively, and, in (i), we plot $u_{5},\dots,u_{10}$. Finally, in (j) we give the pattern for the functions $u_k$ at $t=200,400,500,600$.}
		\label{fig:d}
	\end{figure}
	where $A_{ij}$ is given by \Eqref{Eq:Adjacency_Hyp_1}, and $d(t)=149( 1 + 2\Omega(t,300) )/100$. Recall again that this particular choice of adjacency matrix describes a ``modified star graph''. A plot of the diffusion coefficient $d(t)$ is shown in \subfig{fig:d}{(a)}. For $t\in(0,300)$ we have that $d\approx 1/50$, and for $t>300$ we have that $d(t)\approx 3$. The numerical results for this system are shown in \Figref{fig:d}. In \subfig{fig:d}{(b)} we show the numerically calculated quantity $\ln(w)$. Here we see that $w$ is increasing on the entire numerical domain and hence we expect a pattern to form for all $t\in(0,700)$. For $t<300$, $w$ increases linearly. Conversely, for $t>300$ the function $w$ oscillates as it increases. These oscillations are shown in \subfig{fig:d}{(c)}. This tells us that the systems experiences a Turing instability for $t<300$ and a \emph{Turing-wave} instability for $t>300$. This behaviour is shown in \subfig{fig:h_t}{(d)}. Here, as in \Sectionref{Sec:Pattern_formation_from_hyperbolic_reaction_diffusion_equations}, we find that the values of $u_i$ naturally split into five groups based on how they oscillate. The first four groups contain one element only. Namely, $u_1,u_2,u_3$ and $u_4$. Each of these functions are shown in \subfig{fig:h_t}{(e)},\subfig{fig:h_t}{(f)},\subfig{fig:h_t}{(g)}, and \subfig{fig:h_t}{(h)}, respectively. In \subfig{fig:h_t}{(i)} we show the fifth group, which consist of $u_2--u_{10}$. Finally, in (j) we show the values of $u_{i}$ on the graph at $t=200,400,500,600$.

	\section{Conclusions}
	\label{Sec:Conclusions}
	In this work we have established a method for determining instability, leading to pattern formation, on directed networks. We started by investigating systems of autonomous reaction-diffusion equations on static networks. One of our primary focuses here was to establish a method for determining instability (leading to pattern formation) on non-diagonalizable networks. That is, networks whose Laplacian matrix is non-diagonalizable. In this setting, the eigenvectors of the Laplacian matrix do not form a complete orthonormal basis over the network. Thus, in order to perform an instability analysis, we first had to pick a basis. To do this we symmetrized the adjacency matrix. In doing so we are able to express our original system as the equivalent problem of solving a system of reaction-diffusion equations on an \emph{undirected network} with \emph{global reaction kinetics}. The basis was then constructed as the eigenvectors of the `symmetrized Laplacian', i.e., the Laplacian calculated from the symmetrized adjacency matrix. We were able to show that this approach reduces to the standard Turing analysis in the case of autonomous reaction-diffusion equations on undirected networks with local reaction-kinetics. In addition to static non-diagonalizable networks we also discussed \emph{global reaction kinetics}. This line of investigation rose naturally from our method for constructing a basis. Here we found that global reaction kinetics allowed for Turing-pattern formation to occur on networks that did \emph{not} allow for patterning from reaction-diffusion equations with local reaction kinetics. The approach that we use here has the drawback that our instability condition must be implemented numerically. However, our instability condition can nevertheless be applied to a large class of reaction-diffusion equations on simple networks, without any restrictions on the network structure. In future works it would be interesting to extend our approach here to more complicated networks, such as multiplex networks (see, for example, \cite{Asllani:2014_2}). 
	
	In addition to this we also investigated pattern formation arising from non-autonomous reaction-diffusion equations on \emph{temporal} networks. That is, networks that are allowed to change in time. Our primary focus here was to once again establish a method to determine instability (leading to pattern formation) on temporal networks. In order to determine linear instability we studied the linearized equations (corresponding to the original system) directly, without any basis decompositions. This approach required us to numerically solve linearized equations. This raises the question \emph{why not solve the original, non-linear, system instead?} After all, in either setting one must solve a system of ODE's. One could certainly do this, however, it is worth noting that it is, in general, less computational expensive to solve a system of $m$-linear equations, on an $n$-node network, than a system of $m$-\emph{non-linear} equations. Moreover, this approach is useful if one wants to find what system parameters lead to a Turing-instability. As it is less time consuming to run multiple simulations of the linearized system, than it is to run multiple simulations of the linearized system.

	\newpage
	\bibliographystyle{unsrt}
	\bibliography{bibfile}
	\section*{Statements \& Declarations}
	The author declares that no funds, grants, or other support were received during the preparation of this manuscript. The author has no relevant financial or non-financial interests to disclose.
	\section*{Research Data Availability}
	Data sharing not applicable to this article as no datasets were generated or analysed during the current study.
\end{document}